\documentclass[11pt,reqno]{amsproc}
\usepackage{cite}
\usepackage{color}
\usepackage{float}
\usepackage{psfrag}
\usepackage{caption}
\usepackage{wrapfig}
\usepackage{lettrine}
\usepackage{morefloats}
\usepackage{algorithm}
\usepackage{algpseudocode}
\usepackage[T1]{fontenc}

\usepackage[semicolon,square,authoryear]{natbib}
\numberwithin{equation}{section}
\usepackage[colorinlistoftodos, shadow]{todonotes}
\usepackage{MnSymbol}
\usepackage{stmaryrd}
\usepackage{fullpage}
\usepackage{graphicx}
\usepackage{subfigure}
\usepackage{enumerate}
\usepackage[debug=false, colorlinks=true, pdfstartview=FitV, 
linkcolor=blue, citecolor=blue, urlcolor=blue]{hyperref}
\usepackage{morefloats}

\newlength{\drop}
\definecolor{amethyst}{rgb}{0.6, 0.4, 0.8}
\definecolor{burgundy}{rgb}{0.5, 0.0, 0.13}

\title{A hybrid multi-time-step framework for pore-scale
  and continuum-scale modeling of solute transport in
  porous media}

\author{\textbf{S.~Karimi and K.~B.~Nakshatrala}\\
{\small Department of Civil and Environmental Engineering, 
  University of Houston. \\
  \textbf{Correspondence to:}~knakshatrala@uh.edu}}

\date{\today}

\begin{document}


\begin{titlepage}
    \drop=0.1\textheight
    \centering
    \vspace*{\baselineskip}
    \rule{\textwidth}{1.6pt}\vspace*{-\baselineskip}\vspace*{2pt}
    \rule{\textwidth}{0.4pt}\\[\baselineskip]
         {\LARGE \textbf{\color{burgundy}
    A hybrid multi-time-step framework for pore-scale \\[0.5\baselineskip]
    and continuum-scale modeling of solute transport \\[0.5\baselineskip]
    in porous media}}\\[0.3\baselineskip]
    \rule{\textwidth}{0.4pt}\vspace*{-\baselineskip}\vspace{3.2pt}
    \rule{\textwidth}{1.6pt}\\[\baselineskip]
    \scshape
    
    \vspace*{2\baselineskip}
    Authored by \\[\baselineskip]
    
    {\Large S.~Karimi\par}
    {\itshape Graduate Student, University of Houston.}\\[\baselineskip]
    
    {\Large K.~B.~Nakshatrala\par}
    {\itshape Department of Civil \& Environmental Engineering \\
    University of Houston, Houston, Texas 77204--4003. \\ 
    \textbf{phone:} +1-713-743-4418, \textbf{e-mail:} knakshatrala@uh.edu \\
    \textbf{website:} http://www.cive.uh.edu/faculty/nakshatrala\par}
    
    \vspace*{2\baselineskip}

    \begin{figure*}[h]
      \centering
      \psfrag{O}{$\Omega$}
      \psfrag{Of}{$\Omega_{\mathrm{f}}$}
      \psfrag{Oc}{$\Omega_{\mathrm{c}}$}
      \psfrag{Gc}{$\Gamma_{\mathrm{c} \rightarrow \mathrm{f}}$}
      \psfrag{Gf}{$\Gamma_{\mathrm{f} \rightarrow \mathrm{c}}$}
      \psfrag{Ofc}{$\Omega_{\mathrm{c}} \cap \Omega_{\mathrm{f}} \neq \emptyset$}
      \includegraphics[clip,scale=0.8]{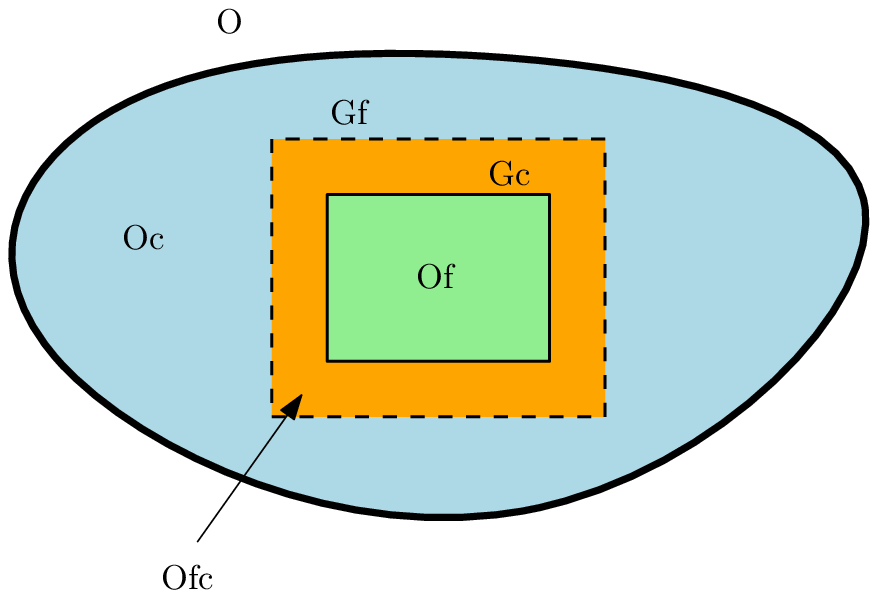}
      \caption*{\emph{The proposed coupling method employs 
          overlapping domain decomposition. This figure illustrates 
          the partitioning of the computational domain $\Omega$ 
          into subdomains in which continuum-scale and pore-scale 
          features are sought after.}}
    \end{figure*}
    \vfill
    {\scshape 2016} \\
    {\small Computational \& Applied Mechanics Laboratory} \par
  \end{titlepage}
  
\begin{abstract}
  Understanding transport processes in porous media is vital
  to many scientific and industrial applications. For instance,
  predicting the fate of chemical contaminants in subsurface,
  and simulating various processes in geological carbon-dioxide
  sequestration are two of the most prominent examples. Considering
  the dimensions of a typical spatial domain in such problems, usually
  in the order of kilometers, continuum models based on averaging
  theories have been the dominant approaches toward simulation
  of the above-mentioned processes. However, a variety of 
  interactions among the involved chemical species at the
  pores (such as dissolution and precipitation) cannot be 
  accounted within the current continuum-scale 
  models. To include these features into a computational 
  model, one needs to opt for length- and time-scales that 
  are much smaller than the ones typically considered  for
  a field-scale model. 
  Capturing such disparate temporal and spatial scales 
  still remains an enduring challenge in computational 
  mechanics. No single numerical method can efficiently 
  bridge the gap between these disparate scales. Hence, 
  designing numerical methodologies that employ different 
  numerical methods in different regions have grown into 
  a major topic of interest among researchers. 
  
  In this paper, we propose a computational framework,
  which is based on a domain decomposition technique, to
  employ both finite element method (which is a popular
  continuum modeling approach) and lattice Boltzmann method
  (which is a popular pore-scale modeling approach) in the same
  computational domain. To bridge the gap across
  the disparate length and time-scales, we first propose a
  new method to enforce continuum-scale boundary conditions
  (i.e., Dirichlet and Neumann boundary conditions) onto
  the numerical solution from the lattice Boltzmann method.
  This method are based on maximization of entropy and 
  preserve the non-negativity of discrete distributions under
  the lattice Boltzmann method.
  The proposed computational framework allows different grid
  sizes, orders of interpolation, and time-steps in different
  subdomains. This allows for different desired resolutions
  in the numerical solution in different subdomains. 
  Through numerical experiments, the effect
  of grid and time-step refinement, disparity of time-steps
  in different subdomains, domain partitioning, and the 
  number of iteration steps on the accuracy and rate 
  of convergence of the proposed methodology are studied. 
  Finally, to showcase the performance of this framework
  in porous media applications, we use it to simulate the 
  dissolution of calcium carbonate in a porous structure.  
\end{abstract}

\maketitle

\clearpage
\newpage 


\section{INTRODUCTION AND MOTIVATION}
\label{Sec:Intro}

\begin{wrapfigure}{r}{0.42\textwidth}
  \centering
  \includegraphics[clip,width=0.4\textwidth]{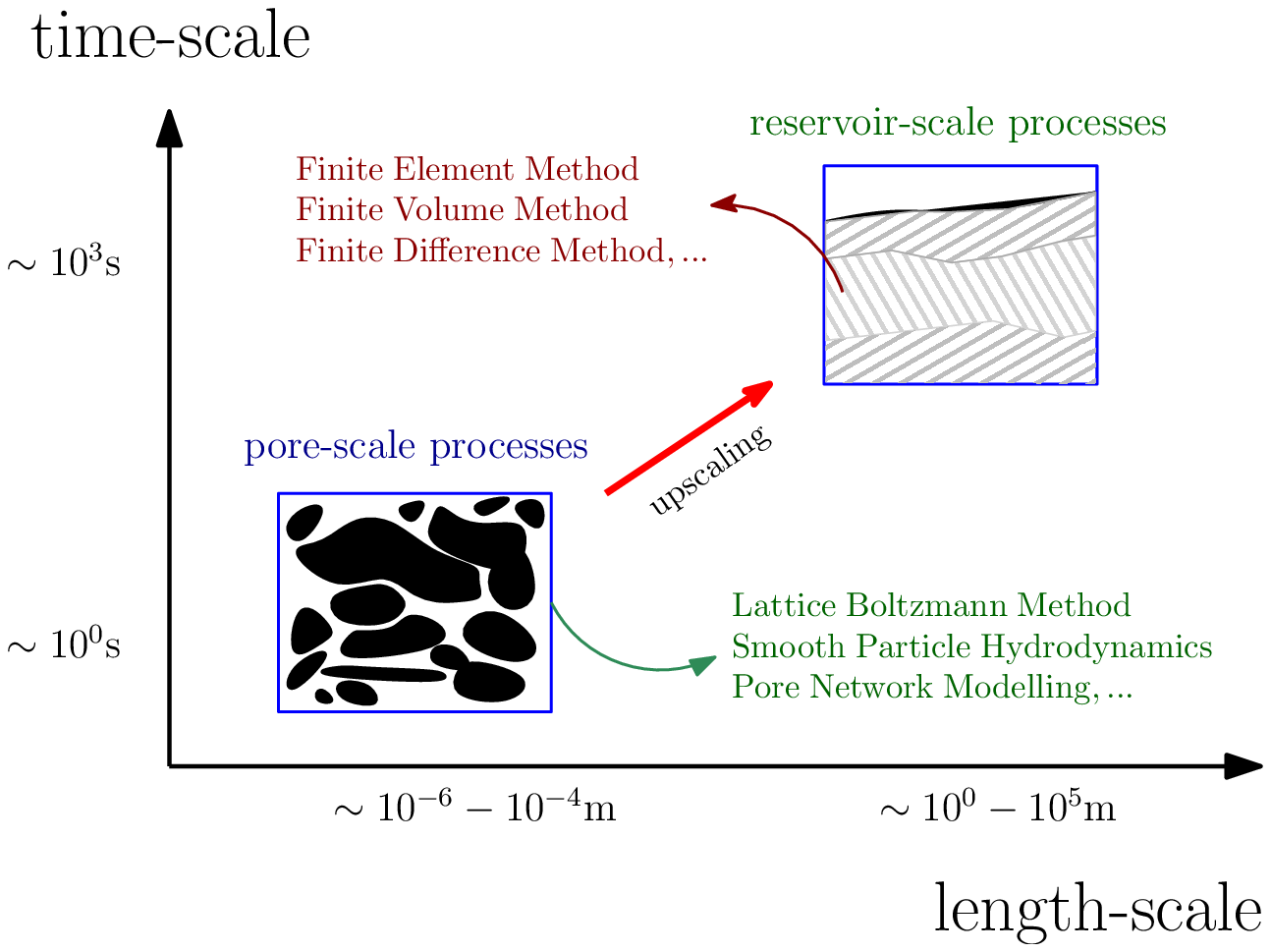}
  \caption{This figure illustrates the disparity
    in time- and length-scales in porous media
    simulations. \label{Fig:Multiscale}}
\end{wrapfigure}

Transport of chemical species in porous media features
a wide variety of time- and length-scales. Reaction and
precipitation at the interface of fluid and solid 
\citep{1995_Adler_CES,2007_Wood_AWR,2008VanNoorden_IMA},
reactive flow and transport \citep{1986_Shapiro_CES}, and
varied dynamics of (bio-)chemical reactions \citep{2000_Ginn_HJ}
are a few of the processes that occur at disparate scales. The
three length-scales that are typically considered in the study
on porous media are pore-scale (also referred to as fine-scale
or micro-scale), meso-scale (also referred to as continuum-scale
or coarse-scale) and macro-scale (also referred to as field-scale).
The properties of interest depend on the scale of observation,
which implies that different modeling approaches are needed at
different scales. Moreover, a numerical method appropriate for
a particular length- or time-scale need not be a viable approach
at a different scale. Due to this scale disparity, as shown in
figure \ref{Fig:Multiscale}, the choice of a particular modeling
approach or a particular numerical methodology that is appropriate
for all the scales of observation is severely limited.

\subsection{Coarse-scale modeling}
Finite Element Method (FEM), Finite Volume Method (FVM) 
and Finite Difference Method (FDM) are commonly practiced 
schemes for coarse-scale fluid dynamics computations. However, 
fine-scale features may not be immediately included into the 
numerical solutions from these methods. Some of the efforts 
towards improving upon this shortcoming are the Variational 
Multi-Scale method in \citep{1998_Hughes_CMAME}, Generalized 
Finite Element Method \citep{1996_Babuska_CMAME}, Multi-Scale 
Finite Element Method \citep{1997_Hou_JCP}, which can include
some fine-scale spatial features into the finite element solution
via manipulation of interpolation functions or the weak formulation.
Although coarse-scale models can be solved in a
computationally efficient manner and can include
some limited fine-scale features, these models
are not capable of capturing all the important
pore-scale processes and their impact at the
meso-scale and field-scale \citep{2002_Gramling_EST}.
The source of this deficiency is, partly, the dependence
of the model parameters on the length-scale. Furthermore,
some processes in reactive-transport (e.g., some pore-scale
reactions) cannot be upscaled from pore-scale to meso-scale
\citep{2009_Battiato_AWR}.

\subsection{Pore-scale modeling}
Methods such as pore network modeling \citep{1956_Fatt}, 
Smooth Particle Hydrodynamics (SPH) \citep{2005_Monaghan_RPP} 
and the Lattice Boltzmann Method (LBM) \citep{1998_Chen_ARFM} 
are amongst the most popular methods for fine-scale simulations. 
In particular, LBM offers great potential in including kinetic 
and atomistic details into the computational model. This fact 
originates from the main purpose of LBM, which is to numerically 
solve the Boltzmann equation \citep{1997_He_Luo_PRE}. This
equation can describe the distribution of particles
of a system in the phase space at any thermodynamic state.
Sophisticated gas-interface interaction models
\citep{1971_Cercignani_Lampis,
1988_Cercignani_book} and kinetic relations can also be included
in the solution of the Boltzmann equation \citep{1988_Cercignani_book,
2013_Cercignani_book}. \emph{Despite the advantages of LBM over 
coarse-scale methods such as FEM or FVM, its application to
real-world problems in subsurface modeling is impractical
due to prohibitive computational cost.}

\subsection{Hybrid modeling}
It is now becoming evident that a viable approach for simulation
of reactive-transport in porous media should consist of both
fine-scale and coarse-scale models; for example, see the
discussion in \citep{2015_Scheib_Groundwater}. The modeling
approaches that employ both fine-scale and coarse-scale models
are collectively referred to as \emph{hybrid} modeling. 
The motivation for hybrid modeling is four-fold: 
\begin{enumerate}[(i)]
\item There is a need for increasing local modeling accuracy
  in certain applications. Some examples include flow and
  transport along thin fractures, and to model processes
  in the well-bore cement that may act as escape passages
  for carbon-dioxide in geological carbon sequestration.
\item The need for hybrid modeling can arise when 
  continuum assumptions locally break down in critical 
  parts of the domain. For example, reactive-transport 
  modeling under advection-dominated or reaction-dominated 
  conditions, as described in \citep{2011_Battiato_AWR}.
\item The need for incorporating effects of the 
  surrounding media on the subdomain for accurate 
  predictions of flow and transport \citep{2012_Sun_EaF}.
\item To achieve a manageable computational cost to
  solve realistic problems arising in subsurface
  applications. 
\end{enumerate}

Recently, there is a surge in research activity in
hybrid modeling. A non-iterative coupling method
for SPH and coarse-scale averaged SPH was proposed
in \citep{2008_Tartakovsky_SIAM} for
advection-diffusion-reaction equation and precipitation in porous 
media. Using SPH for different length-scales allows the mentioned
method to avoid predictor-corrector iterations in each time-step. 
The multi-scale algorithm proposed in \citep{2011_Battiato_AWR}
is based on FVM and uses an iterative approach to resolve the disparate 
length-scales in a transport process. In \citep{2008_Balhoff_CG}
coupling of finite element method and pore network modeling 
for flow problems in porous media, using
the mortar method was introduced for the first time.
This method was then extended in \citep{2014_Balhoff_MSModSim}
to couple FDM and pore network model for simulation of flow and
transport of chemical species. It utilizes the mortar 
finite element spaces to transfer information from one subdomain 
to another. The unknowns are updated iteratively to satisfy continuity
of fluxes at the interface within a user-defined tolerance.
In \citep{2015_Tang_WRR}, these mortar-based methods are used to couple
finite difference and cellular automata methods to model 
the bio-film development in porous media.
Coupling of FDM and LBM for advection-diffusion equation is studied
in \citep{2004_Albuquerque_ICCS}, but non-matching grids and 
disparate time-steps are not considered.  A hybrid method that 
incorporates LBM and FEM for simulation of the diffusion processes
is proposed in \citep{2008_Haslam_CMAME}. A more recent effort
in this direction is given in \citep{2014_Astorino_PhDThesis}
that allows different time-steps and grid sizes for FEM and 
LBM domains.

\subsection{Multiple temporal scales and multi-time-step methods}
Multi-time-step (multi-rate) methods aim at resolving
the disparity in time-scales in a system through use
of appropriate time-steps and time-integrators for each
subsystem. In recent years, development of multi-time-step
methods has received much attention among researchers of
various fields. These include: multi-rate methods based
on Runge-Kutta schemes \citep{2001_Gunther, 2013_Sandu_JSciComp}, adaptive
variational integrators for dynamics \citep{2004_Marsden_IJNME}, 
multi-time-step methods based on non-overlapping domain
decomposition \citep{2014_Karimi_JCP, 2015_Karimi_CMAME},
and symplectic multi-time-step methods for molecular dynamics
simulations \citep{2001_Reich_PLA, 2004_Reich_book}.
Multi-time-step coupling algorithms based on
domain partitioning are often classified as
either \emph{staggered} or \emph{monolithic} 
coupling schemes \citep{2008_Nakshatrala_IJNME}.
Staggered coupling methods update the solution in different
subdomains through a predictor-corrector procedure. Hence, 
there is a time-lag between the solutions at different 
subdomains, which can result in numerical instabilities.
However, this type of algorithms enjoy tremendous popularity
because of modularity; one can employ available solvers
and use them (with different time-steps) in a staggered 
coupling algorithm without any major modification. Unlike 
staggered coupling algorithms, monolithic schemes
update the solution in the entire domain using a single 
iteration. These is no time-lag between the solution of
different subdomains. As a result, monolithic coupling 
algorithms enjoy much better numerical stability
than staggered coupling methods. However, current 
numerical solvers for partial differential equations 
cannot be immediately included in a monolithic coupling 
scheme and a major effort in developing computer codes 
is required. Also, multi-time-step integration requires 
careful design of a coupling algorithm \citep{2014_Karimi_JCP}.
Due to the aforementioned reasons, we shall employ
a staggered coupling approach. 

\subsection{Domain decomposition methods}
A natural way to develop a staggered coupling method is to 
employ domain decomposition techniques, which also offer 
an attractive framework for parallel computing. Over the 
years, a variety of overlapping and non-overlapping domain 
decomposition techniques have been developed \citep{Quarteroni_book,
Toselli_book,Mathew_book}. These methods have the potential to 
employ non-matching computational grids in different subdomains; 
for instance, mortar finite element spaces 
\citep{2000_Wohlmuth_SIAM, 2007_Arbogast_SIAM} and
overlapping methods \citep{1988_Lions, 1996_Cai_NLAA} are among
them. However, having different grid-sizes in different subdomains
may not be enough to account for disparate time-scales that can be
present in the model problem. In order to achieve computational 
efficiency for problems involving multiple temporal scales, one 
needs to employ tailored numerical time-integrators and time-steps 
for each active process. That is, domain decomposition techniques 
and multi-time-stepping schemes go hand in hand. Herein, we employ 
overlapping domain decomposition technique whose advantages will 
be discussed later. 

\subsection{An outline of the paper} 
We provide an overview of our approach in
Section \ref{Sec:S2_Hybrid_Overview}. Section
\ref{Sec:Hybrid_Continuum-scale} provides the
governing equations at the continuum-scale and
the associated numerical modeling. 
Section \ref{Sec:Hybrid_Pore-scale} discusses the modeling 
at the pore-scale using the lattice Boltzmann method. 
An overview of overlapping domain decomposition techniques
and information transfer across non-matching grids is 
given in Section \ref{Sec:Overlapping_DD}.
In Section \ref{Sec:Hybrid_Coupling_Algorithm}, we present a 
robust hybrid multi-time-step coupling method that allows 
to couple pore-scale and continuum-scale subdomains. Section 
\ref{Sec:Hybrid_NR} presents several numerical results using 
the proposed hybrid modeling, and illustrates the robustness 
and utility of the proposed computational framework. Finally, 
conclusions are drawn in Section \ref{Sec:Hybrid_CR} along 
with a discussion on possible future research endeavors in 
the area of hybrid modeling.

\section{AN OVERVIEW OF OUR APPROACH}
\label{Sec:S2_Hybrid_Overview}
In this paper, we present a hybrid method to couple the 
advection-diffusion equation at the continuum-scale with 
the Boltzmann equation at the pore-scale to simulate the 
transport of chemical species. The proposed method can 
capture fine-scale features and processes by solving the 
lattice Boltzmann equation at the pore-scale. The response
at the  continuum-scale is captured by solving the 
advection-diffusion equation using the finite element
method. 

\begin{wrapfigure}{r}{0.27\textwidth}
  \centering
  \includegraphics[clip,width=0.25\textwidth]{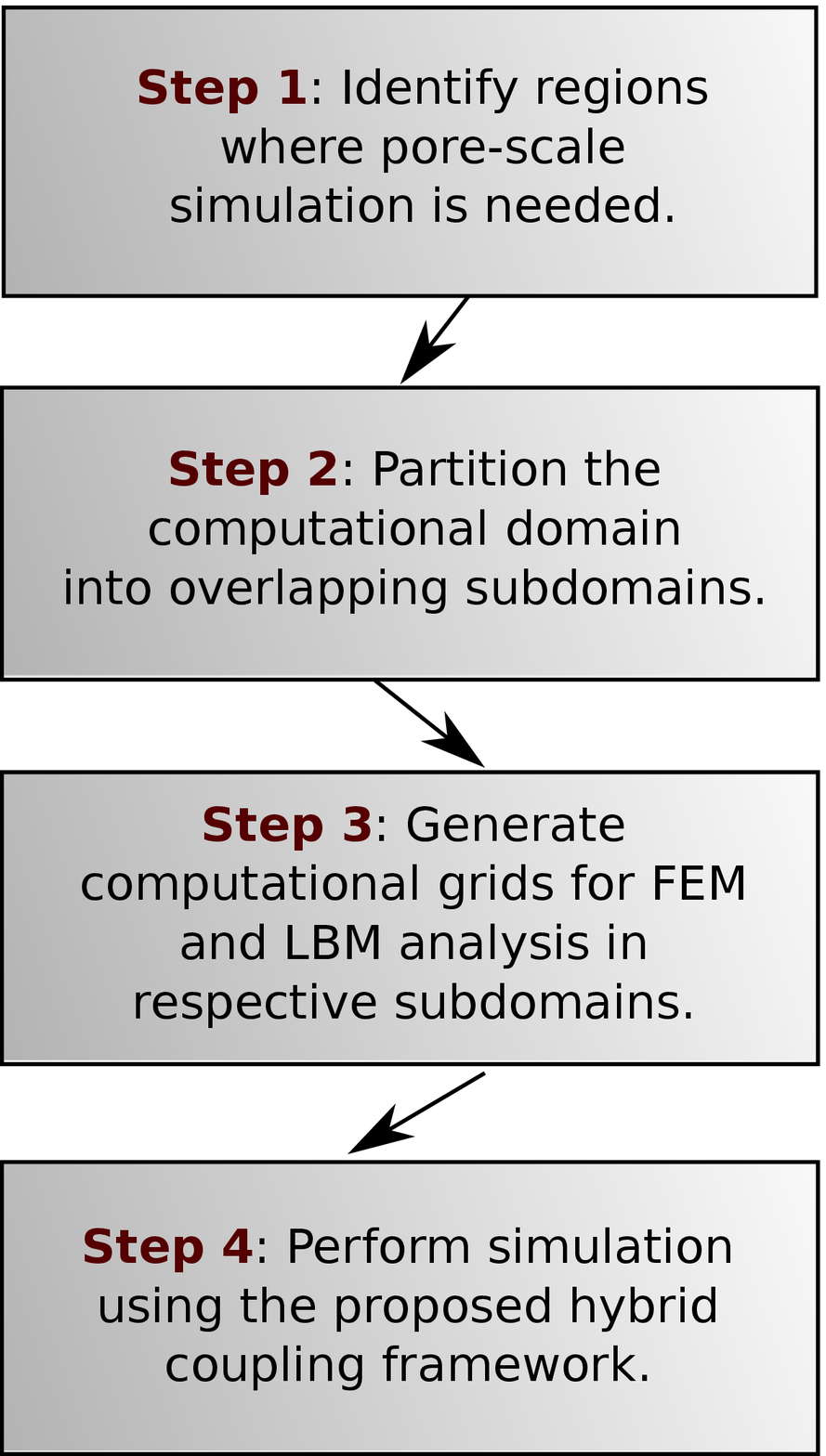}
  \caption{The main steps under the proposed hybrid framework.  
    \label{Fig:MultiscaleSteps}}
\end{wrapfigure}

We use the domain decomposition technique to partition
the computational domain into \emph{fine-scale} and
\emph{coarse-scale} subdomains. For better numerical
stability, we allow the coarse-scale and fine-scale 
computational subdomains to overlap, and appropriate
boundary conditions are designed at the boundary of
the individual computational subdomains. 
To capture disparate time-scale, the proposed computational 
framework allows different time-steps and different time 
integration algorithms in different subdomains. Furthermore,
computational grids and different orders of interpolation can
be employed in different subdomains. This enables the user to
choose appropriate time-step, mesh and interpolation in each
subdomain for stability and desired accuracy. 

The first step in a hybrid simulation using the proposed
framework is to partition the computational domain into
regions for fine-scale and coarse-scale modeling by identifying
the regions where pore-scale analysis is needed. Thanks to the
design of the proposed framework, creating computational
meshes for these two types of subdomains is easy and can
be carried out independent of each other. Finally, the
analysis is carried out by using appropriate models in
different subdomains. The overall procedure is summarized
in figure \ref{Fig:MultiscaleSteps}.
Some of the salient features of the proposed framework
are as follows: 
\begin{enumerate}[(i)]
\item Various transport processes and reactions can be
  incorporated into the analysis. In particular, the
  user can include complex advection velocity field
  (which is encountered frequently in porous media
  applications) and cascade of geochemical reactions
  without any change in the design of the coupling
  framework.
\item One can divide the computational domain into
  multiple subdomains, and can independently employ
  in each subdomain either pore-scale modeling or
  continuum-scale modeling.
\item The computational grids in a subdomain need not
  conform with the computational grid in another subdomain.
  In particular, the finite element mesh in the continuum-scale
  modeling need not match with the lattice structure in the
  lattice Boltzmann method, which is employed in the
  pore-scale analysis. This provides a great degree of
  flexibility for the modelers.
  \begin{enumerate}
  \item One can employ meshes with different degrees
    of approximation independent of other subdomains.
    There is no need for compatibility among the
    interpolation spaces (e.g., mortar spaces)
    along the subdomain interfaces.
  \item This allows to leverage on the existing
    computational methods for modeling at pore-scale
    and continuum-scale. There
    is no need to design new methods just to be compatible
    with the hybrid coupling. 
  \end{enumerate}
\item One can employ different time-steps and different
  time stepping schemes in different subdomains, which
  is an attractive feature to possess for solving problems
  involving multiple temporal scales.
\item An accurate transfer of data across non-matching
  grids has been incorporated into the proposed computational
  framework.
\item A novel way of implementing boundary conditions
  has been developed for the discretization under the
  lattice Boltzmann method. This enhances the accuracy
  at the pore-scale, and hence the overall accuracy of
  a hybrid coupling.
\item No initial guess at the interface of subdomains is 
	needed. Hence, implementation is 
	easier and the simulation procedure can be fully 
	automated. 
\end{enumerate}

In order to achieve aforementioned features,
the computational framework is developed by
integrating the following main ingredients: 
\begin{enumerate}[(a)]
\item A solver for continuum-scale modeling, which
  in our case will be a finite element formulation
  for advection-diffusion/dispersion equations.
\item A solver for pore-scale modeling, which in our
  case will be the lattice Boltzmann method with an
  improved discretization of boundary conditions.
\item An overlapping domain decomposition framework.
\item An accurate algorithm to transfer scalars, vectors
  and tensors across non-matching computational grids.
\item An iterative coupling algorithm to ensure compatibility
  of solution across the overlapping region.
\end{enumerate}

A computational framework with the aforementioned
features, which is essential to gain a fundamental
understanding of subsurface processes, is currently
not available. We therefore strive to design such
a framework in this paper. The details of the
aforementioned ingredients along with the illustration
of the performance of the proposed hybrid multi-time-step
computational framework are provided in subsequent
sections.

\section{CONTINUUM-SCALE MODELING}
\label{Sec:Hybrid_Continuum-scale}
We shall model the transport at the continuum-scale
using unsteady advection-dispersion equations. To this
end, consider a bounded open domain $\Omega_{\mathrm{c}}
\subset \mathbb{R}^{nd}$ on which we seek to perform
continuum modeling, where $\mathbb{R}$ denotes  the
set of real numbers and ``$nd$'' is the number of
spatial dimensions. We assume that the boundary
of this domain, $\partial \Omega_{\mathrm{c}}$, is
comprised of $\Gamma^{\mathrm{N}}_{\mathrm{c}}$ and
$\Gamma^{\mathrm{D}}_{\mathrm{c}}$ such that we have 
\begin{align}
  \partial \Omega_{\mathrm{c}} = \Gamma^{\mathrm{N}}_{\mathrm{c}}
  \cup \Gamma^{\mathrm{D}}_{\mathrm{c}}
  \quad \text{and} \quad
  \Gamma^{\mathrm{D}}_{\mathrm{c}} \cap
  \Gamma^{\mathrm{N}}_{\mathrm{c}} = \emptyset
\end{align}
Dirichlet boundary conditions are enforced on
$\Gamma^{\mathrm{D}}_{\mathrm{c}}$, and Neumann boundary
conditions are enforced on $\Gamma^{\mathrm{N}}_{\mathrm{c}}$.
A spatial point in $\Omega_{\mathrm{c}}$ will be denoted
by $\mathbf{x}$. We use $\mathrm{div}[\cdot]$ and
$\mathrm{grad}[\cdot]$, respectively, to denote 
the spatial divergence and gradient operators. The 
time interval of interest is denoted by $\mathcal{I} 
= (0, \mathrm{T}]$, and the time is denoted by
  $\mathrm{t}$.
The initial boundary value problem at the 
continuum-scale can be written as follows:
\begin{subequations}
  \label{Eqn:AD}
  \begin{alignat}{2}
    \label{Eqn:AD_GE}
    &\frac{\partial \mathrm{u}}{\partial \mathrm{t}} + 
    \mathrm{div}\left[ \mathbf{v} \mathrm{u} - \mathrm{D} 
      \mathrm{grad}\left[ \mathrm{u}\right]\right] = \mathrm{s} 
    &&\quad (\mathbf{x},\mathrm{t}) \in \Omega_{\mathrm{c}} \times
    \mathcal{I} \\
    \label{Eqn:AD_IC}
    &\mathrm{u}(\mathbf{x}, \mathrm{t}=0) = \mathrm{u}_0(\mathbf{x}) 
    &&\quad \mathbf{x} \in \Omega_{\mathrm{c}} \\
    \label{Eqn:AD_DBC}
    &\mathrm{u}(\mathbf{x},\mathrm{t}) = \mathrm{u}^{\mathrm{p}}
    (\mathbf{x},\mathrm{t}) 
    &&\quad (\mathbf{x},\mathrm{t}) \in \Gamma^{\mathrm{D}}_{\mathrm{c}}
    \times \mathcal{I} \\
    \label{Eqn:AD_NBC}
    &\widehat{\mathbf{n}} \cdot \left( \mathbf{v}\mathrm{u} - 
    \mathrm{D} \mathrm{grad}[\mathrm{u}]\right) = \mathrm{q}^{\mathrm{p}}
    &&\quad (\mathbf{x},\mathrm{t}) \in \Gamma^{\mathrm{N}}_{\mathrm{c}}
    \times \mathcal{I}
  \end{alignat}
\end{subequations}
where $\mathrm{u}$ is the concentration, $\mathbf{v}$ 
is the divergence-free advection velocity (i.e., 
$\mathrm{div}[\mathbf{v}] = 0 \; \mathrm{in} \;
\Omega_{\mathrm{c}}$), $\mathrm{D}$ is the dispersion
coefficient, and $\mathrm{s}$ is the source/sink term.
Although the dependence is not explicitly indicated,
all of the mentioned quantities depend on the spatial
coordinates and time. The dispersion coefficient
$\mathrm{D}$ is positive and can be spatially
heterogeneous. The initial concentration in
$\Omega_{\mathrm{c}}$ is denoted by $\mathrm{u}_0$, and
$\mathrm{u}^{\mathrm{p}}$ is the prescribed concentration
on $\Gamma^{\mathrm{D}}_{\mathrm{c}}$. The outward unit normal
to $\partial \Omega_{\mathrm{c}}$ is denoted by
$\widehat{\mathbf{n}}$, and $\mathrm{q}^{\mathrm{p}}$ is
the prescribed flux on $\Gamma^{\mathrm{N}}_{\mathrm{c}}$. 

The solution to the above equations can exhibit disparate 
spatial and temporal scales, which depend on the relative 
strengths of advection, dispersion and reaction processes, 
and volumetric source \citep{2000_Gresho_V1,Hundsdorfer_Verwer}. 
We employ the finite element method for the numerical 
modeling at the continuum-scale.

\subsection{The finite element method}
We shall introduce the following function spaces: 
\begin{subequations}
  \label{Eqn:FEMspaces}
  \begin{align}
    &\mathsf{C} := \left\{ \mathrm{u}:\Omega_{\mathrm{c}}
    \times\mathcal{I} \rightarrow 
    \mathbb{R} \; \big| \; \mathrm{u}(\mathbf{x},\mathrm{t})
    \in H^{1}(\Omega_{\mathrm{c}}) 
    \; \text{and} \; \mathrm{u}(\mathbf{x}\in\Gamma^{\mathrm{D}}_{\mathrm{c}},
    \mathrm{t}) = 
    \mathrm{u}^{\mathrm{p}} \; \forall \mathrm{t}\in \mathcal{I}\right\} \\
    &\mathsf{W} := \left\{ \mathrm{w}:\Omega_{\mathrm{c}} \rightarrow
    \mathbb{R} \; \big| \;
    \mathrm{w}(\mathbf{x}) \in H^{1}(\Omega_{\mathrm{c}}) \;
    \mathrm{and} \; \mathrm{w}(\mathbf{x} 
    \in \Gamma^{\mathrm{D}}_{\mathrm{c}}) = 0\right\}
  \end{align}
\end{subequations}
where $H^{1}(\Omega_{\mathrm{c}})$ is a Sobolev space
defined on $\Omega_{\mathrm{c}}$ \citep{Evans_PDE}.
We shall denote the standard $L_2$ inner product
over a set $\mathcal{K}$ as follows:
\begin{align}
  (\mathrm{w}, \mathrm{u})_{\mathcal{K}} \equiv \int_{\mathcal{K}} 
  \mathrm{w}\cdot\mathrm{u} \; \mathrm{d}\mathcal{K}
\end{align}
For convenience, we shall drop the subscript $\mathcal{K}$ 
if $\mathcal{K} = \Omega_{\mathrm{c}}$. We shall employ the
semi-discrete methodology to decouple the spatial and
temporal discretizations \citep{1977_Zienkiewicz_Book}.
There are a multitude of ways to construct a weak formulation
for equations \eqref{Eqn:AD_GE}--\eqref{Eqn:AD_NBC}. 
In this paper, we shall limit to the Galerkin formulation
and the Streamline Upwind/Petrov-Galerkin formulation
\citep{1992_Franca_CMAME}. However, it should be noted
that any other finite element (or even a finite volume)
formulation can also be employed in the modeling
at the coarse-scale.

\subsubsection{The Galerkin formulation.} 
Find $\mathrm{u}(\mathbf{x},\mathrm{t}) 
\in \mathsf{C}$ such that we have 
\begin{subequations}
  \label{Eqn:Galerkin}
  \begin{align}
    &\left( \mathrm{w}, \partial \mathrm{u}/\partial \mathrm{t}\right)
     + \left( \mathrm{w}, \mathbf{v} \cdot \mathrm{grad}[\mathrm{u}]\right) 
     + \left( \mathrm{grad}[\mathrm{w}], \mathrm{D} \mathrm{grad}[\mathrm{u}]\right)
    = \left(\mathrm{w}, \mathrm{s}\right) + \left( \mathrm{w}, \mathrm{q}^
    {\mathrm{p}}\right)_{\Gamma^{\mathrm{N}}_{\mathrm{c}}} \quad \forall \mathrm{w} \in \mathsf{W} \\
    &\mathrm{u} = \mathrm{u}_0 \quad \forall \mathbf{x}
    \in \Omega_{\mathrm{c}},\;\mathrm{t}=0 \\
    &\mathrm{u} = \mathrm{u}^{\mathrm{p}} \quad
    \forall(\mathbf{x},\mathrm{t})\in\Gamma^
    {\mathrm{D}}_{\mathrm{c}} \times \mathcal{I}
  \end{align}
\end{subequations}
Note that the Dirichlet boundary conditions are enforced
strongly. We shall employ the Galerkin formulation only
for dispersion-dominated problems, as this formulation
is known to perform poorly for advection-dominated
problems. This shortcoming can be partly alleviated
by employing a stabilized formulation instead. 

\subsubsection{The Streamline Upwind/Petrov-Galerkin
  (SUPG) formulation} 
The SUPG formulation is a popular stabilized
formulation, and it reads as follows: Find
$\mathrm{u}(\mathbf{x},\mathrm{t}) \in
\mathsf{C}$ such that we have
\begin{align}
  \label{Eqn:SUPG}
  \left( \mathrm{w}, \partial \mathrm{u}/\partial \mathrm{t}\right)
  &+ \left( \mathrm{w}, \mathbf{v} \cdot \mathrm{grad}[\mathrm{u}]\right) + 
  \left( \mathrm{grad}[\mathrm{w}], \mathrm{D} \mathrm{grad}[\mathrm{u}]\right)
  \nonumber \\
  &+\sum_{e = 1}^{N_{\mathrm{FEM}}} \left( \mathbf{v}\cdot\mathrm{grad}[\mathrm{w}],
  \tau_{e} \mathrm{R} [\mathrm{u}] \right)_{\Omega^{e}_{\mathrm{c}}} 
  = \left(\mathrm{w}, \mathrm{s}\right) + \left( \mathrm{w}, \mathrm{q}^{\mathrm{p}}\right)_
  {\Gamma^{\mathrm{N}}_{\mathrm{c}}} \quad \forall \mathrm{w} \in \mathsf{W} 
\end{align}
where $N_{\mathrm{FEM}}$ is the total number of finite 
elements and $\tau_{e}$ is the stabilization parameter 
for element $e$. The spatial domain contained in element 
$e$ is shown by $\Omega^{e}_{\mathrm{c}}$. The residual
$\mathrm{R}[\mathrm{u}]$ is defined as follows:
\begin{align}
  \mathrm{R}[\mathrm{u}] = \frac{\partial \mathrm{u}}{\partial \mathrm{t}}
  + \mathrm{div}\left[ \mathbf{v} \mathrm{u} - \mathrm{D} \mathrm{grad}
    \left[ \mathrm{u}\right]\right] - \mathrm{s}
\end{align}
The Dirichlet boundary condition and initial condition 
will remain as in equation \eqref{Eqn:Galerkin}. We 
employ the stabilization parameter $\tau_{e}$ as given 
in \citep{2011_Augustin_CMAME}. That is, 
\begin{align}
  \label{Eqn:SUPG_tau}
  \tau_{e} = \frac{h_{e}}{2p \left\| \mathbf{v} \right\|}
  \chi \left(P_{e}^{h}\right), \quad
       {P}_{e}^{h} = \frac{\left\| \mathbf{v} \right\| h_{e}}{2p \mathrm{D}}, \quad
       \chi(\alpha) = \mathrm{coth}(\alpha) - 1/\alpha
\end{align}
where $p$ is the order of finite element interpolation
functions and $\mathrm{D}$ is the \emph{isotropic}
coefficient of dispersion. The element size is denoted
by $h_{e}$, and ${P}_{e}^{h}$ is the element P\'eclet
number.

\section{PORE-SCALE MODELING:~THE LATTICE BOLTZMANN METHOD}
\label{Sec:Hybrid_Pore-scale}
We shall use $\Omega_{\mathrm{f}}$ to denote the region
in which one seeks to perform pore-scale modeling. We
use the Boltzmann equation to describe the transport
processes at the pore-scale. 
The Boltzmann equation provides a statistical description of 
the state of matter away from the thermodynamic equilibrium 
\citep{1988_Cercignani_book}. This equation describes the 
evolution of the distribution of particles in the phase 
space, from which macroscopic variables can be easily 
computed by taking appropriate moments. If one neglects
the external force term, the Boltzmann equation can be
written as:
\begin{align}
  \label{Eqn:Boltzmann}
  \frac{\partial f}{\partial \mathrm{t}} 
  + \mathbf{v} \cdot \mathrm{grad}[f] 
  = \mathcal{Q}\left[ f,f^{\mathrm{eq}}\right]
  \qquad \mathrm{in} \; \Omega_{\mathrm{f}}
\end{align}
where $f$ is the distribution function, $\mathbf{v}$ 
is the macroscopic (or background) velocity, and 
$\mathcal{Q}$ is the collision term. Herein, we 
will consider the Bhatnagar-Gross-Krook (BGK) 
collision model \citep{1954_BGK}, which can 
be written as:
\begin{align}
  \mathcal{Q}\left[f,f^{\mathrm{eq}}\right] 
  = \frac{1}{\lambda}(f^{\mathrm{eq}} - f)
\end{align}
where $f^{\mathrm{eq}}$ is the distribution of particles
in the phase space at the thermodynamics equilibrium. 
The parameter $\lambda$ is called the relaxation time. 
In this paper, we will use the Maxwell-Boltzmann distribution 
for the equilibrium distribution $f^{\mathrm{eq}}$. That is, 
\begin{align}
  \label{Eqn:MxBlt}
  f^{\mathrm{eq}}(\mathbf{x},\boldsymbol{\zeta};\mathbf{v},\mathrm{u})
  = \frac{\mathrm{u}}{\sqrt{2 \pi \mathrm{R}
      \mathrm{T}}} \mathrm{exp}[-(\boldsymbol{\zeta} - 
    \mathbf{v})\cdot (\boldsymbol{\zeta} - \mathbf{v})/2\mathrm{R}\mathrm{T}]
\end{align}
where $\mathrm{u}(\mathbf{x},\mathrm{t})$ is the
concentration, $\mathrm{R}$ is the ideal gas constant,
and $\mathrm{T}$ is the temperature. The velocity of
particles is indicated by $\boldsymbol{\zeta}$.  

In order to provide a complete description of the Boltzmann
equation for a physical problem, appropriate boundary conditions
have to be included. Fortunately, the mathematical theory of 
boundary conditions for Boltzmann equation is rather well-developed;
for example, see \citep{1971_Cercignani_Lampis, 1988_Cercignani_book,
2013_Cercignani_book}. However, to use the existing theories to their
full extent, one has to incorporate detailed dynamics for interaction
between the particle and the surrounding surface of the domain.
Obviously, the continuum model problem given in equation \eqref{Eqn:AD}
lacks such information. Hence, the Boltzmann equation provides a 
much more powerful framework to account for detailed dynamics of
gas-surface interaction that the continuum model is incapable of.
In this paper, we will assume that the user merely intends to replicate
the behavior of the macroscopic solution of equation \eqref{Eqn:AD}
and an in-depth treatment of gas-surface interaction is of no interest.
In the following, we will provide an overview of discretization
of Boltzmann equation.

\subsection{The lattice Boltzmann method}
\label{Sec:Hybrid_Pore_Scale_LBM}
The lattice Boltzmann method is a popular numerical method
to solve the Boltzmann equation \eqref{Eqn:Boltzmann}. This
method offers great potential for parallelization
\citep{2011_Wnag_PC} and simulation in domains with
complex spatial features \citep{2001_Succi_LBM}. 
We will employ standard lattice model $DnQm$ to discretize
the velocity space. These discrete velocities are identified
by vectors $\boldsymbol{e}_i,\;i=1,\cdots,m$.
The discrete population corresponding to the lattice velocity
$\boldsymbol{e}_i$ is denoted by $f_i$. Considering the lattice
cell size of $h$ and a time-step $\Delta t$, the discrete form
of Boltzmann equation can be written as:
\begin{align}
  \label{Eqn:LBE}
  \left| f_i \right\rangle (\boldsymbol{x} + \boldsymbol{e}_i \Delta t,
  t + \Delta t) = \left| f_i \right\rangle (\boldsymbol{x} ,t)
  + \left| \mathcal{Q}_i \right\rangle (\boldsymbol{x},t)
\end{align}
where the column vector of discrete populations
is denoted using the Dirac notation $\left| \cdot
\right \rangle$ \citep{2000_Lallemand_PRE}.
Location of a lattice node is shown by $\boldsymbol{x}$
and $t$ is a discrete time-level. The discrete collision
operator $\mathcal{Q}_i$ is defined as: 
\begin{align}
\label{Eqn:Collision}
	\left| \mathcal{Q}_i \right\rangle (\boldsymbol{x},t) = 
	\frac{1}{\tau} \left( \left| f^{\mathrm{eq}}_i\right\rangle 
	(\boldsymbol{x},t) - \left| f_i \right\rangle (\boldsymbol{x},t)\right)
\end{align}
The non-dimensional relaxation-time is
denoted by $\tau$ and is defined as:
\begin{align}
  \tau = \frac{1}{2} + \frac{D}{c_s^2 \Delta t}
\end{align}
with $c_s$ being the lattice sound velocity (e.g., in the case 
of $D2Q9$ lattice model $c_s = \Delta x/\sqrt{3}\Delta t$).
For the equilibrium distribution in equation \eqref{Eqn:Collision},
we will employ the following approximation to the 
Maxwell-Boltzmann distribution in equation \eqref{Eqn:MxBlt}:
\begin{align}
\label{Eqn:LBM_feq}
	f_i^{\mathrm{eq}} (\boldsymbol{x},t;u) = 
	w_i u \left( 1 + 
	\frac{\boldsymbol{e}_i \cdot \mathbf{v}}{c_s^2} + 
	\frac{1}{2}\frac{(\boldsymbol{e}_i \cdot \mathbf{v})^2}{c_s^4}
	- \frac{\mathbf{v}\cdot\mathbf{v}}{c_s^2}\right)
\end{align}
where $w_i$ is the weight associated with $\boldsymbol{e}_i$ and 
$\mathbf{v}$ is the advection velocity. Concentration is 
denoted by $u$. In the rest of the paper, we will assume that 
$\left\| \mathbf{v} \right\|/c_s \ll 1$ 
(low Mach number assumption). Macroscopic quantities of interest,
in this case concentration and flux, can be obtained by the 
following relations
\begin{subequations}
\label{Eqn:LBM_Macro}
\begin{align}
	&u(\boldsymbol{x},t) = \sum_{i = 1}^{m} f_i(\boldsymbol{x},t) \\
	&\boldsymbol{q}(\boldsymbol{x},t) = \sum_{i = 1}^{m} f_i(\boldsymbol{x},t) \boldsymbol{e}_i
\end{align}
\end{subequations}
Despite ever-growing popularity of lattice Boltzmann methods
for computational fluid dynamics assumptions, these methods are
prone to produce unphysical values for populations $f_i$;
for example, see \citep{2015_Karimi_LBM}. Obviously, for equation 
\eqref{Eqn:Boltzmann} to be meaningful, the value of population $f$ 
needs to be non-negative. Some of the approaches toward resolving
this issue can be found in \citep{2004_Li_JFM,2006_Karlin,2014_Dellacherie_AAM}.
Herein, we will propose a simple condition on the LBM discretization
that guarantees non-negative values for discrete populations.
We assume that the initial values for populations are also
non-negative; for instance, this can be achieved if one takes
$f_i (\boldsymbol{x},t=0) = f_i^{\mathrm{eq}}(\boldsymbol{x},t=0)$.
From equations \eqref{Eqn:LBE} and \eqref{Eqn:Collision} we can
conclude that if $1 - 1/\tau \geq 0$, then the discrete populations
at time-level $t + \Delta t$ will also be non-negative. Note that 
the streaming process does not contribute to negativity (an 
appropriate treatment of boundary conditions comes later).
This condition leads to the following result
\begin{align}
\label{Eqn:NN_Cond}
	\Delta t \leq \frac{2 \mathrm{D}}{c_s^2}
\end{align}
For instance, in the case of a $D2Q9$ lattice model, one should
have 
\begin{align}
	h^2 \leq 6\Delta t \mathrm{D}
\end{align}
The advantage of this method compared to methods such as 
entropic lattice Boltzmann method is that one does not 
need to solve a non-linear equation at each lattice node.
Hence, it is much easier to code and computationally more 
efficient. Furthermore, the standard collision and streaming
steps in the lattice Boltzmann method have remained untouched
and no further modification is necessary. 

To complete the description of lattice Boltzmann method for 
advection-diffusion equation, we need to demonstrate how to 
apply boundary conditions so that the resulting discrete
populations are non-negative.
In the following we will provide new methods for 
enforcing Dirichlet- and Neumann-type boundary 
conditions in equation \eqref{Eqn:AD}.

\subsubsection{Boundary conditions for the lattice Boltzmann method}
Over the past few decades, a multitude of methods for
enforcing macroscopic boundary conditions in  the context
of lattice Boltzmann methods for flow and transport equations
have been proposed. For example, see\citep{1993_Skordos_PRE,
  1997_Zou_PF,2013_Chen_PRE}.
However, note that the boundary conditions 
typically considered for flow or transport problems
in a macroscopic framework do not imply a unique 
configuration of particles in phase space. Another 
drawback of these methods can be that they may 
result in negative values for discrete populations.
Other physical properties of the solution, such
as monotonicity of entropy production may also be
lost following enforcement of boundary conditions.
Designing a numerical method to enforce boundary
conditions appropriately, is indeed a challenging 
topic. Herein, to partially rectify the aforementioned 
problems, we propose a new framework for enforcing 
Dirichlet and Neumann boundary conditions for 
lattice Boltzmann method. These methods are 
based on the assumption that the system encompassed
in domain $\Omega$ is connected to a bath of particles
that reside in a specific state of thermodynamics 
equilibrium. This thermodynamic state can be identified
by maximizing the entropy function, subject to 
a hydrodynamic constraint. 

In this paper, we will utilize the Boltzmann's
$\mathcal{H}$ function to find the state of
equilibrium. Obviously, the entropy $\mathcal{S}$
is related to the $\mathcal{H}$ function via the
following relation:
\begin{align}
  \mathcal{S} = - \mathcal{H}
\end{align}
Hence, maximization of entropy $\mathcal{S}$ is 
mathematically equivalent to minimizing $\mathcal{H}$.
The $\mathcal{H}$ function at each point is defined as: 
\begin{align}
  \label{Eqn:BH}
  \mathcal{H}(\left| f_i\right\rangle)\big|_{(\boldsymbol{x},t)} = 
  \sum_{i = 1}^{m} f_i(\boldsymbol{x},t) \; 
  \log\left[\frac{f_i(\boldsymbol{x},t)}{w_i}\right]
\end{align}
For brevity, we will use the following notation:
\begin{align}
  \label{Eqn:Boundary_Vel}
  \mathsf{M}^{-} (\mathbf{x}) := \left\{ 
  i \big| \; \boldsymbol{e}_i \cdot \widehat{\mathbf{n}}(\mathbf{x})
  < 0,\; i=1,\cdots,m \; \text{and} \; \mathbf{x}\in \partial \Omega \right\}
\end{align}
where $\widehat{\mathbf{n}}(\mathbf{x})$ is the unit outward 
normal to $\partial \Omega$. Obviously, the unknown populations near 
the boundary are $f_i$ with $\boldsymbol{e}_i \in \mathsf{M}^{-}(\mathbf{x})$ 
for every point $\mathbf{x}$ on $\partial \Omega$. The rest of the discrete 
populations are known from the collision and streaming steps 
prior to enforcement of boundary conditions.
\begin{enumerate}[(a)]
\item \emph{Dirichlet boundary condition}:~Let $\mathbf{x} \in \Gamma^
{\mathrm{D}}$ and $\mathrm{u}^{\mathrm{p}}(\mathbf{x},t)$ be the prescribed 
concentration at that point. The unknown populations are calculated
by solving the following constrained optimization problem:
\begin{subequations}
  \begin{align}
    \label{Egn:LBM_Dir}
    &\min_{f_j,\;j\in \mathsf{M}^{-}(\mathbf{x})} \; \mathcal{H}
    \left( \left. \big| f_i \right \rangle \right) \big|_{(\mathbf{x},t)} \\
    &\text{subject to} \; \sum_{i = 1}^{m} f_i (\mathbf{x},t) = \mathrm{u}^{\mathrm{p}}(\mathbf{x},t)
  \end{align}
\end{subequations}
where the function $\mathcal{H}$ is defined in equation
\eqref{Eqn:BH}. This minimization problem will result in
the following relation: 
\begin{align}
  &f_i (\mathbf{x},t) = \frac{w_{i}}{\sum_{j\in \mathsf{M}^{-}
      (\mathbf{x})} w_j}
  \left( \mathrm{u}^{\mathrm{p}}(\mathbf{x},t) - \sum_{k \nin 
    \mathsf{M}^{-}(\mathbf{x})}
  f_k (\mathbf{x},t)\right) \quad i \in \mathsf{M}^{-}(\mathbf{x}) 
  \nonumber \\
  &f_i(\mathbf{x},t) \geq 0 \quad \forall i = 1,\cdots,m
\end{align}
Note that the function $\mathcal{H}$ is only
defined for non-negative arguments. 
\item \emph{Neumann boundary condition}:~Let $\mathbf{x}\in \Gamma^{\mathrm{N}}$ 
and $\mathrm{q}^{\mathrm{p}}(\mathbf{x},t)$ be the prescribed flux at that point. 
The unknown populations are governed by the
following constrained optimization problem:
\begin{subequations}
  \begin{align}
    \label{Eqn:LBM_Neu}
    &\min_{f_j,\;j\in \mathsf{M}^{-}(\mathbf{x})} \; 
    \mathcal{H}(\left. \big| f_i \right \rangle)
    \big|_{(\mathbf{x},t)} \\
    &\text{subject to} \; \sum_{i = 1}^{m} f_i(\mathbf{x},t)
    \boldsymbol{e}_i \cdot \widehat{\mathbf{n}}(\mathbf{x})
    = \mathrm{q}^{\mathrm{p}}(\mathbf{x},t)
  \end{align}
\end{subequations}
with $\widehat{\mathbf{n}}(\mathbf{x})$ being the unit outward normal defined 
earlier. This minimization problem will result in a non-linear equation in 
terms of Lagrange multiplier for the hydrodynamic constraint:
\begin{align}
  \sum_{i \in \mathsf{M}^{-}(\mathbf{x})} w_i (\boldsymbol{e}_i 
  \cdot \widehat{ \mathbf{n}}(\mathbf{x}))
  \mathrm{exp}\left[ -1 - \gamma \boldsymbol{e}_i \cdot
    \widehat{\mathbf{n}}(\mathbf{x}) \right]
  = \mathrm{q}^{\mathrm{p}} (\mathbf{x},t) - \sum_{j \nin 
	\mathsf{M}^{-}(\mathbf{x})} (\boldsymbol{e}_j \cdot 
	\widehat{\mathbf{n}}(\mathbf{x})) f_j (\mathbf{x},t)
\end{align}
where $\gamma$ is the Lagrange multiplier. Once the value of $\gamma$ is 
known, the populations can be found using the following relation:
\begin{align}
  f_i(\mathbf{x},t) = w_i \mathrm{exp}\left[ -1 - \gamma 
    \boldsymbol{e}_i \cdot \widehat{\mathbf{n}}(\mathbf{x})
    \right] \quad i\in\mathsf{M}^{-}(\mathbf{x})
\end{align}
which guarantees non-negative values for $f_i$. In the case of one-dimensional
lattice models (e.g. the $D1Q2$ lattice) and the $D2Q4$ or $D2Q5$ models, this method reduces to the conventional bounce-back method. But, in general
this method is different than bounce-back or specular reflection methods.
\end{enumerate}

Through this method of enforcing boundary conditions,
which is based on maximization of entropy, we ensure
monotonic increase in entropy. The physical interpretation
of this method is that the system in $\Omega$ is connected
to systems in thermodynamic equilibrium. The adjacent systems
connected to $\Omega$ through $\Gamma^{\mathrm{D}}$ and
$\Gamma^{\mathrm{N}}$ are in different states of thermodynamic
equilibrium. Minimization of the function $\mathcal{H}$ ensures 
that the equilibrium condition for particles near the boundary
is respected. The constraints in equations \eqref{Egn:LBM_Dir}
and \eqref{Eqn:LBM_Neu} are the macroscopic hydrodynamic conditions
of the system at the respective points of the domain.

\subsubsection{A numerical example}
\begin{figure}
  \centering
  \subfigure[concentration at $t=0.25$]{
    \includegraphics[clip,scale=0.35]{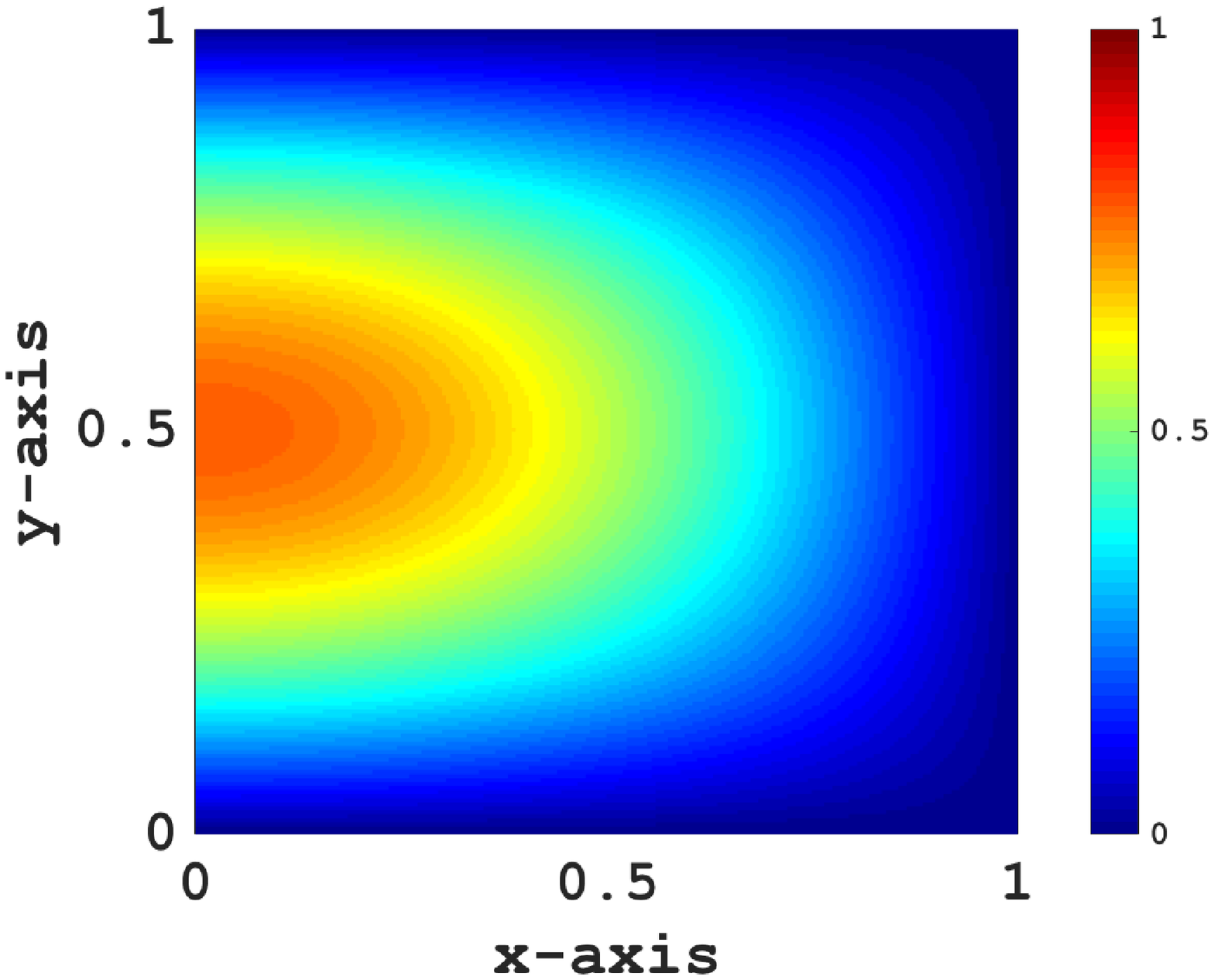}}
  \subfigure[difference between the numerical and exact solutions 
  at $t = 0.25$]{
    \includegraphics[clip, scale=0.35]{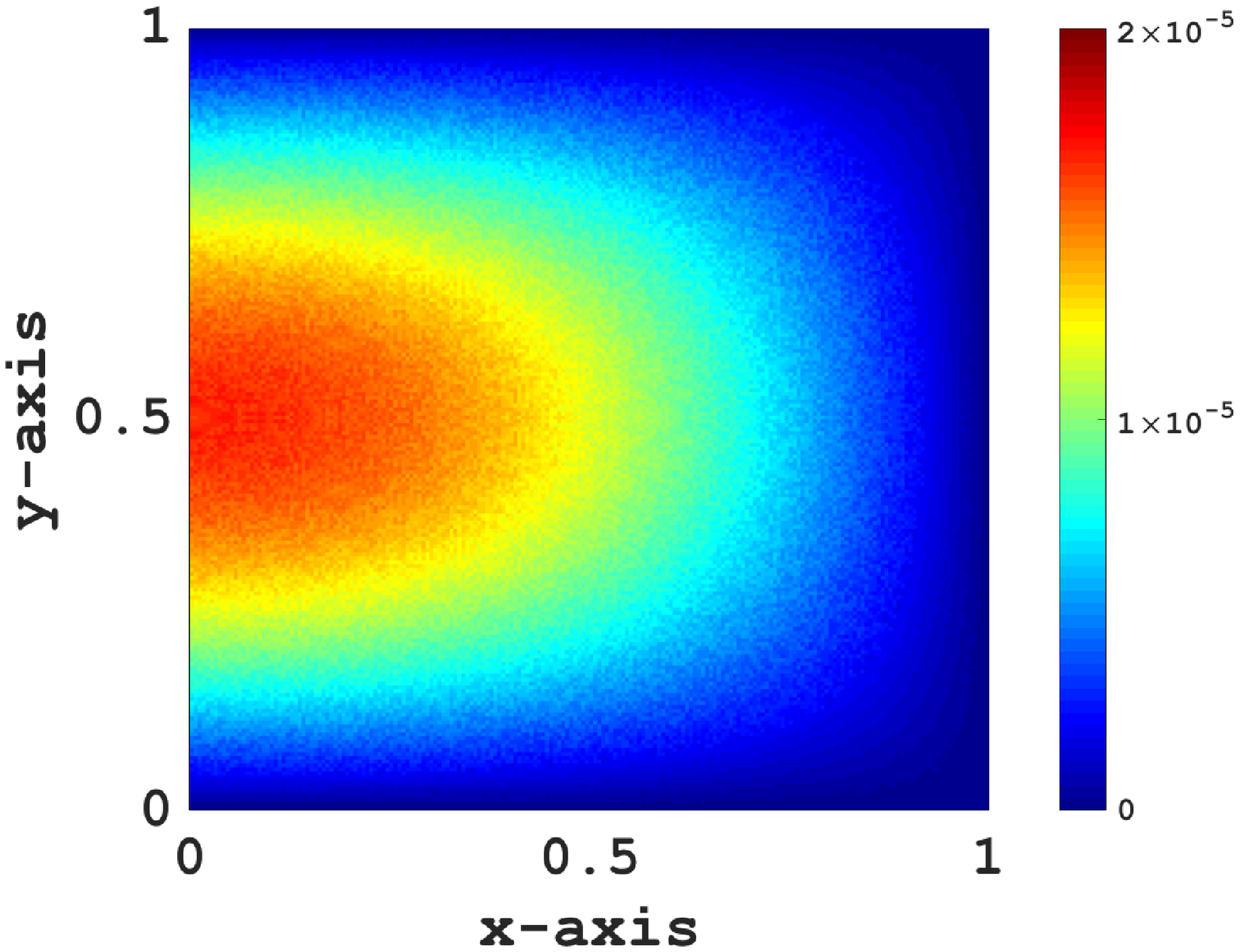}}
  \caption{\textsf{Numerical results for LBM}:~Concentration 
    and error in concentration are shown over the computational
    domain. LBM along with the new method for enforcing boundary
    conditions provide accurate numerical solutions, as illustrated 
    in this example. These results correspond to Case 4 in 
    Table \ref{Tbl:LBM_p1}.\label{Fig:LBM_P1}}
\end{figure}

\begin{figure}
  \centering
  \psfrag{h}{$h$}
  \psfrag{E}{$\mathcal{E}(t = 0.25)$}
  \psfrag{slope}{slope=2}
  \includegraphics[clip,scale=0.4]{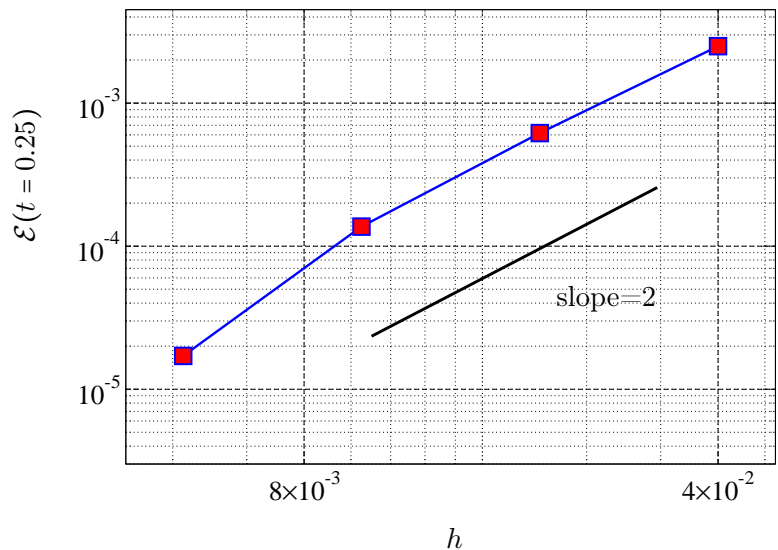}
  \caption{\textsf{Numerical results for LBM}:~In this 
    figure, the error in the numerical solution is shown 
    against the lattice cell size. With the proposed 
    method for the boundary conditions, second-order 
    convergence of LBM is obtained. 
    \label{Fig:LBM_convergence_P1}}
\end{figure}

We now assess the accuracy of the proposed methods
for boundary conditions under LBM. Consider the
domain $\Omega = (0,1)\times(0,1)$, with the diffusion
coefficient $\mathrm{D} = \frac{4}{5\pi^2}$ and advection
velocity being $\mathbf{v}=\mathbf{0}$. The source term is
also zero throughout the domain. The initial concentration
is taken as:
\begin{align}
  \mathrm{u}_{0}(\mathrm{x},\mathrm{y})
  = \sin (\pi \mathrm{y}) \cos(\pi\mathrm{x}/2)
\end{align}
and the boundary conditions are 
\begin{subequations}
  \begin{alignat}{2}
    -&\mathrm{D} \mathrm{grad}[\mathrm{u}] \cdot 
    \widehat{\mathbf{n}} = 0
    &&\quad \text{on}\;\Gamma^{\mathrm{N}} \\
    &\mathrm{u} = 0
    &&\quad \text{on}\;\Gamma^{\mathrm{D}}
  \end{alignat}
\end{subequations}
where $\Gamma^{\mathrm{N}} = \{0\}\times(0,1)$ and $\Gamma^
{\mathrm{D}} = \partial \Omega - \Gamma^{\mathrm{N}}$. We will 
employ the $D2Q9$ lattice model with grid spacing of $h$. The
time-step will be chosen according to equation \eqref{Eqn:NN_Cond}
to avoid negative values for discrete distributions. This problem
is solved using several choices of discretization parameters
as given in Table \ref{Tbl:LBM_p1}. We will use the following
definition for calculation of error:
\begin{align}
  \mathcal{E}(t) = \max_{i}\left\{ | u(\boldsymbol{x}_i,t) - 
  \mathrm{u}_{\mathrm{exact}}(\boldsymbol{x}_i, t)|\right\}
\end{align}
where $u(\boldsymbol{x}_i,t)$ is the computed numerical
value at $i$-th node and time-level $t$. The exact solution
is denoted by $\mathrm{u}_{\mathrm{exact}}$. Numerical results
from LBM with the proposed methods for boundary conditions
are shown in figure \ref{Fig:LBM_P1}. The variation of error
with respect to the cell-size has been documented in Table
\ref{Tbl:LBM_p1} and figure \ref{Fig:LBM_convergence_P1},
which show a second-order convergence.
\begin{table}
\caption{\textsf{Numerical results for LBM}:~In this table,
numerical values for discretization parameters and the 
calculated error at time-level $t = 0.25$ are shown.}
\label{Tbl:LBM_p1}
\begin{tabular}{| c | c | c | c |}
	\hline
	Case & $h$ & $\Delta t$ & $\mathcal{E}(t = 0.25)$ \\ \hline
	1 	 & $4\times10^{-2}$& $3.3\times10^{-3}$& $2.5\times10^{-3}$\\ \hline
	2 	 & $2\times10^{-2}$& $8.2\times10^{-4}$& $6.2\times10^{-4}$ \\ \hline
	3 	 & $10^{-2}$& $2.1 \times 10^{-4}$& $1.4\times10^{-4}$\\ \hline
	4 	 & $5\times10^{-3}$& $5.2\times10^{-5}$& $1.7\times10^{-5}$\\ \hline
\end{tabular}
\end{table}
\subsubsection{Comparison with other methods}
\begin{figure}
  \centering
  \psfrag{time}{time}
  \psfrag{H}{$\mathcal{H}(t)$}
  \subfigure[Concentration at $t = 0.5$]{
    \includegraphics[scale=0.25,clip]{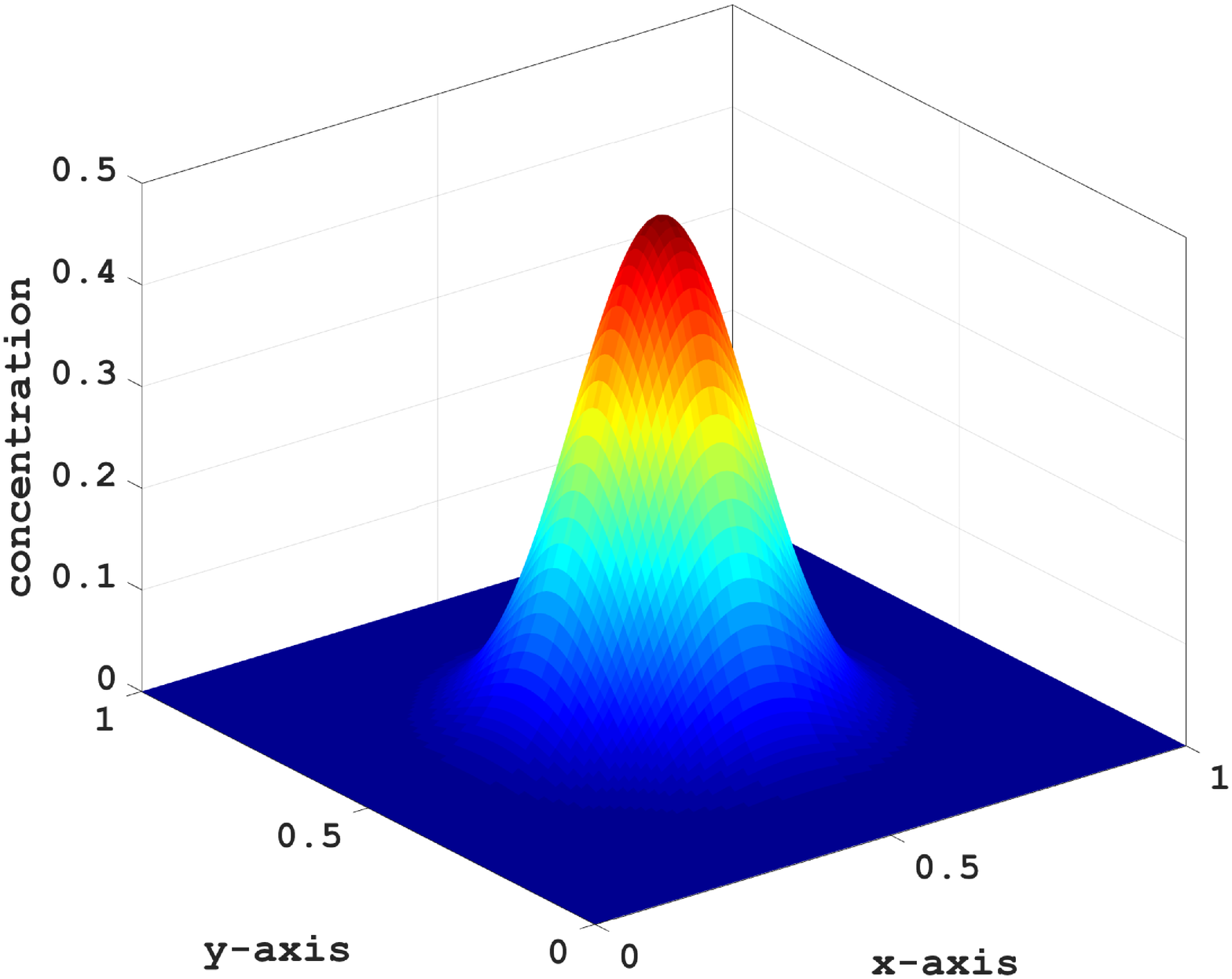}}
  \subfigure[Boltzmann's $\mathcal{H}$ function]{
    \includegraphics[scale=0.25, clip]{Figures/LBM/LBM_Hfunc_p2.eps}}
  \caption{\textsf{Numerical results for LBM}:~This figure 
    presents sample numerical results under the lattice 
    Boltzmann method with the new boundary conditions. 
    Figure (a) shows the concentration at time $t=0.5$, 
    which is non-negative throughout the domain. 
    In figure (b), the value of the Boltzmann's 
    $\mathcal{H}$ function is plotted against time. 
    Clearly, the value of $\mathcal{H}$ is decreasing 
    monotonically with time. \label{Fig:LBM_sample}}
\end{figure}

Consider the domain $\Omega = (0,1)\times(0,1)$
with zero-flux boundary conditions enforced on
$\partial \Omega$. The initial condition is taken
as: 
\begin{align}
  \mathrm{u}_{0}(\mathbf{x}) = \left\{ \begin{array}{l l}
    1 & \mathbf{x}\in[a,b]\times[a,b] \\
    0 & \text{otherwise}
  \end{array} \right.
\end{align}
where we take $a = 0.4$ and $b = 0.6$. The diffusion
coefficient is $\mathrm{D} = 10^{-2}$ and the advection
velocity is zero. The $D2Q9$ lattice model is used.
Figure \ref{Fig:LBM_sample} shows the numerical result
from
the lattice Boltzmann method, along with the proposed methods for 
enforcing boundary conditions. The bound given by equation \eqref{Eqn:NN_Cond} 
for cell-size and time-step is respected. Hence, all discrete 
populations, and consequently, concentration at all nodes are 
non-negative. The change in the Boltzmann $\mathcal{H}$ function 
is monotonic, which means that the $\mathcal{H}$-theorem is satisfied.

\begin{figure}
  \centering
  \subfigure[proposed boundary condition vs. bounce-back method]{
    \includegraphics[clip,scale=0.35]{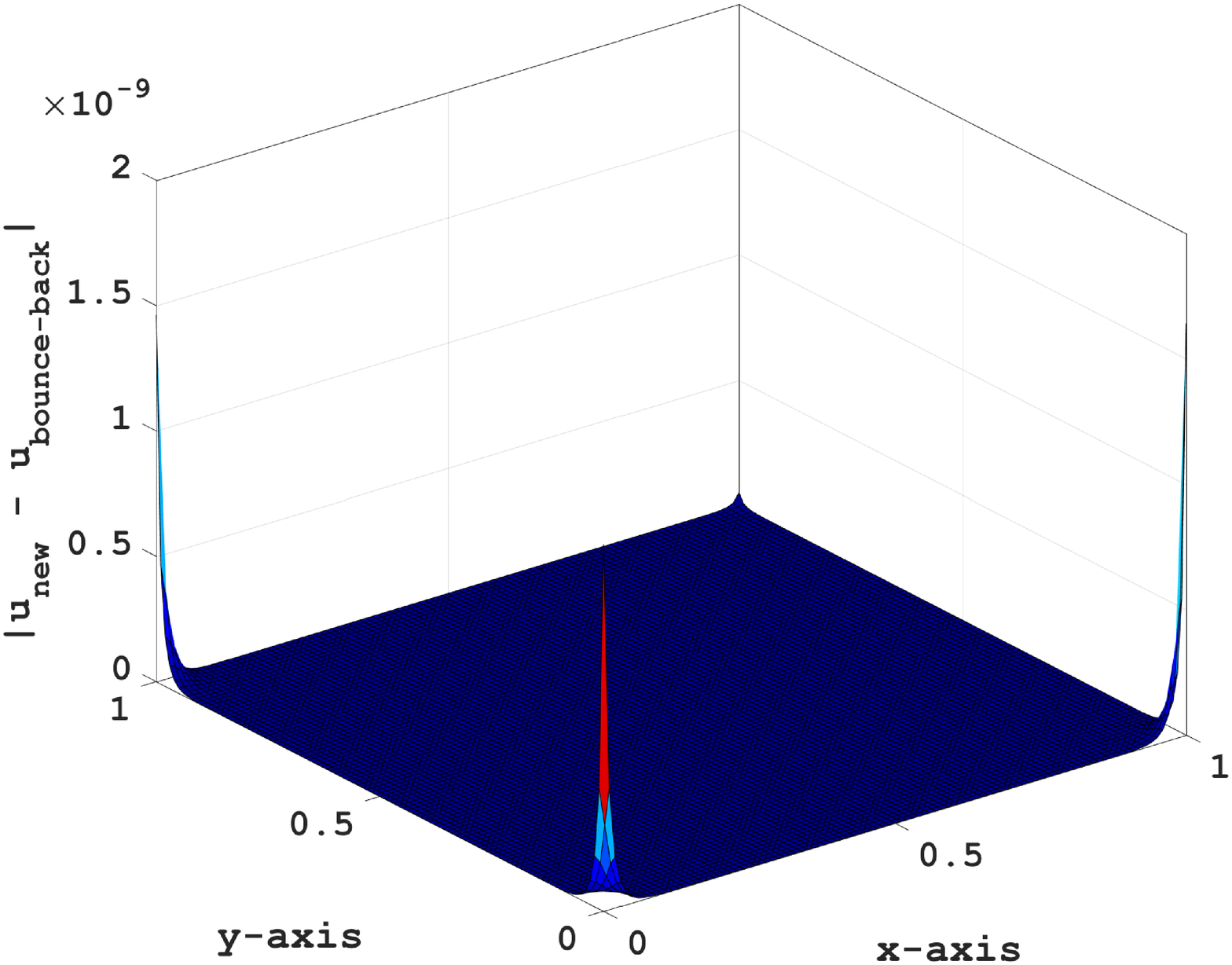}
  }
  \subfigure[proposed boundary condition vs. specular reflection method]{
    \includegraphics[clip,scale=0.35]{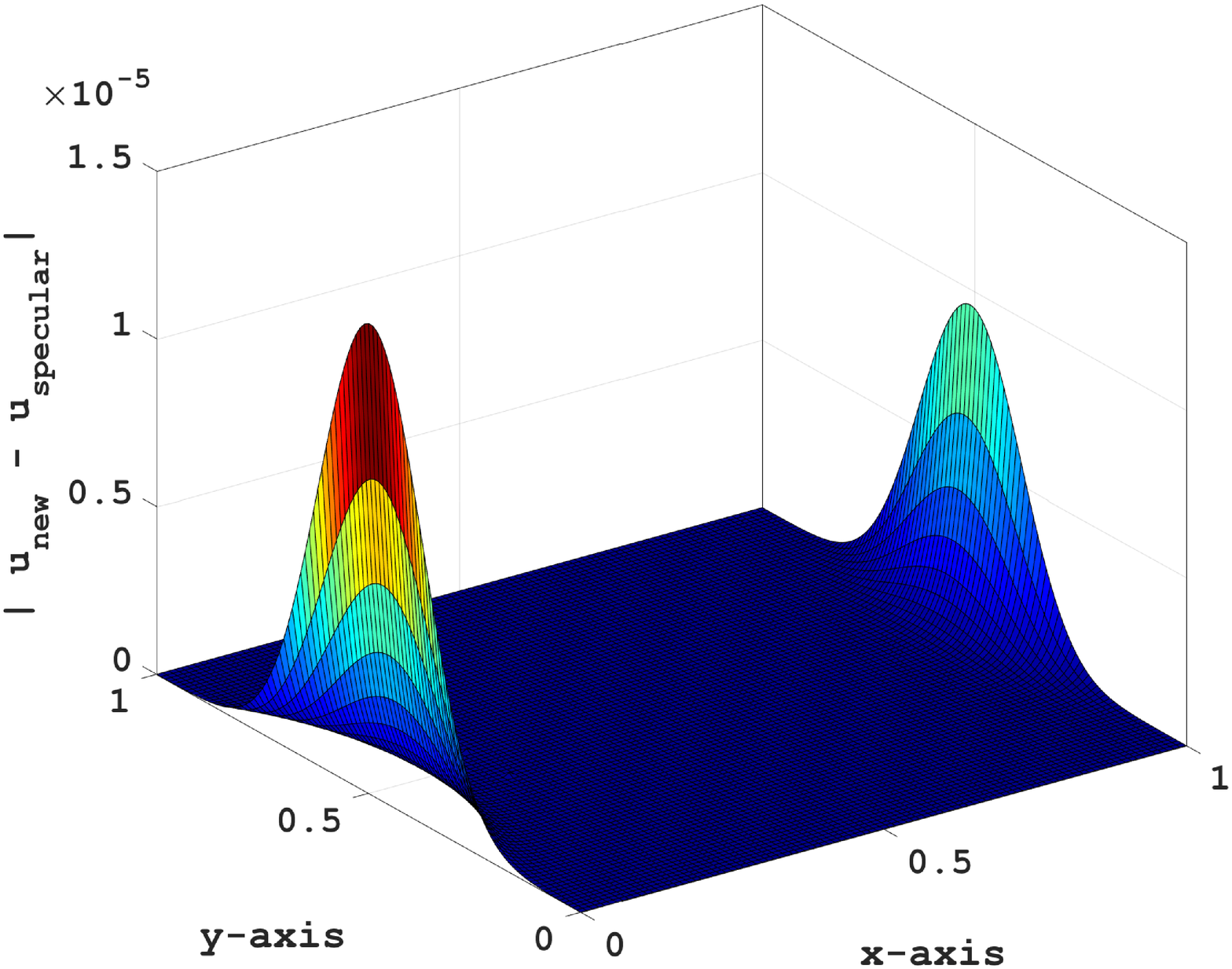}		
  }
  \caption{\textsf{Numerical results for LBM}:~
    In this figure, the difference between the numerical 
    solution due to different treatment of zero-flux boundary
    is shown. This implies that a macroscopic boundary condition
    (i.e., Dirichlet or Neumann) can be interpreted in 
    different ways in the context of lattice Boltzmann scheme.
    \label{Fig:LBM_sample_2}}
\end{figure}

Note that a zero-flux boundary (or any other macroscopic boundary
condition) can lead to various interpretations in the context of 
kinetic theory. For instance, a rigid and impermeable wall can 
lead to a zero-flux condition. Also, zero-flux can mean that there
is a bath of particles at a Maxwell-Boltzmann equilibrium state
with background velocity $\mathbf{v} = \mathbf{0}$. These 
interpretations are all valid in their own right. One needs to 
account for more physical details and choose the right method
for enforcing those conditions. To show the difference in the 
numerical results due to different treatment of boundaries under
lattice Boltzmann method, the given numerical example is solved 
using bounce-back and specular reflection methods \citep{1993_Ziegler_JSP,
2003_Yu_PAS}. The difference in the solution is shown in figure 
\ref{Fig:LBM_sample_2}. This difference should not be taken as a 
drawback of lattice Boltzmann method. It is in fact one of the 
advantages of kinetics-based methods over continuum-based methods. 
Extra information on the nature of interaction of particles with 
the boundary can be included in the numerical model. A continuum-based 
method may not be able to account for such details.

\section{AN OVERLAPPING DOMAIN DECOMPOSITION METHOD}
\label{Sec:Overlapping_DD}
Domain decomposition methods are powerful methods
for obtaining numerical solutions for partial
differential equations \citep{Quarteroni_book,
  Toselli_book}. These methods are particularly
effective in a parallel computing setting. The
basic idea is to split the computational domain
into an arbitrary number of subdomains and seek
the numerical solution in different subdomains
separately. These subdomains can be \emph{overlapping}
or  \emph{non-overlapping}. In a non-overlapping domain
decomposition scheme, one needs to account for an
interface equation to enforce compatibility of 
numerical solutions near the interface between
subdomains. Two of the more popular methods 
for constructing interface compatibility  conditions
are Lagrange multiplier framework, 
Steklov-Poincar\'e framework \citep{Quarteroni_book}. 
Introduction of such an interface condition may lead to 
higher complexity in the algorithm design but
is also shown to give accurate numerical solutions. 
Overlapping domain decomposition do not require
addition of a new interface constraint equation.
In the proposed hybrid coupling method, we shall
employ the overlapping domain decomposition
approach. We now describe the iterative Schwartz
method for numerical solution of a partial
differential equation in an overlapping domain
decomposition scheme. 

Consider a domain $\Omega$ with boundary $\Gamma =
\partial \Omega$. Consider the following equation
defined on this domain:
\begin{subequations}
  \begin{alignat}{2}
    &\mathcal{L}[\mathrm{u}] = \mathrm{f}
    &&\quad \text{in} \; \Omega \\
    &\mathrm{u} = \mathrm{u}^{\mathrm{p}}
    &&\quad \text{on} \; \Gamma
  \end{alignat}
\end{subequations}
For simplicity, we assume that the boundary condition is
purely Dirichlet, and employ two overlapping subdomains
$\Omega_{1}$ and $\Omega_{2}$ (i.e., $\Omega_{1}\cap
\Omega_{2} \neq \emptyset$ and $\Omega_1 \cup \Omega_2 
= \Omega$). The governing partial differential equations 
in each subdomain will be as follows
\begin{align}
\label{Eqn:DD_example}
	\left\{ \begin{array}{l l}
		\mathcal{L}[\mathrm{u}_{1}] = \mathrm{f}_{1} \quad 
		&\text{in} \; \Omega_{1} \\
		\mathrm{u}_{1} = \mathrm{u}_{2} \quad 
		&\text{on} \; \Gamma_{1} \cap \Omega_{2}\\
		\mathrm{u}_{1} = \mathrm{u}^{\mathrm{p}} \quad 
		&\text{on} \; \Gamma_{1} \cap \Gamma 
	\end{array}\right., \quad
	\left\{ \begin{array}{l l}
		\mathcal{L}[\mathrm{u}_{2}] = \mathrm{f}_{2} \quad 
		&\text{in} \; \Omega_{2} \\
		\mathrm{u}_{2} = \mathrm{u}_{1} \quad 
		&\text{on} \; \Gamma_{2} \cap \Omega_{1}\\
		\mathrm{u}_{2} = \mathrm{u}^{\mathrm{p}} \quad 
		&\text{on} \; \Gamma_{2} \cap \Gamma 
	\end{array}\right.
\end{align}
where the subindex is used to show the restriction of 
that quantity to the respective subdomain. The numerical
solution to the system given in \eqref{Eqn:DD_example}
can be found as
\begin{align}
	\left\{ \begin{array}{l l}
		\tilde{\mathcal{L}}[u_{1}^{k}] = f_{1} \quad 
		&\text{in} \; \Omega_{1} \\
		u_{1}^{k} = u_{2}^{k-1} \quad 
		&\text{on} \; \Gamma_{1} \cap \Omega_{2}\\
		u_{1}^{k} = \mathrm{u}^{\mathrm{p}} \quad 
		&\text{on} \; \Gamma_{1} \cap \Gamma 
	\end{array}\right., \quad
	\left\{ \begin{array}{l l}
		\tilde{\mathcal{L}}[u_{2}^{k+1}] = f_{2} \quad 
		&\text{in} \; \Omega_{2} \\
		u_{2}^{k+1} = u_{1}^{k} \quad 
		&\text{on} \; \Gamma_{2} \cap \Omega_{1}\\
		u_{2}^{k+1} = \mathrm{u}^{\mathrm{p}} \quad 
		&\text{on} \; \Gamma_{2} \cap \Gamma 
	\end{array}\right.
\end{align}
where $\tilde{\mathcal{L}}$ is the discrete differential
operator and super-indices $k-1$, $k$ and $k+1$ are used to
show consecutive iterations. The numerical solution in
one subdomain, from the previous iteration, is used to
determine the Dirichlet condition on the boundary of 
other subdomain. This approach can be extended to the
case were more than two subdomains are involved.

The advantages of overlapping domain decomposition methods
compared to non-overlapping methods are simpler algorithm
design, increased flexibility in choice of numerical solver 
in different subdomains, and easy incorporation of non-matching 
grids and multi-time-stepping. In the following section, we will 
further scrutinize the methods for projecting data from a 
coarse mesh to a fine grid, and vice versa.
\subsection{Transfer of information across non-matching grids}
Typically, the grid-size for a coarse-scale simulation is much larger 
than the grid-size for a fine-scale simulation. Under lattice 
Boltzmann method, small cell-size can help accounting for complex
spatial features of the computational domain in a fine-scale simulation.
Upscaled (averaged) models for flow and transport in porous media
such as Darcy's model do not need any such details of the pore 
structure, hence, the computational mesh for numerical solution of
these models can be coarse. Under the proposed hybrid coupling 
method and the domain decomposition schemes introduced above, 
interaction among different subdomains occurs through the 
interface between any two subdomains. Transfer of information, 
consequently, needs to be done between non-matching grids that 
are disparate in size. This issue has been an active area of
study in recent years. For instance, in simulation of fluid-structure
interaction problems, traction at the interface of fluid and solid
needs to be interpolated between non-matching grids 
\citep{2004_Heath_IJNME,2006_Geubelle_JCP, 2007_deBoer_CMAME}. 
In the context of overlapping domain decomposition schemes,
numerical methods for flow and transport simulation on
overlapping grids in \citep{1990_Henshaw_JCP,2003_Henshaw_JCP} 
and a study of stability of interpolation at the interface of 
subdomains in \citep{1996_Olsson_CF} can be mentioned. However,
in this paper, since we intend to use different numerical methods
in different subdomains (i.e., FEM or LBM), the interpolation 
for concentration is not alike. Our approach to transfer 
the values of concentration at the interface and across non-matching
grids is described next.

\begin{figure}
  \centering
  \includegraphics[clip, scale=0.75]{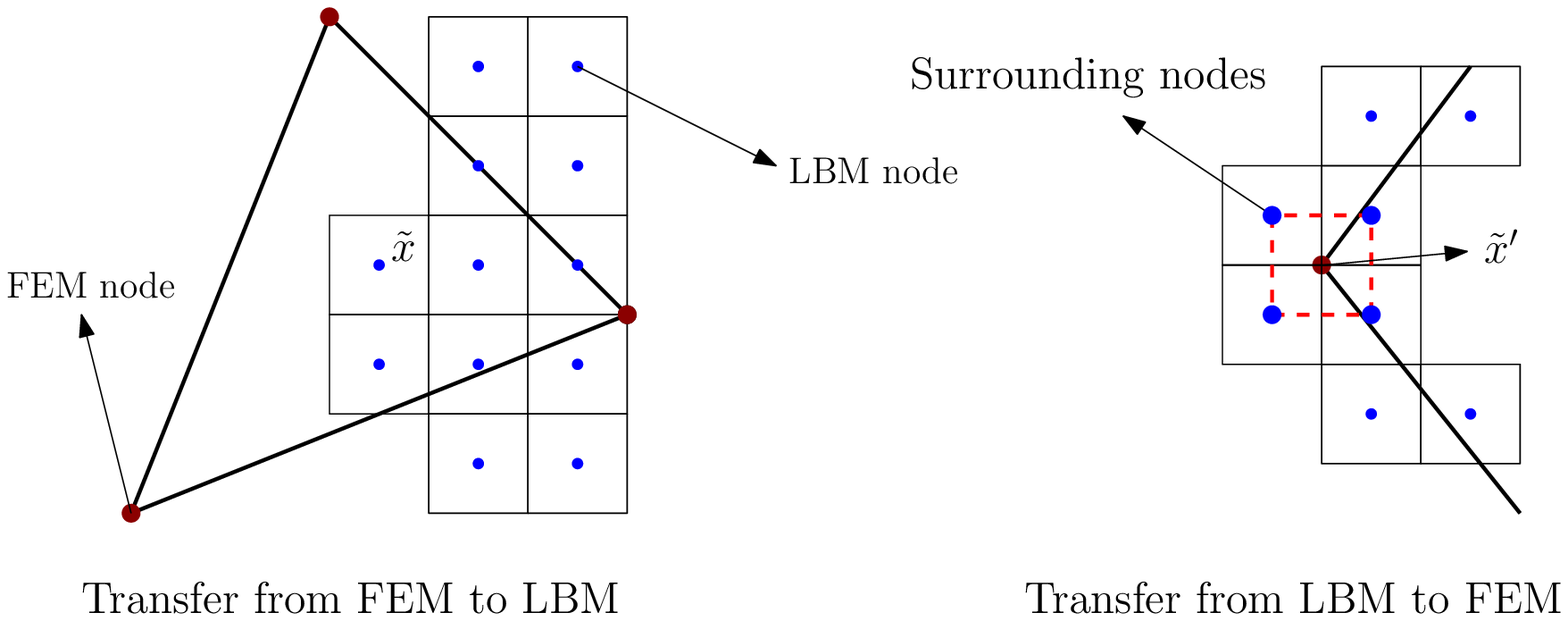}
  \caption{\textsf{Transfer of information across non-matching grids}:~
    A pictorial description for interpolation across non-matching 
    grids is provided. In transferring information from FEM to LBM,
    one only needs to use the FEM interpolation already in use to
    approximate concentration at the lattice node. In transferring 
    from LBM to FEM however, the surrounding lattice nodes need to 
    be located to form a patch.
    \label{Fig:InfoTransfer}}
\end{figure}

Consider a two-dimensional domain and let $\tilde{\boldsymbol{x}}$ 
be the coordinates of a cell lying on the boundary of a subdomain 
of LBM discretization. Then, the values of the concentration at 
this point can be approximated via the finite element interpolation 
on the element that contains the point $\tilde{\boldsymbol{x}}$. 
To approximate concentration at a point $\tilde{\boldsymbol{x}}'$ 
that lies on the boundary of the subdomain with FEM discretization, 
one needs to locate the surrounding cells (of the LBM solution). 
Hence, the point $\tilde{\boldsymbol{x}}'$ is inside the square 
patch with the surrounding LBM nodes at the corners. Concentration
at this finite element node can be approximated using the 
values of concentration at the surrounding points of the square
patch. For instance, one can use a four-node quadrilateral finite element
interpolation (figure \ref{Fig:InfoTransfer} provides a pictorial description). 
Obviously, three-dimensional cases can be handled
similarly, however, the choice of interpolation function can 
be more varied (e.g., one can use interpolation functions over 
hexagonal or tetrahedral elements identified with surrounding
LBM nodes). 

\begin{figure}[h]
  \centering
  \subfigure[approximation of $\mathrm{g}(\mathrm{x},\mathrm{y})$
    on coarse grid:~$\mathrm{g}(\mathrm{x},\mathrm{y}) \rightarrow \mathrm{g}_{\mathrm{c}}(\mathrm{x},\mathrm{y})$]{
    \includegraphics[clip, scale=0.3]{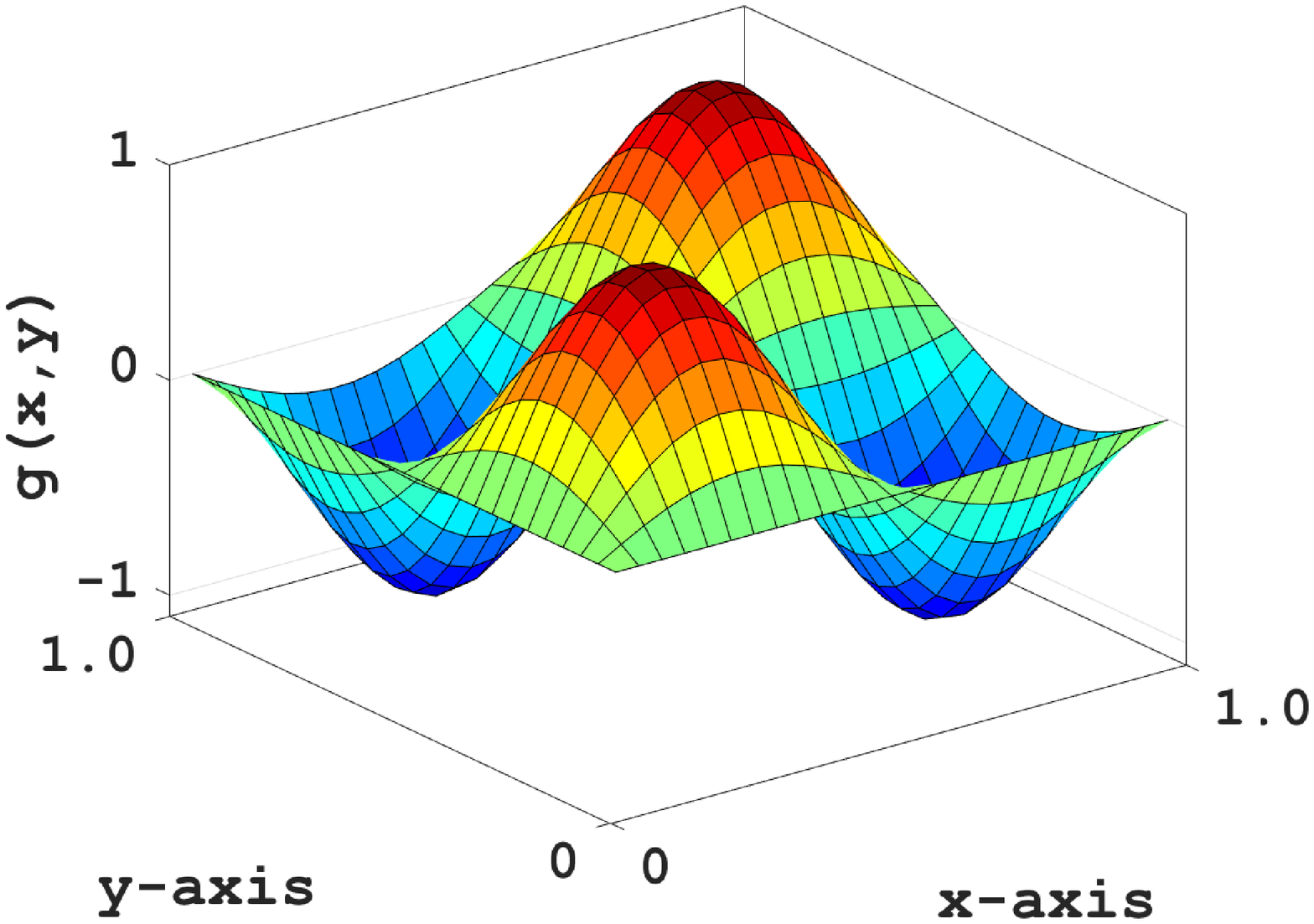}
  }
  \subfigure[transfer of information from coarse grid
    to fine grid:~$\mathrm{g}_{\mathrm{c}}(\mathrm{x},\mathrm{y})
    \rightarrow \mathrm{g}_{\mathrm{f}}(\mathrm{x},\mathrm{y})$]{
    \includegraphics[clip, scale=0.3]{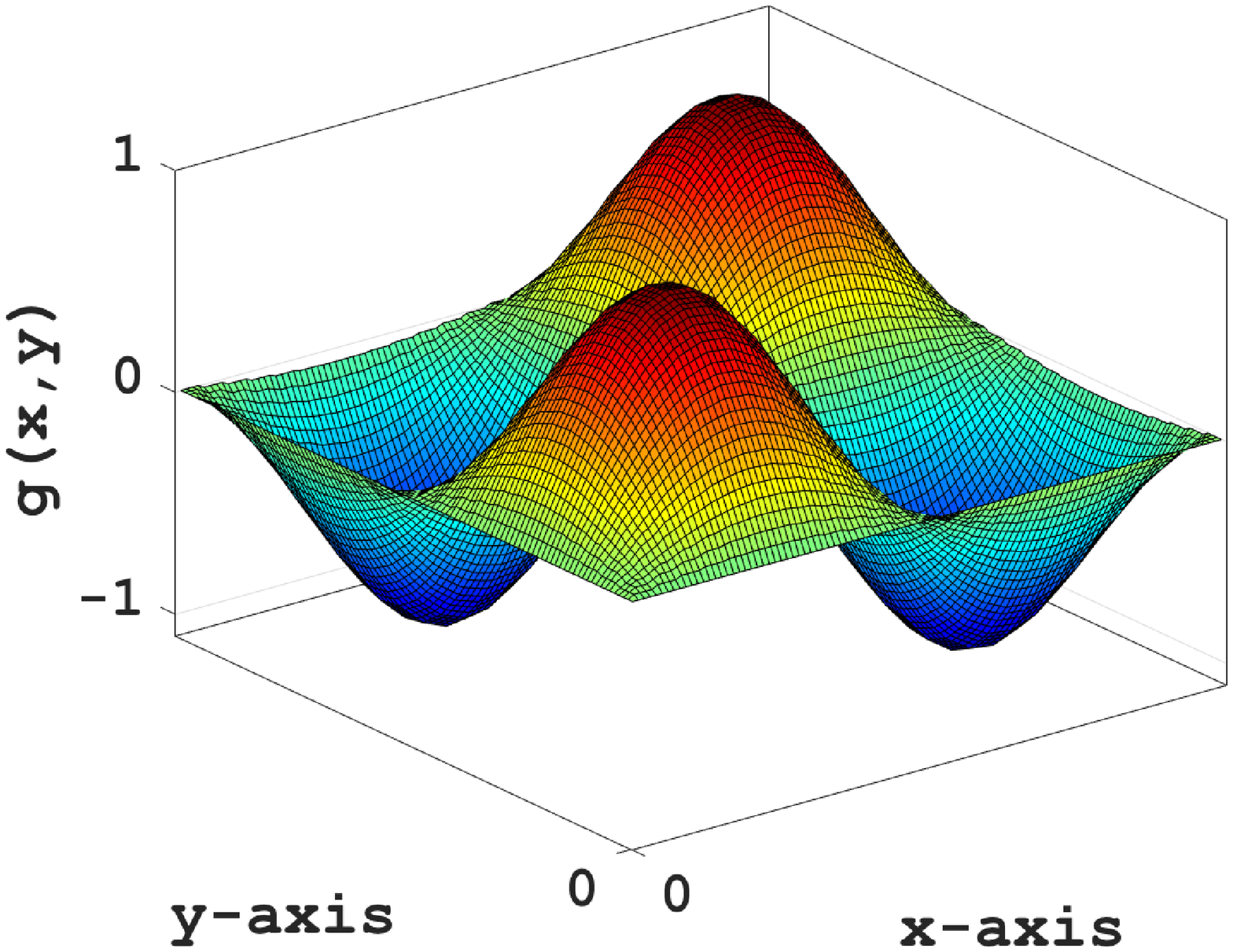}
  }
  \subfigure[error on the fine grid:~$|\mathrm{g}_{\mathrm{f}}(\mathrm{x}, \mathrm{y})
    - \mathrm{g}(\mathrm{x},\mathrm{y})|$]{
    \includegraphics[clip, scale=0.3]{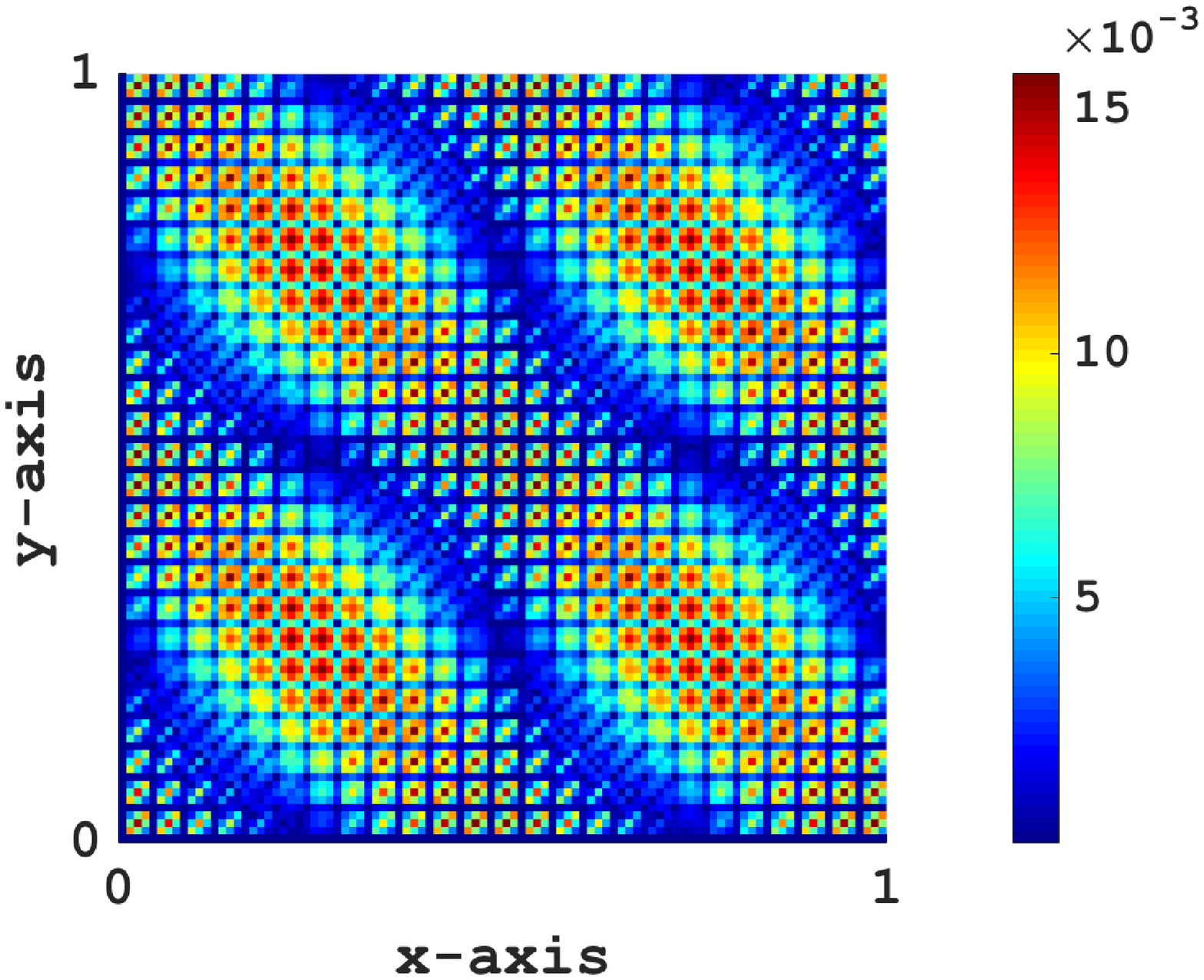}
  }
  \caption{\textsf{Transfer of information across non-matching grids}:~
    The function $\mathrm{g}(\mathrm{x},\mathrm{y})$ is first approximated
    on a coarse grid. Then, it is mapped onto a fine grid. The error in the 
    final approximation (on fine grid) is also shown. The data corresponds to 
    Case 2 in Table \ref{Tbl:FEM2LBM}.
    \label{Fig:FEM2LBM}}
\end{figure}

\begin{figure}[h]
  \centering
  \subfigure[approximation of $\mathrm{g}(\mathrm{x},\mathrm{y})$ on fine grid:
  $\mathrm{g}(\mathrm{x},\mathrm{y}) \rightarrow \mathrm{g}_{\mathrm{f}}(\mathrm{x},\mathrm{y})$]{
    \includegraphics[clip, scale=0.3]{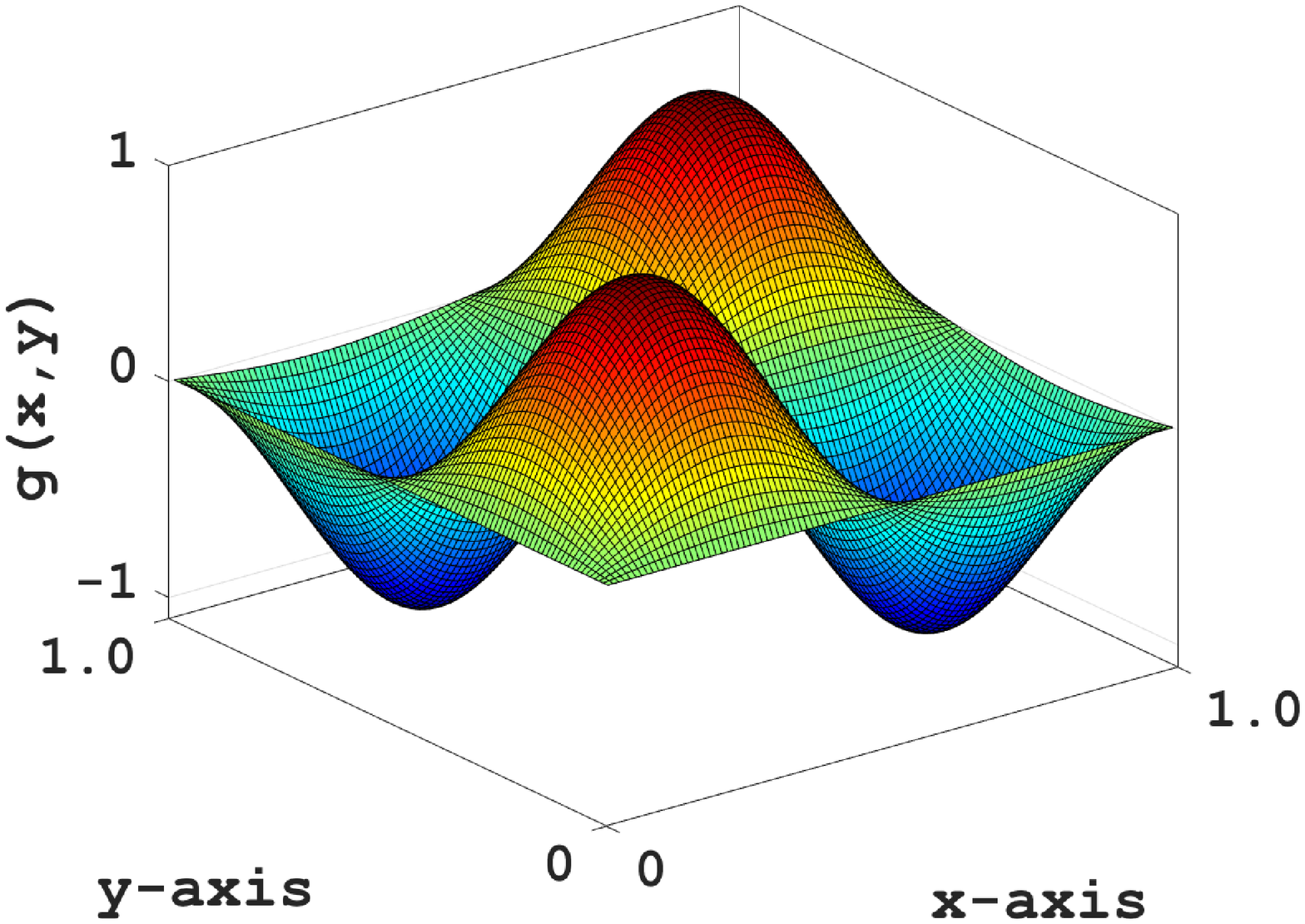}
  }
  \subfigure[transfer of information from fine grid to coarse grid:
  ~$\mathrm{g}_{\mathrm{f}}(\mathrm{x},\mathrm{y})
    \rightarrow \mathrm{g}_{\mathrm{c}}(\mathrm{x},\mathrm{y})$]{
    \includegraphics[clip, scale=0.3]{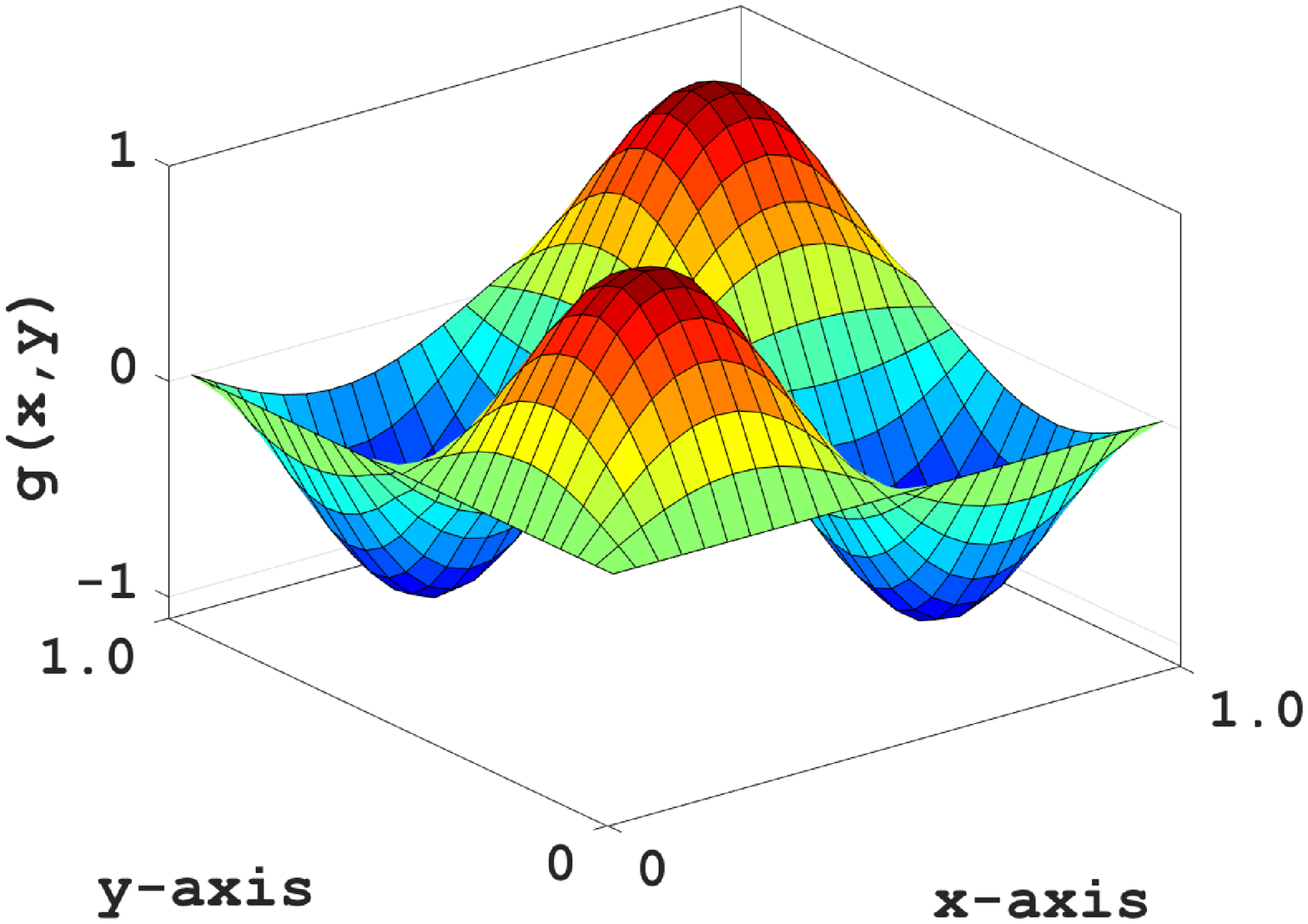}
  }
  \subfigure[error on coarse grid:~$| \mathrm{g}_{\mathrm{c}}(\mathrm{x}, \mathrm{y})
    - \mathrm{g}(\mathrm{x},\mathrm{y}) |$]{
    \includegraphics[clip, scale=0.3]{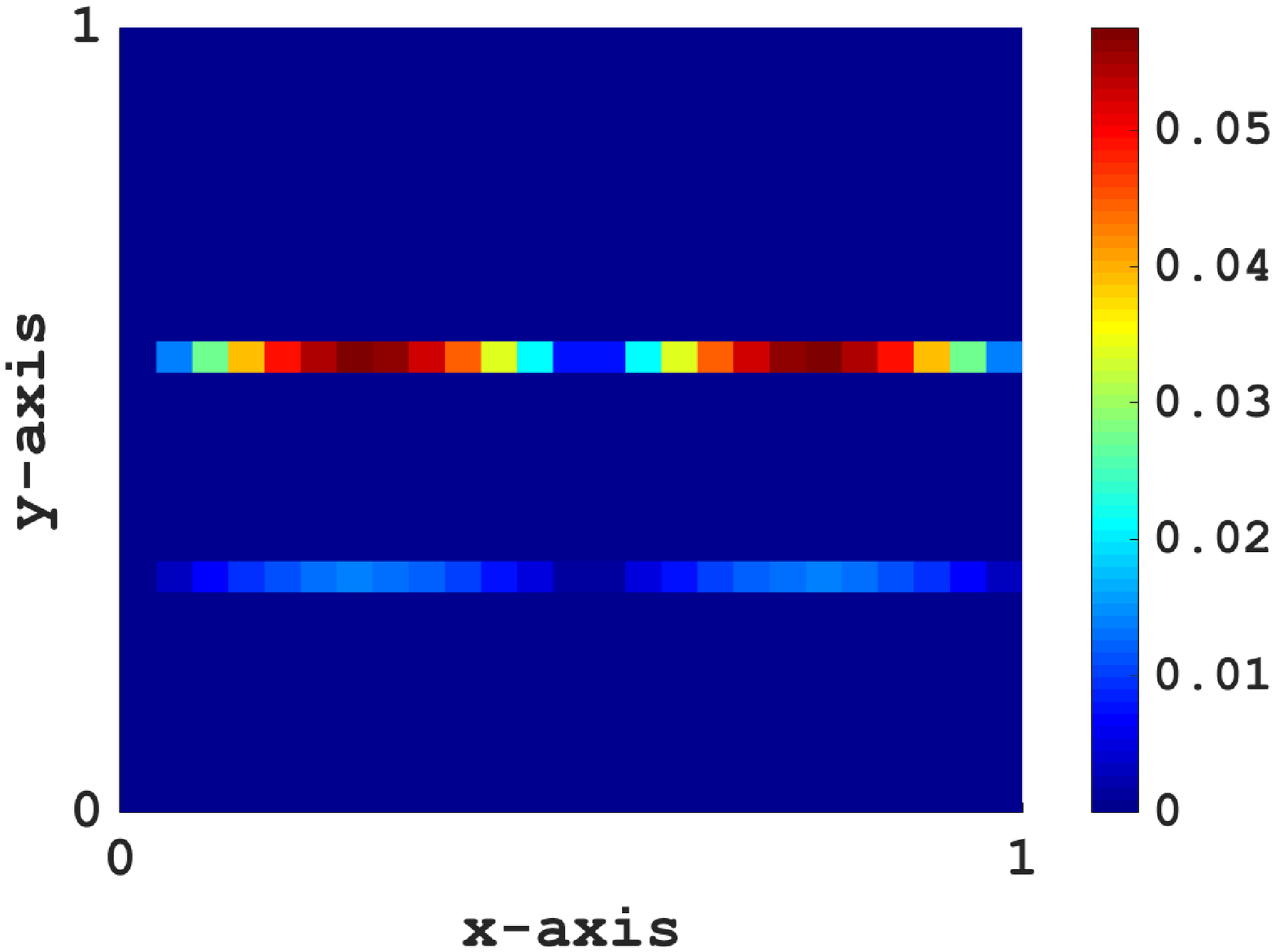}
  }
  \caption{\textsf{Transfer of information across non-matching grids}:~
    The function $\mathrm{g}(\mathrm{x},\mathrm{y})$ is
    first approximated on a fine grid. Then, it is mapped
    onto a coarse grid using the method described in figure
    \ref{Fig:InfoTransfer}. The error on the coarse grid is
    shown as well. The data corresponds to Case 2 in Table
    \ref{Tbl:LBM2FEM}. \label{Fig:LBM2FEM}}
\end{figure}

To demonstrate this technique, consider the following function
defined over domain $\Omega = (0,1)\times(0,1)$:
\begin{align}
	\mathrm{g}(\mathrm{x}, \mathrm{y}) = \sin(2\pi\mathrm{x}) \sin(2\pi\mathrm{y})
\end{align}
The coarse grid size will be denoted by $h$ (linear three-node 
triangular elements used) and the fine grid size
is shown by $h'$. The maximum error in the domain is denoted 
by $\mathcal{E}_{\max}$ and is defined by:
\begin{align}
	\mathcal{E}_{\max} = \max_{(x,y)} \left| \mathrm{g}(x,y) - g(x,y)\right|
\end{align}
where $(x,y)$ is a point on grid and $g(x,y)$ is the approximation
of function $\mathrm{g}$ on a computational grid (either coarse or fine).
Numerical results for transferring information 
across non-matching grids is given in Tables \ref{Tbl:FEM2LBM} and 
\ref{Tbl:LBM2FEM}. From Table \ref{Tbl:FEM2LBM}, we can conclude that
the accuracy on the fine grid changes as $\mathcal{O}(h^2)$, which is 
expected, as it complies with the convergence rate of finite element
approximation \citep{1987_Babuska_SIAM}. However, the error on the coarse 
grid, with information transferred to it from the fine domain, is 
$\mathcal{O}(h')$. Figures \ref{Fig:FEM2LBM} and \ref{Fig:LBM2FEM} show 
some demonstrative numerical results and outline the process given above. 
From this numerical experiment we conclude that, a bottleneck in convergence 
of the proposed coupling method can be the accuracy of fine to coarse grid 
information transfer.

\begin{table}
  \caption{\textsf{Transfer of information across non-matching
      grids}:~The numerical result for transfer of information
    from coarse grid to fine grid is given. The error is
    $\mathcal{O}(h^{2})$, as expected.}
  \label{Tbl:FEM2LBM}
  \begin{tabular}{|c|c|c|c|}
    \hline 
    Case & $h$ & $h'$ & $\mathcal{E}_{\max}$ \\ \hline
    1&$10^{-1}$ & $10^{-2}$ & $9.55\times10^{-2}$ \\ \hline
    2&$4.0\times10^{-2}$ & $10^{-2}$ & $1.57\times10^{-2}$ \\ \hline
    3&$2.0\times10^{-2}$ & $10^{-2}$ & $3.94\times10^{-3}$ \\ \hline 
  \end{tabular}
\end{table}

\begin{table}
  \caption{\textsf{Transfer of information across
      non-matching grids}:~In this table, numerical
    values for transferring information from fine
    grid to coarse grid is presented. The error
    in the values of the coarse-grid approximation
    behaves as $\mathcal{O}(h')$.}
  \label{Tbl:LBM2FEM}
  \begin{tabular}{|c|c|c|c|}
    \hline 
    Case&$h$ & $h'$ & $\mathcal{E}_{\max}$ \\ \hline
    1&$4.0\times10^{-2}$ & $2.0\times10^{-2}$ & $1.13\times10^{-1}$ \\ \hline
    2&$4.0\times10^{-2}$ & $10^{-2}$ & $5.75\times10^{-2}$ \\ \hline
    3&$4.0\times10^{-2}$ & $5.0\times10^{-3}$ & $2.90\times10^{-2}$ \\ \hline 
  \end{tabular}
\end{table}

\section{A NEW HYBRID MULTI-TIME-STEP COUPLING}
\label{Sec:Hybrid_Coupling_Algorithm}
In this section, we shall present a robust coupling 
method that allows hybrid modeling to be able to 
couple pore- and continuum-scale subdomains with 
disparate time-scales for solute transport in
porous media.
The spatial domain of interest $\Omega$ is partitioned
into overlapping subdomains. The subdomains where
\emph{fine-scale} features of the solution are sought
are denoted by $\Omega_{\mathrm{f}}$. 
Subdomains in which \emph{coarse-scale} features are solved for 
are shown by $\Omega_{\mathrm{c}}$. Figure \ref{Fig:DomainDec} 
provides a pictorial description of this partitioning scheme.
In this paper, we will employ a finite element discretization 
in subdomain $\Omega_{\mathrm{c}}$. This finite element method 
is applied to the equation \eqref{Eqn:AD}. The fine-scale features in 
subdomain $\Omega_{\mathrm{f}}$ are solved for using the lattice 
Boltzmann method, which solves the Boltzmann's transport equation in 
equation \eqref{Eqn:Boltzmann} in the mentioned region. Compatibility 
of the solutions is enforced using a Dirichlet condition 
at $\Gamma_{\mathrm{f} \rightarrow \mathrm{c}}$ and $\Gamma_
{\mathrm{c}\rightarrow\mathrm{f}}$. The time- and space-continuous 
partial differential equations in each subdomain, along with 
their respective boundary conditions, are as follows: 
\begin{subequations}
\label{Egn:CouplingPDE}
\begin{align}
	&\text{in} \; \Omega_{\mathrm{c}}:\; \left\{ \begin{array}{l l}
		\partial \mathrm{u}/\partial \mathrm{t} + \mathrm{div}\left[ 
		\mathbf{v}\mathrm{u} - \mathbf{D} \mathrm{grad}\left[ \mathrm{u} 
		\right]\right]
		= \mathrm{s} \quad &(\mathbf{x},\mathrm{t}) \in \Omega_{\mathrm{c}} 
		\times \mathcal{I} \\
		\mathrm{u}(\mathbf{x},\mathrm{t}=0) = \mathrm{u}_{0}(\mathbf{x}) 
		\quad &\mathbf{x} \in 						\Omega_{\mathrm{c}} \\
		\mathrm{u}(\mathbf{x},\mathrm{t}) = \mathrm{u}^{\mathrm{p}}
		(\mathbf{x},\mathrm{t}) \quad &(\mathbf{x},\mathrm{t}) \in 
		\Gamma^{\mathrm{D}} \times \mathcal{I} \\
		\left( \mathbf{v} \mathrm{u} - \mathbf{D} \mathrm{grad}\left[ 
		\mathrm{u}\right]\right) 
		\cdot \widehat{\mathbf{n}}(\mathbf{x}) = \mathrm{q}^{\mathrm{p}}
		(\mathbf{x},\mathrm{t}) \quad
		&(\mathbf{x},t) \in \Gamma^{\mathrm{N}} \times \mathcal{I} \\
		\emph{coupling condition}:\; \mathrm{u}(\mathbf{x},\mathrm{t}) =
			\tilde{\mathrm{u}}_{\mathrm{f} \rightarrow \mathrm{c}} 
			(\mathbf{x},\mathrm{t}) \quad
		&(\mathbf{x},\mathrm{t}) \in \Gamma_{\mathrm{f} \rightarrow \mathrm{c}} \times 
		\mathcal{I} 
	\end{array} \right. \\
	&\text{in}\;\Omega_{\mathrm{f}}: \; \left\{ \begin{array}{l l}
		\partial f/\partial \mathrm{t} + \mathbf{v}\cdot\mathrm{grad}[f] = 
		\left( f^{\mathrm{eq.}} - 
		f\right)/\lambda \quad &(\mathbf{x}, \boldsymbol{\zeta}, \mathrm{t}) \in 
		\Omega_{\mathrm{f}} \times\mathbb{R}^{n}  \times \mathcal{I} \\
		f(\mathbf{x},\boldsymbol{\zeta},t=0) = f^{\mathrm{eq}}(\mathbf{x},
		\boldsymbol{\zeta},\mathrm{t}=0;\mathrm{u}_{0}(\mathbf{x}),\mathbf{v}) 
		\quad &(\mathbf{x},\boldsymbol{\zeta}) \in \Omega_{\mathrm{f}} \times 
		\mathbb{R}^{n} \\
	    \int f \mathrm{d}\boldsymbol{\zeta} = \mathrm{u}^{\mathrm{p}}
	    (\mathbf{x},\mathrm{t}) 
		\quad &(\mathbf{x},\mathrm{t})\in\Gamma^{\mathrm{D}}\times \mathcal{I}\\
		\left( \int f \boldsymbol{\zeta} \mathrm{d}\boldsymbol{\zeta} \right)
		\cdot \widehat{\mathbf{n}}(\mathbf{x}) = \mathrm{q}^{\mathrm{p}}
		(\mathbf{x},\mathrm{t}) &
		(\mathbf{x},t)\in\Gamma^{\mathrm{N}}\times \mathcal{I}\\
		\emph{coupling condition}:\; \int f  \mathrm{d}\boldsymbol{\zeta} = 
		\tilde{\mathrm{u}}_{\mathrm{c}\rightarrow\mathrm{f}}(\mathbf{x},\mathrm{t})  
		\quad &(\mathbf{x},\mathrm{t})\in\Gamma_{\mathrm{c} \rightarrow\mathrm{f}}\times 
		\mathcal{I}
	\end{array} \right.
\end{align}
\end{subequations}
This set of equations provides a basis to employ numerical 
methods of different origins in the same computational domain. 
The advection-diffusion equation is rooted in the continuum theory. 
The Boltzmann's equation however, is based on the kinetic theory. 
Using equation \eqref{Egn:CouplingPDE}, one can solve for physical 
features at different temporal and spatial scales (macroscopic vs. 
mesoscopic), in a single computational framework. In the following, 
we will provide the temporal and spatial discretization of equation 
\eqref{Egn:CouplingPDE}.
\subsection{Space and time discretization}
\subsubsection{Coarse-scale problem}
\begin{figure}
  \centering
  \psfrag{O}{$\Omega$}
  \psfrag{Of}{$\Omega_{\mathrm{f}}$}
  \psfrag{Oc}{$\Omega_{\mathrm{c}}$}
  \psfrag{Gc}{$\Gamma_{\mathrm{c} \rightarrow \mathrm{f}}$}
  \psfrag{Gf}{$\Gamma_{\mathrm{f} \rightarrow \mathrm{c}}$}
  \psfrag{Ofc}{$\Omega_{\mathrm{c}} \cap \Omega_{\mathrm{f}} \neq \emptyset$}
  \includegraphics[clip, scale=1]{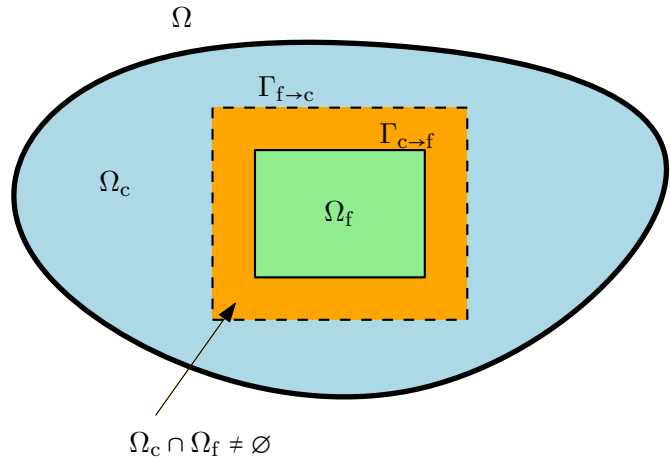}
  \caption{\textsf{Overlapping domain partitioning}:~The proposed 
    coupling method employs overlapping domain partitioning. This 
    figure illustrates the decomposition of the computational domain
    $\Omega$ into the subdomains where coarse-scale and fine-scale 
    features are sought after. These subdomains are denoted by 
    $\Omega_{\mathrm{c}}$ and $\Omega_{\mathrm{f}}$, respectively.
    The portions of the boundary of the mentioned subdomains where
    transfer of information occurs are denoted by $\Gamma_
    {\mathrm{f}\rightarrow\mathrm{c}}$ and $\Gamma_{\mathrm{c}\rightarrow \mathrm{f}}$.
    Obviously, $\Gamma_{\mathrm{f}\rightarrow\mathrm{c}} \subseteq \partial \Omega_{\mathrm{c}}$
    and $\Gamma_{\mathrm{c}\rightarrow\mathrm{f}} \subseteq \partial \Omega_{\mathrm{f}}$.
    An attractive feature of this partitioning technique is that the grid sizes
    in subdomains $\Omega_{\mathrm{c}}$ and $\Omega_{\mathrm{f}}$ need not be 
    conforming. \label{Fig:DomainDec}}
\end{figure}

The coarse-scale problem is defined by equation \eqref{Eqn:AD}, 
over domain $\Omega_{\mathrm{c}}$ in Figure \ref{Fig:DomainDec}. 
We will use the semi-discrete methodology \citep{1977_Zienkiewicz_Book} 
, which gives the following time-continuous equation for the 
coarse-scale problem
\begin{align}
\label{Eqn:FEM_semi}
	\boldsymbol{M} \dot{\boldsymbol{u}} + \boldsymbol{K} \boldsymbol{u}
	= \boldsymbol{s} 
\end{align}
where $\boldsymbol{M}$ is the capacity matrix, $\boldsymbol{K}$ is 
the transport matrix and $\boldsymbol{u}$ is the nodal concentration.
The superposed dot denotes the time derivative. The discretized
right-hand-side of the finite element weak formulation is shown by 
$\boldsymbol{s}$. For time discretization, we will use the following 
notation
\begin{align}
\label{Eqn:FEM_notation}
	t^{(n)} = n \Delta t_{\mathrm{c}}, \quad
	\boldsymbol{u}(t^{(n)}) \approx \boldsymbol{d}^{(n)}, \quad
	\dot{\boldsymbol{u}}(t^{(n)}) \approx \boldsymbol{v}^{(n)}
\end{align}
where $\Delta t_{\mathrm{c}}$ is the time-step used for integrating 
the coarse-scale problem. Using the trapezoidal method for time 
integration yields the following system of equations
\begin{subequations}
\label{Eqn:FEM_discrete}
\begin{align}
	&\boldsymbol{M} \boldsymbol{v}^{(n+1)} + 
	\boldsymbol{K} \boldsymbol{d}^{(n+1)} = \boldsymbol{s}^{(n+1)}\\
	&\boldsymbol{d}^{(n+1)} = \boldsymbol{d}^{(n)} + 
		\Delta t_{\mathrm{c}} (1-\vartheta) \boldsymbol{v}^{(n)}
		+ \Delta t_{\mathrm{c}} \vartheta \boldsymbol{v}^{(n+1)}
\end{align}
\end{subequations}
where $\vartheta \in [0,1]$ is the time-integration parameter 
\citep{1990_Wood}. In this paper, we will use $\vartheta=1/2$, 
which gives a second-order accurate and unconditionally stable 
time-integrator (the midpoint rule). Once the the value of
flux $\tilde{\mathrm{u}}_{\mathrm{c} \rightarrow \mathrm{f}}$ is 
known, the values for nodal concentrations 
$\boldsymbol{d}^{(n+1)}$ and the rate variable $\boldsymbol{v}^{(n+1)}$ 
can be found. In the following section we will briefly overview the 
discretization of the fine-scale problem.
\subsubsection{Fine-scale problem}
Our objective is to solve for the distribution of particles in the
phase space defined by $\Omega_{\mathrm{f}} \times \mathbb{R}^{n}$.
This goal can be achieved by solving the Boltzmann equation 
\eqref{Eqn:Boltzmann} numerically. The lattice Boltzmann
method, introduced in Section \ref{Sec:Hybrid_Pore-scale}, can 
provide relevant numerical results for simulation of the 
advection-diffusion process.
 
Consider a uniform grid, with the spacing between the cells equal
to $h_{\mathrm{f}}$, defined over the domain $\Omega_{\mathrm{f}}$.
We will denote the time-step for the fine-scale problem 
by $\Delta t_{\mathrm{f}}$, and the ratio $\eta = \Delta t_{\mathrm{c}}
/\Delta t_{\mathrm{f}}$. The procedure for updating the discrete 
populations over a time-step is the same as what was outlined earlier 
in Section \ref{Sec:Hybrid_Pore_Scale_LBM}. 
In the following section, we will describe the new computational 
framework in detail and point out the transfer of data from 
fine-scale to coarse-scale domain.
\subsection{The proposed hybrid computational framework}
\begin{figure}
	\centering
	\includegraphics[clip,scale=1.5]{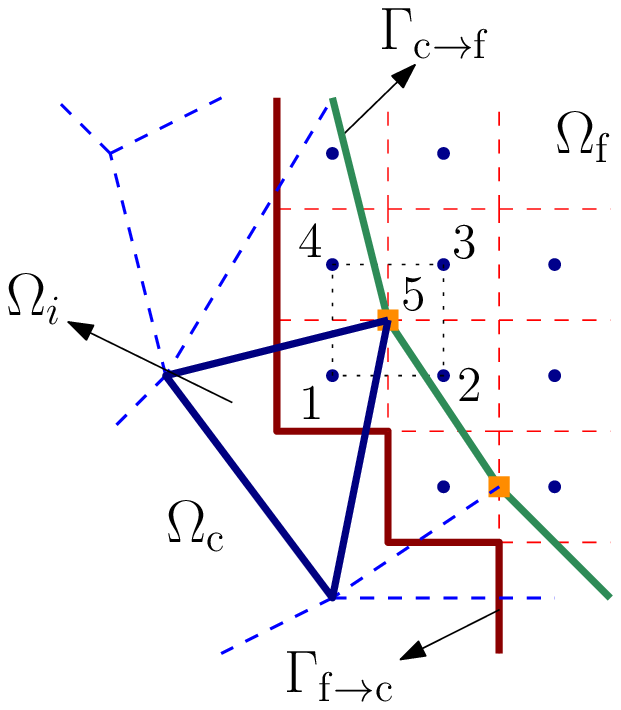}
	\caption{{\sf{Information transfer across non-matching grids}}:~
		In this figure, transfer of information at the interface
		of subdomains is depicted. The proposed coupling method
		allows non-conforming grids, which is a basic requirement
		of any multi-scale computational framework. Nodes 1 to 4
		and 6 denote representative LBM cells and node 5 is a 
		finite element node. The spatial domain inside $i$-th
		element is denoted by $\Omega_{i}$, which is an arbitrary
		element near $\Gamma_{\mathrm{c}\rightarrow\mathrm{f}}$.
	\label{Fig:Interface}}
\end{figure}

\begin{figure}
  \centering
  \includegraphics[clip, scale=1.0]{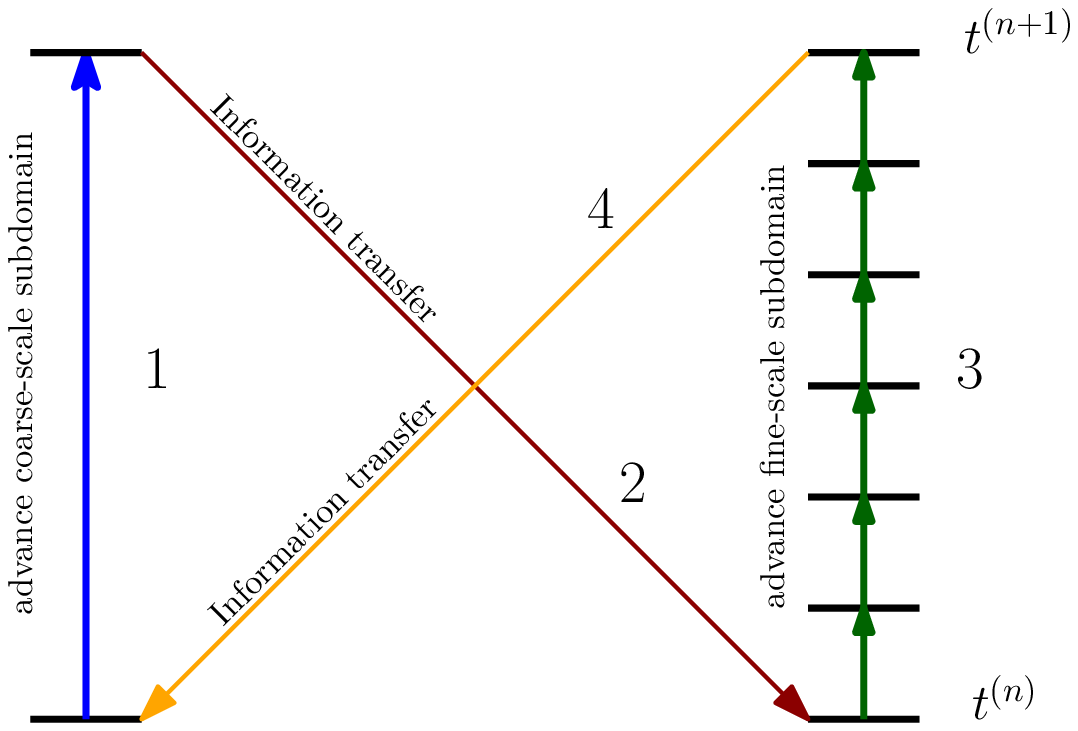}
  \caption{{\sf{Proposed coupling method}}:~ In this figure,
  	use of multiple time-steps for time-integration under the 
	proposed coupling framework is demonstrated. In step 1, 
	the solution in the coarse-scale domain advances in 
	time by $\Delta t_{\mathrm{c}}$. In step 2, interface
	information is transferred to the fine-scale domain,
	using the updated values from the coarse-scale domain.
	The solution in the fine-scale domain progresses 
	in time in step 3. In step 4, updated solution in the
	fine-scale domain is used to alter the interface
	condition for the coarse-scale domain.
  \label{Fig:TimeInt}}
\end{figure}

Before providing a step-by-step procedure for a numerical simulation
using the proposed framework, we need to introduce a set of tools that
will be useful. These tools will enable multi-time-step integration and
information transfer across non-matching grids. The details are as
below:
\begin{enumerate}[(i)]
\item \textsf{Initializing the discrete unknowns}:~
	In $\Omega_{\mathrm{c}}$ we utilize a finite element discretization.
	The nodal concentrations $\boldsymbol{d}$ can be simply initialized
	according to $\mathrm{u}_{0}(\mathbf{x})$. In $\Omega_{\mathrm{f}}$
	however, we assume that for the given initial concentration, the 
	discrete populations $f_{i}$ are given as
	\begin{align}
		f_{i}(\boldsymbol{x}, t=0) = 
		f_{i}^{\mathrm{eq}}(\boldsymbol{x},t=0;\mathrm{u}_{0}(\boldsymbol{x}))
	\end{align}
	Other methods for initializing the discrete populations can also
	be considered.
\item \textsf{Information transfer across the interface}:~
	To identify values of prescribed concentration on interfaces 
	$\Gamma_{\mathrm{f}\rightarrow\mathrm{c}}$ and 
	$\Gamma_{\mathrm{c}\rightarrow\mathrm{f}}$, we need to approximate 
	the concentration at nodes lying on these boundaries. Figure 
	\ref{Fig:Interface} is an illustrative example of lattice and 
	finite elements at the boundary of each subdomain. We will denote
	the coordinates of the point $j$, numbered in figure \ref{Fig:Interface},
	as $(x_j,y_j)$ and the concentration at that node as $u_i$.
	For given concentrations at nodes 1 to 4, the concentration at node
	5, which belongs to a finite element in $\Omega_{\mathrm{c}}$, can
	be approximated as follows:
	\begin{align}
		u_{5} \approx u_1(1 - \gamma_{x})(1 - \gamma_y) + 
			u_2 \gamma_x (1 - \gamma_y)
			+ u_3 \gamma_x \gamma_y + u_4 (1 - \gamma_x) \gamma_y
	\end{align}
	where $\gamma_x = (x_5 - x_1)/h_{\mathrm{f}}$ and 
	$\gamma_y = (y_5 - y_1)/h_{\mathrm{f}}$. This method is obviously
	synonymous to approximation via a four-node quadrilateral element
	with its vertices lying on nodes 1 to 4. To transfer information 
	from $\Omega_{\mathrm{c}}$ to $\Omega_{\mathrm{f}}$, for instance 
	at node number 1, one can use the finite element approximation in
	the element that includes the coordinates of node 1 (element $i$ 
	shown in figure \ref{Fig:Interface}). This value will serve as a
	Dirichlet-type condition on $\Gamma_{\mathrm{c}\rightarrow\mathrm{f}}$
	and can be enforced using equation \ref{Egn:LBM_Dir}. Three-dimensional
	cases can be handled similarly.
\item \textsf{Multi-time-step integration}:~
	The solution in $\Omega_{\mathrm{c}}$ advances in time with a 
	time-step of $\Delta t_{\mathrm{c}}$. This time-step is typically
	much larger than the time-step needed for fine-scale problem in 
	subdomain $\Omega_{\mathrm{f}}$. However, to perform time-integration
	in $\Omega_{\mathrm{f}}$, we need to know the concentration on 
	$\Gamma_{\mathrm{c}\rightarrow\mathrm{f}}$, which can only be 
	determined by the numerical values in $\Omega_{\mathrm{c}}$. To
	approximate the concentration on $\Gamma_{\mathrm{c}\rightarrow\mathrm{f}}$
	at intermediate time-level $\jmath$, between $t$ and 
	$t + \Delta t_{\mathrm{c}}$, we will use the following interpolation 
	in time:
	\begin{align}
		\label{Eq:MTS}
		u_{\mathrm{c}}(\boldsymbol{x},t + \frac{\jmath}{\eta}\Delta 
		t_\mathrm{c}) \approx \left( \frac{\jmath}{\eta}\right) 
		u_{\mathrm{c}}(\boldsymbol{x},t + \Delta t_\mathrm{c}) + 
		\left( 1 - \frac{\jmath}{\eta}\right) u_{\mathrm{c}}(\boldsymbol{x},t)
		\quad \boldsymbol{x}\in \Gamma_{\mathrm{c}\rightarrow\mathrm{f}}
	\end{align}
	where $u_{\mathrm{c}}$ is the concentration in subdomain 
	$\Omega_{\mathrm{c}}$. Here, we have assumed that the rate of change in
	concentration remains constant in a time-step $\Delta t_{\mathrm{c}}$. 
	This value for $u_{\mathrm{c}}(\boldsymbol{x},
	t + \frac{\jmath}{\eta}\Delta t_{\mathrm{c}})$ will be enforces 
	as a Dirichlet condition on the solution in $\Omega_{\mathrm{f}}$.
\item \textsf{Sub-iterations at each time-step}:~
	In order to ensure convergence of the proposed algorithm, one needs
	to transfer information between the subdomains iteratively. Compatibility
	of the numerical solutions from the pore and fine-scale problems
	at the overlap region $\Omega_{\mathrm{f}}\cap\Omega_{\mathrm{c}}$ has a 
	vital role in accuracy of the numerical solution in the entire 
	domain $\Omega$. Figure \ref{Fig:TimeInt} illustrates one iteration 
	in a time-step $\Delta t_{\mathrm{c}}$. The solution of the 
	coarse-scale problem advances by $\Delta t_{\mathrm{c}}$ in step 1.
	Using the updated values of solution in $\Omega_{\mathrm{c}}$,
	boundary conditions onto subdomain $\Omega_{\mathrm{f}}$ at intermediate
	time-levels is determined. The solution of the fine-scale domain 
	advances by time-step $\Delta t_{\mathrm{f}}$ successively. The new
	numerical values are then used to find the concentrations on 
	boundary $\Gamma_{\mathrm{f}\rightarrow\mathrm{c}}$, which is used
	to update the solution in the coarse-scale domain in the next 
	iteration. This procedure is repeated an arbitrary number of times
	in order to satisfy accuracy requirements defined by the user.
\end{enumerate}
Given the tools described above, one can implement the proposed 
coupling method in a systematic manner. A step-by-step procedure
is given in Algorithm \ref{Alg:Hybrid_Alg}. In the following 
section, we will provide numerical examples to showcase the 
performance of this framework.
\begin{algorithm}
\caption{\textsf{Hybrid multi-time-step coupling framework}:~
Outline of the algorithm for proposed framework.}
\label{Alg:Hybrid_Alg}
\begin{algorithmic}[1]
	\State Initialize $\boldsymbol{u}$ in $\Omega_{\mathrm{c}}$ and 
		$f_{i}$ in $\Omega_{\mathrm{f}}$ for $t = 0$.
	\State Set $t \leftarrow 0$
	\While{$t < \mathrm{T}$}
		\State Set $t \leftarrow t + \Delta t$
		\State Set $\mathrm{Iter} \leftarrow 0$.
		\While{$\mathrm{Iter} \leq \mathrm{MaxIter}$}
			\State Set $\mathrm{Iter} \leftarrow \mathrm{Iter} + 1$.
			\State Find $\tilde{u}_{\mathrm{f}\rightarrow\mathrm{c}}$
				defined on $\Gamma_{\mathrm{f}\rightarrow\mathrm{c}}$.
			\State Advance the solution in $\Omega_{\mathrm{c}}$ 
				by $\Delta t_{\mathrm{c}}$.
			\State Find $\tilde{u}_{\mathrm{c}\rightarrow\mathrm{f}}$
				defined on $\Gamma_{\mathrm{c}\rightarrow\mathrm{f}}$
				at time-levels $t - \Delta t_{c}$ and $t$.
			\State Set $\jmath \leftarrow 0$
			\While{$\jmath \leq \eta$}
				\State Set $\jmath \leftarrow \jmath + 1$
				\State Advance the solution in $\Omega_{\mathrm{f}}$
					by $\Delta t_{\mathrm{f}}$ to find $f_i(\boldsymbol{x},
					t + \jmath \Delta t_{\mathrm{f}})$ (stream and collide).
				\State Impose Dirichlet boundary condition on 
					$\Gamma_{\mathrm{c}\rightarrow\mathrm{f}}$ with 
					\begin{align*}
						\tilde{u}_{\mathrm{c}\rightarrow\mathrm{f}} 
						(\boldsymbol{x},t+\jmath\Delta t_{\mathrm{f}}) = 
						(1 - \jmath/\eta)\tilde{u}_{\mathrm{c}\rightarrow\mathrm{f}}
						(\boldsymbol{x},t - \Delta t_{\mathrm{c}}) + (\jmath/\eta)
						\tilde{u}_{\mathrm{c}\rightarrow\mathrm{f}}(\boldsymbol{x},t )
					\end{align*}
			\EndWhile
			\State From the new numerical values in $\Omega_{\mathrm{f}}$
					find $\tilde{u}_{\mathrm{f}\rightarrow\mathrm{c}}$.
		\EndWhile
	\EndWhile
\end{algorithmic}
\end{algorithm}
\subsection{The case of many subdomains}
Thus far, the proposed coupling algorithm is presented for 
the case of only two subdomains, a coarse-scale subdomain 
$\Omega_{\mathrm{c}}$ and a fine-scale domain $\Omega_{\mathrm{f}}$.
However, in practical applications decomposition into multiple
subdomains may be required. In this section, we will present 
the proposed coupling method for cases where there are multiple
coarse and fine-scale subdomains.

Suppose that the domain $\Omega \subset \mathbb{R}^{n}$ is 
partitioned into coarse and fine-scale subdomains, given
as follows:
\begin{align}
	\label{Eqn:ManyDD}
	\Omega = \underbrace{\left( \bigcup_{i = 1}^{N_{\mathrm{c}}} \Omega_{\mathrm{c}, i}\right)}_{
	\mathrm{coarse-scale \; subdomains}} 
	\bigcup 
	\underbrace{\left( \bigcup_{j = 1}^{N_{\mathrm{f}}} \Omega_{\mathrm{f}, j} \right)}_{
	\mathrm{fine-scale \; subdomains}}
\end{align}
where all coarse and fine-scale subdomains are overlapping.
The number of coarse-scale subdomains is shown by $N_{\mathrm{c}}$
and $N_{\mathrm{f}}$ is the number of fine-scale subdomains.
Each subdomain $\Omega_{\mathrm{f},j}$ (for $j = 1,\cdots,N_{\mathrm{f}}$) 
is a fine-scale subdomain and will be integrated using the lattice 
Boltzmann method with grid size $h_{\mathrm{f},j}$ and time-step 
$\Delta t_{j}$. Coarse-scale subdomains $\Omega_{\mathrm{f},i}$
(for $i = 1,\cdots,N_{\mathrm{c}}$) are solved using the 
finite element method with mesh-size $h_{\mathrm{c},i}$ and 
$\Delta t_{\mathrm{c},i}$.
The details regarding multi-time-stepping and transferring
data from coarse-scale grid to fine-scale grid (and vise versa)
remains the same as before. Since discretization parameters 
for coarse-scale domains are much larger than the ones used in 
the fine-scale subdomains, the solution in coarse-scale domains
advances first, then the updated values near the interface of 
coarse-scale/fine-scale subdomains are used for multi-time-step
integration. Obviously, even coarse-scale subdomains can be 
integrated with different time-steps. Multi-time-step integration
for the coarse-scale subdomains can be done in the same spirit as
for the coarse-scale subdomains presented earlier. However, an
alternative approach would be to use the method presented in 
\citep{2015_Karimi_CMAME} to solve the coarse-scale subdomains
(that share an interface), and then use 
the updated solution to transfer to fine-scale domains. We will
not follow this procedure here, but it can be explored in
future research endeavors. We will denote the system time-step,
the same definition used in \citep{2015_Karimi_CMAME}, by $\Delta t$.
The proposed coupling framework for the case of multiple subdomains 
is given in Algorithm \ref{Alg:Hybrid_Alg_ManyDD}. 

\begin{algorithm}
\caption{\textsf{Hybrid multi-time-step coupling framework for many subdomains}:~
	The algorithmic procedure for the proposed framework is outlined.}
\label{Alg:Hybrid_Alg_ManyDD}
\begin{algorithmic}[1]
	\State Set $t \leftarrow 0$
	\While{$t + \Delta t < \mathrm{T}$}
		\For{$\mathrm{Iter} = 1, \cdots$}
		\For{$i = 1, \cdots , N_{\mathrm{c}}$}
			\State Advance the solution in subdomain $\Omega_{\mathrm{c},i}$ by 
				one system time-step, subject to boundary values from the 
				solutions from previous iteration.
		\EndFor
		\For{$j = 1 , \cdots , N_{\mathrm{f}}$}
			\State Advance the solution in subdomain $\Omega_{\mathrm{f},j}$ by
				one system time-step, subject to boundary values approximated by
				equation \eqref{Eq:MTS}.
		\EndFor
		\EndFor
	\EndWhile
\end{algorithmic}
\end{algorithm}

\section{REPRESENTATIVE NUMERICAL RESULTS}
\label{Sec:Hybrid_NR}
In this section, we will apply the proposed coupling
algorithm to one- and two-dimensional problems. The 
performance of the new method with respect to discretization
parameters will be studied. Computer implementation is
done using NumPy \citep{Numpy} and FEniCS \citep{FENICS} 
software packages.

\subsection{Advection and diffusion of one-dimensional Gaussian hill}
Consider $\Omega = (0,1)$ with zero-flux condition imposed on 
both ends. The initial concentration is given as
\begin{align}
	\mathrm{u}_0 (\mathrm{x}) = \frac{\phi}{\sqrt{2\pi\sigma_0^2}} 
		e^{-\left( \mathrm{x} - \mathrm{x}_0 \right)^2/2\sigma_0^2}
\end{align} 
where $\phi = 10^{-1}$ and $\sigma_0 = 10^{-2}$. The initial
location of the tip of the Gaussian hill is at $\mathrm{x}_0$
and is set to be $3 \times 10^{-1}$. The advection velocity in
the entire domain is taken to be $\mathrm{v} = 1$ and the 
diffusion coefficient is $\mathrm{D} = 10^{-2}$. The source
term is taken to be zero and the time-interval of 
interest is $\mathrm{T} = 4\times10^{-2}$. We will use 
the proposed hybrid coupling method to numerically solve
this problem. We will use the finite element method with 
the Galerkin formulation in $\Omega_{\mathrm{c}}$ and 
the lattice Boltzmann method in $\Omega_{\mathrm{f}}$.
To showcase the performance of the proposed method, we 
will use the following definition for error (error in 
$\infty$-norm)
\begin{align}
\label{Egn:Error1D}
	\mathcal{E}(t) = \max_{i = 1, \cdots , N} 
	\big| u(\boldsymbol{x}_i,t) - 
	\mathrm{u}_{\mathrm{exact}}(\boldsymbol{x_i},t)\big|
\end{align}
where $N$ is the number of nodes for numerical solution,
$u (\boldsymbol{x},t)$ is the approximate solution at
point $\boldsymbol{x}_i$ and time $t$. The exact solution
is represented by $\mathrm{u}_{\mathrm{exact}}$. Following
the definition given in \eqref{Egn:Error1D}, the error
in $\Omega_{\mathrm{c}}$ and $\Omega_{\mathrm{f}}$ will be
denoted by $\mathcal{E}_{\mathrm{c}}$ and $\mathcal{E}_{\mathrm{f}}$
respectively.
We will denote the length of the overlap region $\Omega_{
\mathrm{f}}\cap \Omega_{\mathrm{c}}$ by $L_{\mathrm{overlap}}$. 
The domain partitioning is as follows
\begin{align}
  \Omega_{\mathrm{c}} = \left(0,\frac{1}{2} + \frac{L_{\mathrm{overlap}}}{2}\right),
  \quad
  \Omega_{\mathrm{f}} = \left(\frac{1}{2} - \frac{L_{\mathrm{overlap}}}{2},1\right)
\end{align}
We will employ two-node linear finite elements of equal lengths 
$h_{\mathrm{c}}$ to discretize $\Omega_{\mathrm{c}}$. The time-step
is set to be $\Delta t_{\mathrm{c}} = h_{\mathrm{c}}^2/2\mathrm{D}$. 
Subdomain $\Omega_{\mathrm{f}}$ is discretized using a uniform grid
with spacing $h_{\mathrm{f}}$ and a time-step of $\Delta t_{\mathrm{f}}
= h_{\mathrm{f}}^2/2\mathrm{D}$. The $D1Q2$ lattice model
will be used in $\Omega_{\mathrm{f}}$. The number of sub-iterations in
each time-step is shown by MaxIter. 

\begin{figure}
  \centering
  \psfrag{x}{$x$-axis}
  \psfrag{u}{$u(x)$}
  \psfrag{exact}{Exact solution}
  \psfrag{fem}{FEM solution}
  \psfrag{lbm}{LBM solution}
  \subfigure[Concentration at $t = 0.2$.]{
    \includegraphics[scale=0.27,clip]{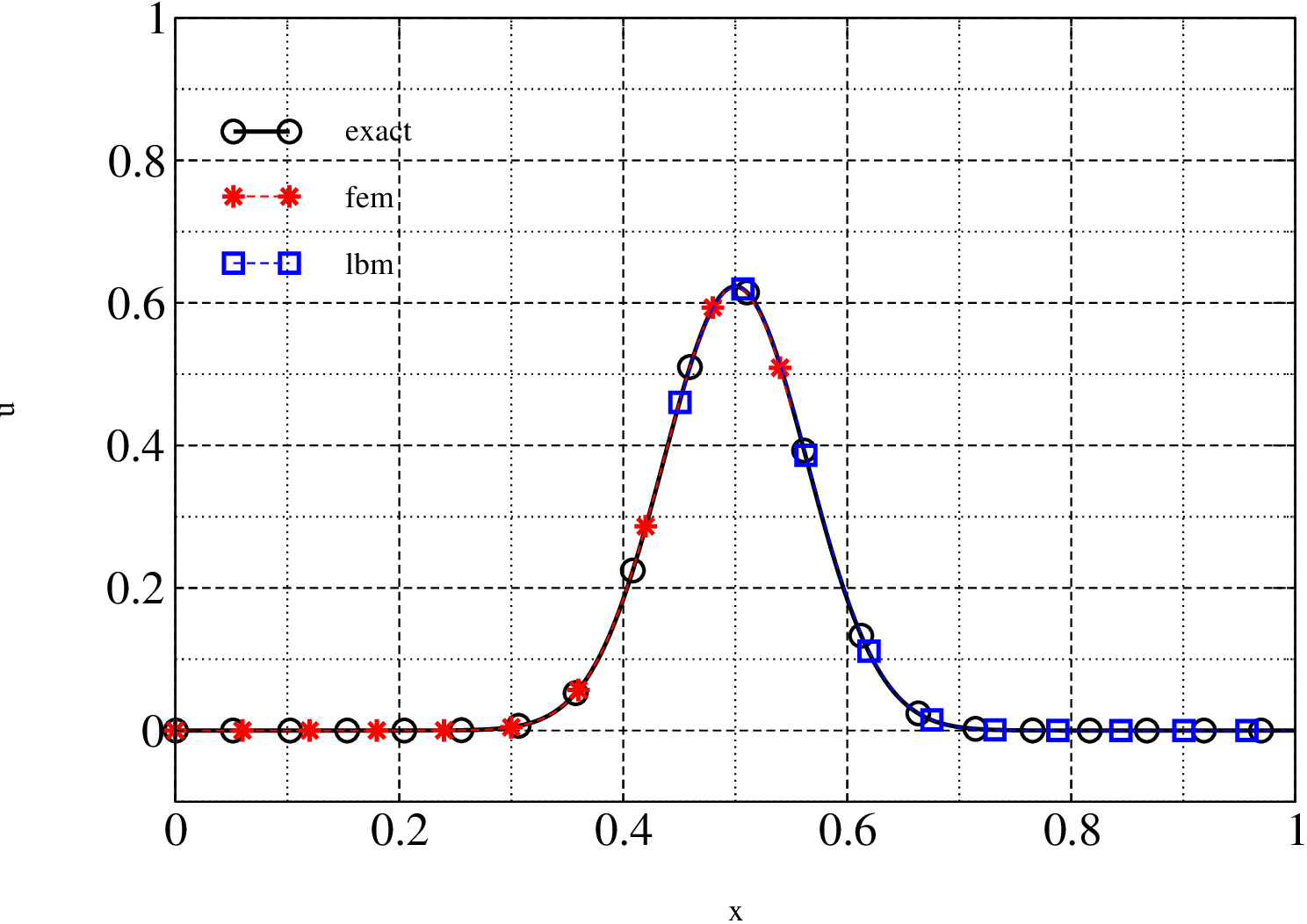}}
  \subfigure[Concentration at $t = 0.4$.]{
    \includegraphics[scale=0.27,clip]{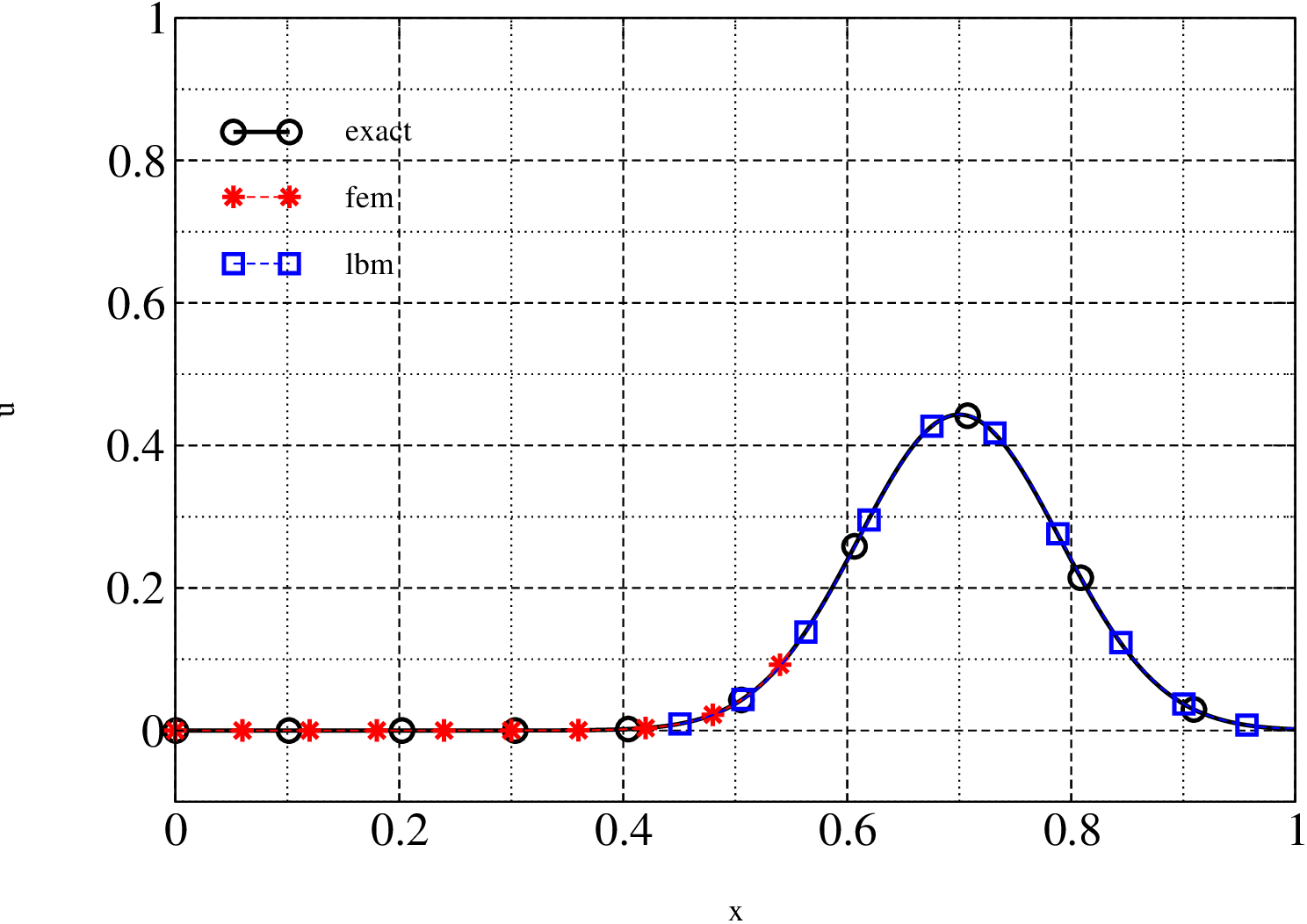}}
  \caption{\textsf{Advection and diffusion of one-dimensional 
      Gaussian hill}:~This figure shows the exact and numerical 
    concentration profiles at two different time-levels. At $t 
    = 0.2$, the front passes through the overlap region. The 
    numerical solution shows good agreement with the exact 
    solution.
    We have taken $h_{\mathrm{c}} = 10^{-2}$ and $\Delta t_{\mathrm{c}} 
    = 5 \times 10^{-3}$. In the fine-scale domain, $h_{\mathrm{f}} 
    = 1.25\times10^{-3}$ and $\Delta t_{\mathrm{f}} = 7.81\times10^{-5}$. 
    Length of the overlap region is $L_{\mathrm{overlap}} = 0.1$.
    \label{Fig:1D_Gauss_Concentration}}
\end{figure}

Figure \ref{Fig:1D_Gauss_Concentration} shows a comparison between
the numerical 
solution from the hybrid coupling framework and the exact solution. The
concentration profile is shown when the front is passing through
the overlap region and afterwards. In both cases, the numerical solution
is in accordance with the exact solution.

\begin{figure}
  \centering
  \psfrag{E}{$\mathcal{E}(\mathrm{T})$}
  \psfrag{h}{$1/h_{\mathrm{f}}$}
  \psfrag{ef}{$\mathcal{E}_{\mathrm{f}}$}
  \psfrag{ec}{$\mathcal{E}_{\mathrm{c}}$}
  \psfrag{slope}{$\mathrm{slope}=-1$}
  \includegraphics[scale=0.35,clip]{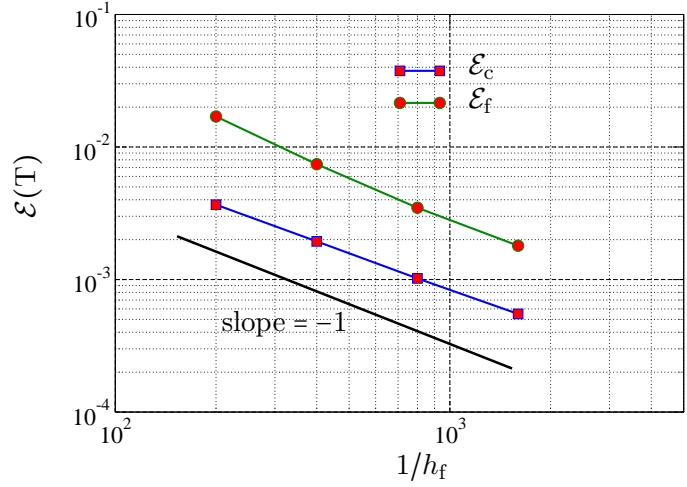}
  \caption{\textsf{Advection and diffusion of one-dimensional
      Gaussian hill}:~This figure shows the error in the
    coarse-scale and fine-scale subdomains against refinement
    in the fine-scale region. Table \ref{Tbl:1DGauss_1} provides
    the numerical values employed in this numerical simulation.
    \label{Fig:1D_Error_1}}
\end{figure}

\begin{figure}
	\centering
	\psfrag{h}{$h_{\mathrm{c}}$ or $h_{\mathrm{f}}$}
	\psfrag{E}{$\mathcal{E}(\mathrm{T})$}
	\psfrag{lbm}{$\mathcal{E}_{\mathrm{f}}$}
	\psfrag{fem}{$\mathcal{E}_{\mathrm{c}}$}
	\psfrag{slope}{slope$=1$}
	\includegraphics[clip, scale=0.35]{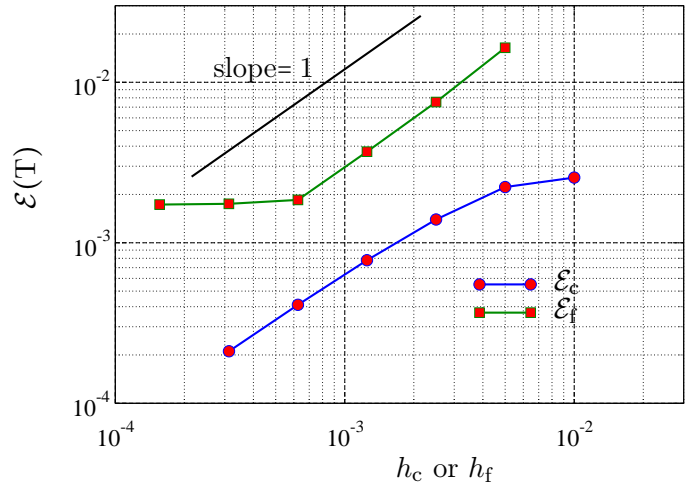}
	\caption{\textsf{Advection and diffusion of one-dimensional Gaussian hill}:~
	The error in the fine and coarse-scale subdomains is plotted against
	grid-size. In all cases $h_{\mathrm{c}} = 2 h_{\mathrm{f}}$.
	Grid refinement is done simultaneously in both subdomains. The ratio
	between the time-steps in the two subdomains is constant for all
	cases, $\eta = 4$. The length of the overlap region is $L = 0.04$.
	Parameters are given in Table \ref{Tbl:1DGauss_4}.
	\label{Fig:1D_Error_2}}
\end{figure}

The numerical experiments discussed here show that the proposed 
hybrid coupling framework gives an accurate solution to the 
advection-diffusion equation and is indeed a converging scheme
(see figures \ref{Fig:1D_Error_1} and \ref{Fig:1D_Error_2}).
From these numerical experiments, we conclude that the 
convergence of the numerical solution under the proposed framework
is $\mathcal{O}(h)$. In the following, the effect
of discretization in coarse and fine-scale subdomains, effect of 
length of overlap region and the number of sub-iterations 
on the accuracy of the numerical solution are described.

\begin{figure}
  \centering
  \psfrag{h}{$h_{\mathrm{c}}$ or $h_{\mathrm{f}}$}
  \psfrag{E}{$\mathcal{E}(\mathrm{T})$}
  \psfrag{lbm}{$\mathcal{E}_{\mathrm{f}}$}
  \psfrag{fem}{$\mathcal{E}_{\mathrm{c}}$}
  \includegraphics[clip, scale=0.35]{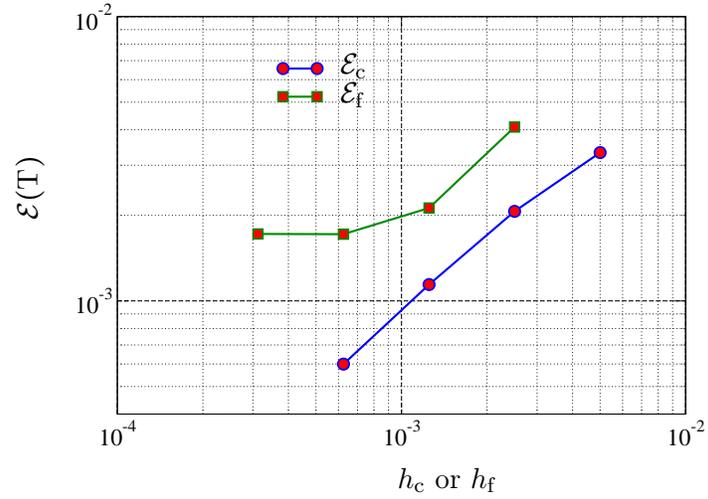}
  \caption{\textsf{Advection and diffusion of one-dimensional
      Gaussian hill}:~In this figure, the error in the coarse-
    and fine-scale subdomains is shown. In this case the length
    of the overlap region is $L_{\mathrm{overlap}} = 10^{-2}.$
    \label{Fig:1D_Error_3}}
\end{figure}

\begin{figure}
  \centering
  \psfrag{h}{$h_{\mathrm{c}}$ or $h_{\mathrm{f}}$}
  \psfrag{E}{$\mathcal{E}(\mathrm{T})$}
  \psfrag{lbm}{$\mathcal{E}_{\mathrm{f}}$}
  \psfrag{fem}{$\mathcal{E}_{\mathrm{c}}$}
  \includegraphics[clip, scale=0.35]{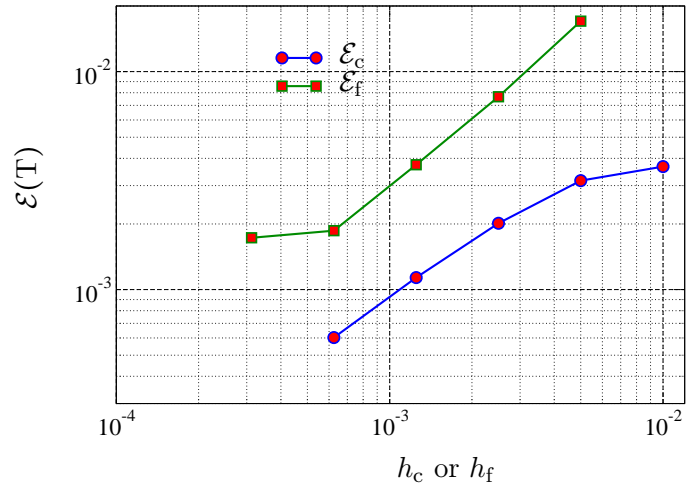}
  \caption{\textsf{Advection and diffusion of one-dimensional Gaussian hill}:~
    Error in the fine and coarse-scale domains with respect to mesh
    size in each 
    subdomain is shown. Here, the length of the overlap region is 
    $L_{\mathrm{overlap}} = 10^{-1}$.
    \label{Fig:1D_Error_4}}
\end{figure}

\begin{figure}
  \centering
  \psfrag{x}{$x$-axis}
  \psfrag{E}{Error}
  \psfrag{p1}{$p = 1$}
  \psfrag{p2}{$p = 2$}
  \psfrag{p5}{$p = 5$}
  \subfigure[$h_{\mathrm{c}}=10^{-2}$, $h_{\mathrm{f}} = 5 \times 10^{-3}$,
    $\Delta t_{\mathrm{c}} = 5 \times 10^{-3}$, $\Delta t_{\mathrm{f}} = 1.25 \times 10^{-3}$ (subcycling)]{
    \includegraphics[clip, scale=0.3]{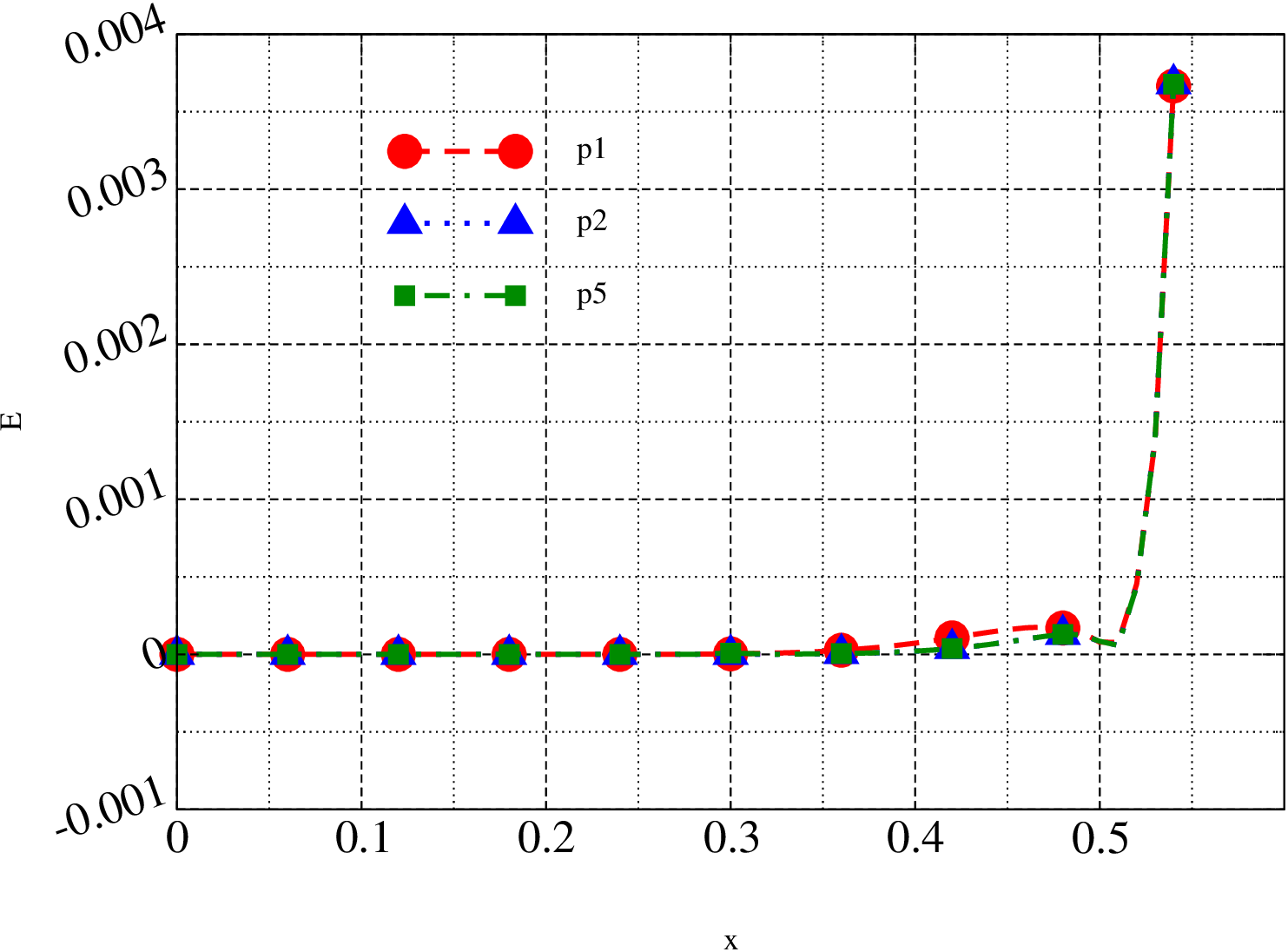}}
  \subfigure[$h_{\mathrm{c}} = h_{\mathrm{f}} = 10^{-2}$,
    $\Delta t_{\mathrm{c}} = \Delta t_{\mathrm{f}} = 5 \times 10^{-3}$
  (no subcycling)]{
    \includegraphics[clip, scale=0.3]{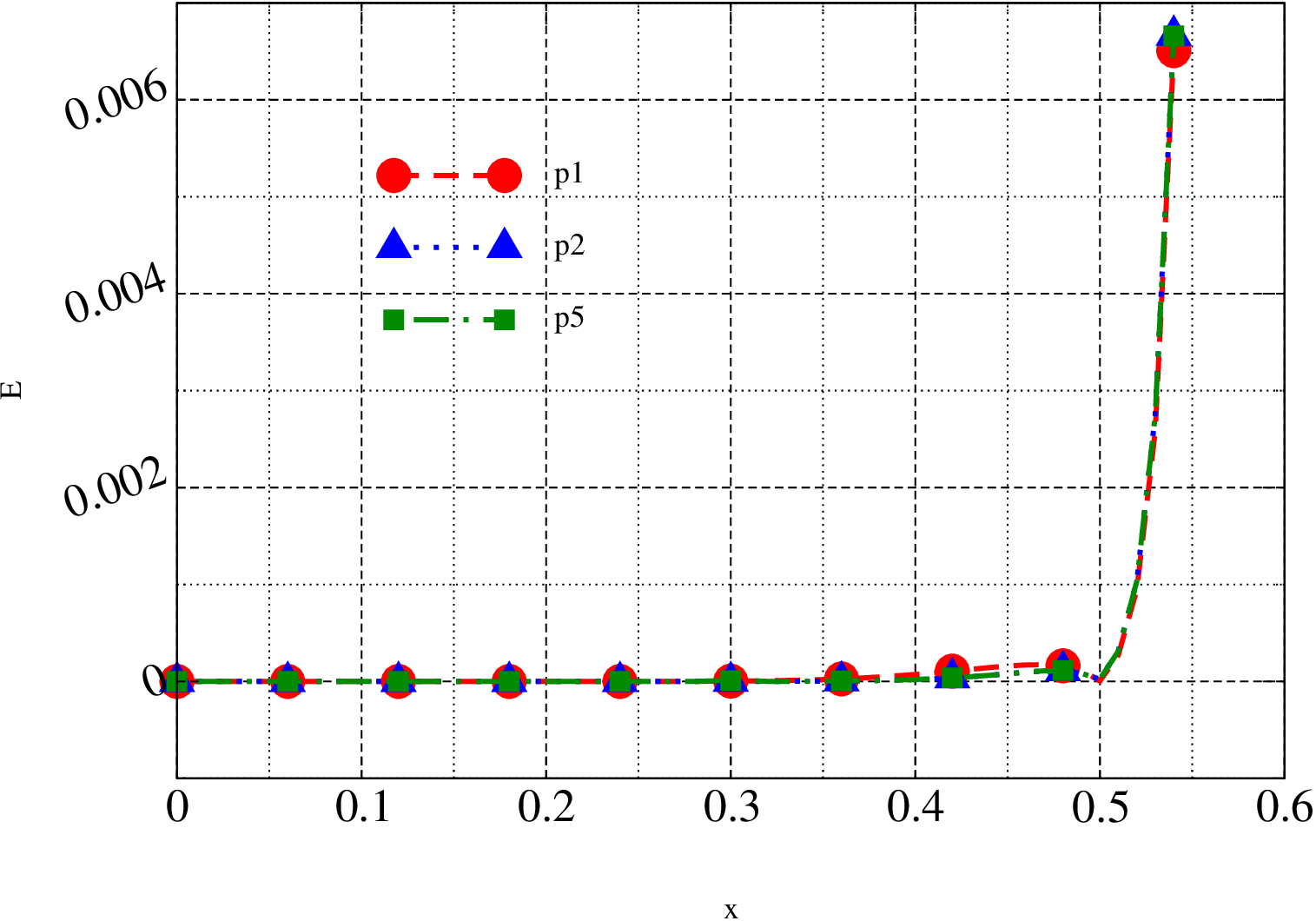}}
  \caption{\textsf{Advection and diffusion of one-dimensional
      Gaussian hill}:~In this figure, point-wise error in the
    coarse-scale domain at time $t = \mathrm{T}$ is shown.
    Different orders of interpolation (denoted by $p$ here)
    in the finite elements are used. The length of the overlap
    region is $L_{\mathrm{overlap}} = 10^{-1}$. It can be observed
    that increasing
    the order of finite element interpolation does not improve
    accuracy near the overlap region. Moreover, the relative
    accumulation of the error near the overlapping region is
    not associated with multi-time-stepping (i.e., subcycling).
    \label{Fig:1D_Error_FE_p}}
\end{figure}

\begin{enumerate}[(1)]
\item Discretization in fine-scale domain:
	Our numerical experiments indicate that for a given discretization
	in the coarse-scale domain (i.e., $h_{\mathrm{c}}$ and 
	$\Delta t_{\mathrm{c}}$), refinement of parameters $h_{\mathrm{f}}$
	and $\Delta t_{\mathrm{f}}$ improves the overall accuracy of 
	numerical solution. The results presented in Table \ref{Tbl:1DGauss_1}
	show that the mentioned refinement reduces the error in both 
	fine-scale and coarse-scale subdomains. 
\item Discretization in coarse-scale domain:
	Considering the numerical results presented in Table \ref{Tbl:1DGauss_2},
	one can conclude that for a given discretization in fine-scale domain
	(i.e., $h_{\mathrm{f}}$ and $\Delta t_{\mathrm{f}}$), refinement of 
	respective parameters in the coarse-scale domain does not necessarily
	improve accuracy. This behavior can be attributed to the fact that
	the lower accuracy in the fine-scale domain (due to use of lattice 
	Boltzmann method), results in a less accurate estimation of the
	concentration on $\Gamma_{\mathrm{f}\rightarrow\mathrm{c}}$. Hence,
	the numerical solution in the coarse-scale region converges to a
	solution other than the exact solution.
\item Length of overlap region:
	For a given discretization in subdomains $\Omega_{\mathrm{f}}$ and 
	$\Omega_{\mathrm{c}}$, increase in the length of the overlap region
	results in reduction of overall accuracy. This conclusion can be
	drawn from the numerical experiments presented in Table 
	\ref{Tbl:1DGauss_3}. However, if the grid-size and 
	time-step in both subdomains change simultaneously, convergence 
	rate of the numerical solution to the exact solution may slow down.
	Following the numerical results given in Tables \ref{Tbl:1DGauss_4}, 
	\ref{Tbl:1DGauss_5} and \ref{Tbl:1DGauss_6}, as well as figures 
	\ref{Fig:1D_Error_2}, \ref{Fig:1D_Error_3} and \ref{Fig:1D_Error_4} 
	shows that convergence under simultaneous refinement in both subdomains 
	has an inverse relation to the length of the overlap region. 
\item Number of sub-iterations in each time-step:
	In the numerical experiments performed, increasing the maximum
	number of sub-iterations to values greater than 4 did not result
	in a significant improvement in accuracy. However, compatibility,
	especially near the overlap region, can be improved by increasing 
	the number of sub-iterations.
\item Order of interpolation in the coarse-scale subdomain:
	Figure \ref{Fig:1D_Error_FE_p} shows the point-wise error in 
	the coarse-scale subdomain, for different orders of interpolation 
	in finite elements and under multi-time-stepping. For different
        cases, the error in the fine-scale 
	subdomain remains largely unchanged from one case to another.
	Error in the coarse-scale subdomain decreases by increasing the
	order of interpolation, however, the error near the overlap region
	remains unchanged. The figure the error accumulates near the
        overlapping region under both multi-time-stepping and under
        single time-step in all the subdomains.
\end{enumerate}
These observations regarding the effect of number of sub-iterations and
length of the overlap region are in accordance with the theory of 
overlapping domain decomposition methods \citep{1997_Nataf_NM,2015_Chang_SIAM}.
It seems that, generally, decrease in the size of the overlap region 
reduces the rate of convergence and the error decreases proportional to 
the inverse of square root of number of sub-iterations.
One key observation from these numerical experiments is that majority 
of error in the numerical solution accumulates near the overlap region.
This error can be much higher than the error in the rest of the domain 
and refinement in either of the subdomains may not improve it. Hence,
a topic for future research can be designing efficient methods for 
removing the accumulated error in the overlap region under the 
proposed coupling framework.

Here, we showed that one can use highly disparate mesh-size and 
time-steps in different subdomains. 
Furthermore, we showed that to improve accuracy throughout the 
computational domain, grid refinement in the fine-scale 
domain is sufficient. We also demonstrated that mesh refinement
only in the coarse-scale domain may not lead to a more
accurate numerical solution. 
\begin{table}
\caption{\textsf{Advection and diffusion of one-dimensional Gaussian hill}:~
In this table, the accuracy of the numerical solution using the proposed
coupling framework is shown. Here, only cell size and time-step in 
the fine-scale domain are refined. Note that despite the refinement in 
the fine-scale domain only, the accuracy of the solution in the 
entire computational domain is improving.}
\label{Tbl:1DGauss_1}
\begin{tabular}{|c|c|c|c|c|c|c|c|c|}
	\hline
	$h_{\mathrm{c}}$ & $\Delta t_{\mathrm{c}}$ & $h_{\mathrm{f}}$ & 
	$\Delta t_{\mathrm{f}}$& $\eta$ & $L_{\mathrm{overlap}}$ & MaxIter &
	$\mathcal{E}_{\mathrm{c}}(\mathrm{T})$ & $\mathcal{E}_{\mathrm{f}}(\mathrm{T})$ \\ \hline
	$10^{-2}$ & $5.00 \times 10^{-3}$ & $5.00 \times 10^{-3}$ & $1.25\times10^{-3}$ &
	4 & $10^{-1}$ & 4 & $3.67\times10^{-3}$ & $1.70\times10^{-2}$ \\ \hline
	$10^{-2}$ & $5.00 \times 10^{-3}$ & $2.50 \times 10^{-3}$ & $3.13\times10^{-4}$ &
	16 & $10^{-1}$ & 4 & $1.94\times10^{-3}$ & $7.42\times10^{-3}$ \\ \hline
	$10^{-2}$ & $5.00 \times 10^{-3}$ & $1.25 \times 10^{-3}$ & $7.81\times10^{-5}$ &
	64 & $10^{-1}$ & 4 & $1.02\times10^{-3}$ & $3.48\times10^{-3}$ \\ \hline
	$10^{-2}$ & $5.00 \times 10^{-3}$ & $6.25 \times 10^{-4}$ & $1.95\times10^{-5}$ &
	256 & $10^{-1}$ & 4 & $5.50\times10^{-4}$ & $1.80\times10^{-3}$ \\ \hline
\end{tabular}
\end{table}
\begin{table}
\caption{\textsf{Advection and diffusion of one-dimensional Gaussian hill}:~
In this table, performance of the proposed method for numerical solution
of the one-dimensional problem is shown. In this case, element size and time-step
refinement are done only in the coarse-scale domain. The discretization parameters
in the fine-scale domain remain unchanged in the fine-scale domain. It can be
observed that refinement, merely in the coarse-scale domain, has adverse effect
on the accuracy of numerical solution. This experiment shows that
the numerical method with the slowest convergence has the dominant role 
in overall accuracy.}
\label{Tbl:1DGauss_2}
\begin{tabular}{|c|c|c|c|c|c|c|c|c|}
	\hline
	$h_{\mathrm{c}}$ & $\Delta t_{\mathrm{c}}$ & $h_{\mathrm{f}}$ & 
	$\Delta t_{\mathrm{f}}$& $\eta$ & $L_{\mathrm{overlap}}$ & MaxIter &
	$\mathcal{E}_{\mathrm{c}}(\mathrm{T})$ & $\mathcal{E}_{\mathrm{f}}(\mathrm{T})$ 
	\\ \hline
	$10^{-2}$ & $5.00 \times 10^{-3}$ & $1.25 \times 10^{-3}$ & $7.81\times10^{-5}$ &
	64 & $10^{-1}$ & 4 & $1.02\times10^{-3}$ & $3.48\times10^{-3}$ \\ \hline
	$5.00 \times 10^{-3}$ & $1.25 \times 10^{-3}$ & $1.25 \times 10^{-3}$ & $7.81\times10^{-5}$ &
	16 & $10^{-1}$ & 4 & $1.65\times10^{-3}$ & $3.71\times10^{-3}$ \\ \hline
	$2.50 \times 10^{-3}$ & $3.13 \times 10^{-4}$ & $1.25 \times 10^{-3}$ & $7.81\times10^{-5}$ &
	4 & $10^{-1}$ & 4 & $2.01\times10^{-3}$ & $3.74\times10^{-3}$ \\ \hline
	$1.25 \times 10^{-3}$ & $7.81 \times 10^{-5}$ & $1.25 \times 10^{-3}$ & $7.81\times10^{-5}$ &
	1 & $10^{-1}$ & 4 & $2.22\times10^{-3}$ & $3.74\times10^{-3}$ \\ \hline
\end{tabular}
\end{table}
\begin{table}
  \caption{\textsf{Advection and diffusion of one-dimensional
      Gaussian hill}:~This numerical experiment indicates that
    increasing the length of the overlapping region could have
    adverse effect on the accuracy of the numerical solution.}
\label{Tbl:1DGauss_3}
\begin{tabular}{|c|c|c|c|c|c|c|c|c|}
\hline
	$h_{\mathrm{c}}$ & $\Delta t_{\mathrm{c}}$ & $h_{\mathrm{f}}$ & 
	$\Delta t_{\mathrm{f}}$& $\eta$ & $L_{\mathrm{overlap}}$ & MaxIter &
	$\mathcal{E}_{\mathrm{c}}(\mathrm{T})$ & $\mathcal{E}_{\mathrm{f}}(\mathrm{T})$ 
	\\ \hline
	$10^{-2}$ & $5.00 \times 10^{-3}$ & $1.25 \times 10^{-3}$ & $7.81\times10^{-5}$ &
	64 & $2.00\times10^{-2}$ & 4 & $5.78\times10^{-4}$ & $3.08\times10^{-3}$ \\ \hline
	$10^{-2}$ & $5.00 \times 10^{-3}$ & $1.25 \times 10^{-3}$ & $7.81\times10^{-5}$ &
	64 & $4.00\times10^{-2}$ & 4 & $5.85\times10^{-4}$ & $3.43\times10^{-3}$ \\ \hline
	$10^{-2}$ & $5.00 \times 10^{-3}$ & $1.25 \times 10^{-3}$ & $7.81\times10^{-5}$ &
	64 & $8.00\times10^{-2}$ & 4 & $8.63\times10^{-4}$ & $3.47\times10^{-3}$ \\ \hline
	$10^{-2}$ & $5.00 \times 10^{-3}$ & $1.25 \times 10^{-3}$ & $7.81\times10^{-5}$ &
	64 & $10^{-1}$ & 4 & $1.02\times10^{-3}$ & $3.48\times10^{-3}$ \\ \hline
\end{tabular}
\end{table}

\begin{table}
  \caption{\textsf{Advection and diffusion of
      one-dimensional Gaussian hill}:~In this
    table, values of discretization parameters and errors in each subdomain
	are provided. In all the cases, $\eta = 4$ and $L_{\mathrm{overlap}} = 4\times10^{-2}$.
	The number of sub-iterations in each time-step is 10.}
\label{Tbl:1DGauss_4}
\begin{tabular}{| c | c | c | c | c | c |}
	\hline
	$h_{\mathrm{c}}$ & $\Delta t_{\mathrm{c}}$ & 
		$h_{\mathrm{f}}$ & $\Delta t_{\mathrm{f}}$ & 
		$\mathcal{E}_{\mathrm{c}}(\mathrm{T})$ &
		$\mathcal{E}_{\mathrm{f}}(\mathrm{T})$ \\ \hline 
		$1.00\times10^{-2}$&$5.00\times10^{-3}$&$5.00\times10^{-3}$&$1.25\times10^{-3}$&$2.55\times10^{-3}$&$1.64\times10^{-2}$ \\ \hline
		$5.00\times10^{-3}$&$1.25\times10^{-3}$&$2.50\times10^{-3}$&$3.13\times10^{-4}$&$2.22\times10^{-3}$&$7.54\times10^{-3}$ \\ \hline
		$2.50\times10^{-3}$&$3.13\times10^{-4}$&$1.25\times10^{-3}$&$7.81\times10^{-5}$&$1.40\times10^{-3}$& $3.70\times10^{-3}$\\ \hline
		$1.25\times10^{-3}$&$7.81\times10^{-5}$&$6.25\times10^{-4}$&$1.95\times10^{-5}$&$7.79\times10^{-4}$&$1.85\times10^{-3}$ \\ \hline
		$6.25\times10^{-4}$&$1.95\times10^{-5}$&$3.13\times10^{-4}$&$4.88\times10^{-6}$&$4.11\times10^{-4}$&$1.75\times10^{-3}$ \\ \hline
		$3.13\times10^{-4}$&$4.88\times10^{-6}$&$1.56\times10^{-4}$&$1.22\times10^{-6}$&$2.11\times10^{-4}$&$1.73\times10^{-3}$ \\ \hline
\end{tabular}
\end{table}

\begin{table}
\caption{\textsf{Advection and diffusion of one-dimensional Gaussian hill}:~
Values for the discretization parameters and errors in fine and coarse-scale
subdomains are given. In all cases, $\eta = 4$, $L_{\mathrm{overlap}} = 10^{-2}$ 
and the number of sub-iterations is 10.}
\label{Tbl:1DGauss_5}
\begin{tabular}{| c | c | c | c | c | c |}
	\hline
	$h_{\mathrm{c}}$ & $\Delta t_{\mathrm{c}}$ & 
		$h_{\mathrm{f}}$ & $\Delta t_{\mathrm{f}}$ & 
		$\mathcal{E}_{\mathrm{c}}(\mathrm{T})$ &
		$\mathcal{E}_{\mathrm{f}}(\mathrm{T})$ \\ \hline 
		$5.00\times10^{-3}$&$1.25\times10^{-2}$&$2.50\times10^{-3}$&$3.13\times10^{-4}$&$3.32\times10^{-3}$&$4.09\times10^{-3}$ \\ \hline
		$2.50\times10^{-3}$&$3.13\times10^{-4}$&$1.25\times10^{-3}$&$7.81\times10^{-5}$&$2.07\times10^{-3}$& $2.12\times10^{-3}$ \\ \hline
		$1.25\times10^{-3}$&$7.81\times10^{-5}$&$6.25\times10^{-4}$&$1.95\times10^{-5}$&$1.14\times10^{-3}$&$1.72\times10^{-3}$ \\ \hline
		$6.25\times10^{-4}$&$1.95\times10^{-5}$&$3.13\times10^{-4}$&$4.88\times10^{-6}$&$6.00\times10^{-4}$&$1.72\times10^{-3}$ \\ \hline
\end{tabular}
\end{table}

\begin{table}
\caption{\textsf{Advection and diffusion of one-dimensional Gaussian hill}:~
  Discretization and errors in fine-scale and coarse-scale domains
  are given in this table. The number of sub-iterations in each
  time-step is 10. In all the cases, $\eta = 4$ and $L_{\mathrm{overlap}}
  = 10^{-1}$.}
\label{Tbl:1DGauss_6}
\begin{tabular}{| c | c | c | c | c | c |}
\hline
	$h_{\mathrm{c}}$ & $\Delta t_{\mathrm{c}}$ & 
		$h_{\mathrm{f}}$ & $\Delta t_{\mathrm{f}}$ & 
		$\mathcal{E}_{\mathrm{c}}(\mathrm{T})$ &
		$\mathcal{E}_{\mathrm{f}}(\mathrm{T})$ \\ \hline
		$1.00\times10^{-2}$&$5.00\times10^{-3}$&$5.00\times10^{-3}$&$1.25\times10^{-3}$&$3.67\times10^{-3}$&$1.70\times10^{-2}$ \\ \hline
		$5.00\times10^{-3}$&$1.25\times10^{-3}$&$2.50\times10^{-3}$&$3.13\times10^{-4}$&$3.16\times10^{-3}$&$7.67\times10^{-3}$ \\ \hline
		$2.50\times10^{-3}$&$3.13\times10^{-4}$&$1.25\times10^{-3}$&$7.81\times10^{-5}$&$2.01\times10^{-3}$&$3.74\times10^{-3}$ \\ \hline
		$1.25\times10^{-3}$&$7.81\times10^{-5}$&$6.25\times10^{-4}$&$1.95\times10^{-5}$&$1.13\times10^{-3}$&$1.86\times10^{-3}$ \\ \hline
		$6.25\times10^{-4}$&$1.95\times10^{-5}$&$3.13\times10^{-4}$&$4.88\times10^{-6}$&$6.02\times10^{-4}$&$1.73\times10^{-3}$ \\ \hline
\end{tabular}
\end{table}

\subsection{Simulation of fast bimolecular reaction using multiple subdomains}
This example will be used to demonstrate the application
of the proposed hybrid framework for bimolecular fast
reactions and its ability to handle multiple subdomains.
To this end, we simulate the evolution of the concentrations
of the participating chemical species in the following
bimolecular reaction:
\begin{align}
  \mathrm{n}_{\mathrm{A}} \mathrm{A}
  + \mathrm{n}_{\mathrm{B}} \mathrm{B}
  \rightarrow \mathrm{n}_{\mathrm{C}} \mathrm{C}
\end{align}
where $\mathrm{n}_{\mathrm{A}}$, $\mathrm{n}_{\mathrm{B}}$
and $\mathrm{n}_{\mathrm{C}}$ are the stoichiometry
coefficients. Here, we have chosen $\mathrm{n}_{\mathrm{A}}
= 1$, $\mathrm{n}_{\mathrm{B}} = 2$ and $\mathrm{n}_{\mathrm{C}}
= 1$. The computational domain $\Omega = (0,1)$
is partitioned into the following two coarse-scale and one
fine-scale subdomains:
\begin{align*}
  \Omega_{\mathrm{c},1} = (0,0.40), \;
  \Omega_{\mathrm{f}} = (0.39,0.61)
  \quad \mathrm{and} \quad  
  \Omega_{\mathrm{c},2} = (0.6,1.0)
\end{align*}
The time-interval of interest is $\mathrm{T} = 0.5$. The 
coefficient of diffusion is $\mathrm{D} = 10^{-2}$ and the 
advection velocity is zero throughout the domain. We will
enforce zero-flux boundary conditions at $\mathrm{x}=0$ and 
$\mathrm{x}=1$. The initial values for each of the species is
as follows:
\begin{align}
  \mathrm{u}_{0,i}(\mathrm{x}) = 
  \frac{\phi_{0,i}}{\sqrt{2\pi \sigma^{2}}}
  \mathrm{exp}\left[ -(\mathrm{x} - \mathrm{x}_{0,
      i})^{2}/2 \sigma^{2}\right]
  \quad i = \mathrm{A}, \mathrm{B}, \mathrm{C}
\end{align}
where $\sigma = 0.1$, $\phi_{0,\mathrm{A}} = 0.1$,
$\mathrm{x}_{0,\mathrm{A}} = 0.3$, $\phi_{0,\mathrm{B}}
= 0.05$ and $\mathrm{x}_{0,\mathrm{B}} = 0.7$. The
initial concentration of the species $\mathrm{C}$
is zero in the entire domain.
To solve the problem numerically, it is convenient
to introduce the following invariants:
\begin{align}
  \alpha = \mathrm{u}_{\mathrm{A}}
  + \frac{\mathrm{n}_{\mathrm{A}}}{\mathrm{n}_{\mathrm{C}}}
  \mathrm{u}_{\mathrm{C}}
  \quad \mathrm{and} \quad 
  \beta = \mathrm{u}_{\mathrm{B}}
  + \frac{\mathrm{n}_{\mathrm{B}}}{\mathrm{n}_{\mathrm{C}}}
  \mathrm{u}_{\mathrm{C}}
\end{align}
where $\mathrm{u}_{\mathrm{A}}$, $\mathrm{u}_{\mathrm{B}}$
and $\mathrm{u}_{\mathrm{C}}$ are the concentrations of
the chemical species $\mathrm{A}$, $\mathrm{B}$ and
$\mathrm{C}$, respectively. Once numerical values for
$\alpha$ and $\beta$ are found, the concentrations of
the participating chemical species can be calculated
as follows:
\begin{subequations}
\begin{align}
  &\mathrm{u}_{\mathrm{A}} = \max \left\{ \alpha - 
  \frac{\mathrm{n}_{\mathrm{A}}}{\mathrm{n}_{\mathrm{B}}}
  \beta , \; 0\right\} \\
  &\mathrm{u}_{\mathrm{B}} = \frac{\mathrm{n}_{\mathrm{B}}}{\mathrm{n}_{\mathrm{A}}}\max 
  \left\{ -\alpha + \frac{\mathrm{n}_{\mathrm{A}}}{\mathrm{n}_{\mathrm{B}}}
  \beta , \; 0\right\} \\
  &\mathrm{u}_{\mathrm{C}} = \frac{\mathrm{n}_{\mathrm{C}}}{\mathrm{n}_{\mathrm{A}}}
  \left( \alpha - \mathrm{u}_{\mathrm{A}}\right)
\end{align}
\end{subequations}
Subdomains $\Omega_{\mathrm{c},1}$ and $\Omega_{\mathrm{c},2}$ are
discretized using the finite element method, with a mesh size
of $h_{\mathrm{c},1} = h_{\mathrm{c},2} = 10^{-2}$ and time-step
of $\Delta t_{\mathrm{c},1} = \Delta t_{\mathrm{c},2} = 5\times10^{-3}$. 
Subdomain $\Omega_{\mathrm{f}}$ is solved using the lattice Boltzmann method
with cell size of $h_{\mathrm{f}} = 10^{-3}$ and time-step 
$\Delta t_{\mathrm{f}} = 2\times10^{-5}$. The number of 
sub-iterations at each time-level is set to 10.

\begin{figure}
	\centering
	\psfrag{x}{$x$-axis}
	\psfrag{u}{concentration}
	\psfrag{fe1}{$u_{\mathrm{c},1}$}
	\psfrag{fe2}{$u_{\mathrm{c},2}$}
	\psfrag{lb}{$u_{\mathrm{f}}$}
	\subfigure[$t = 0.1$]{\includegraphics[clip,scale=0.35]{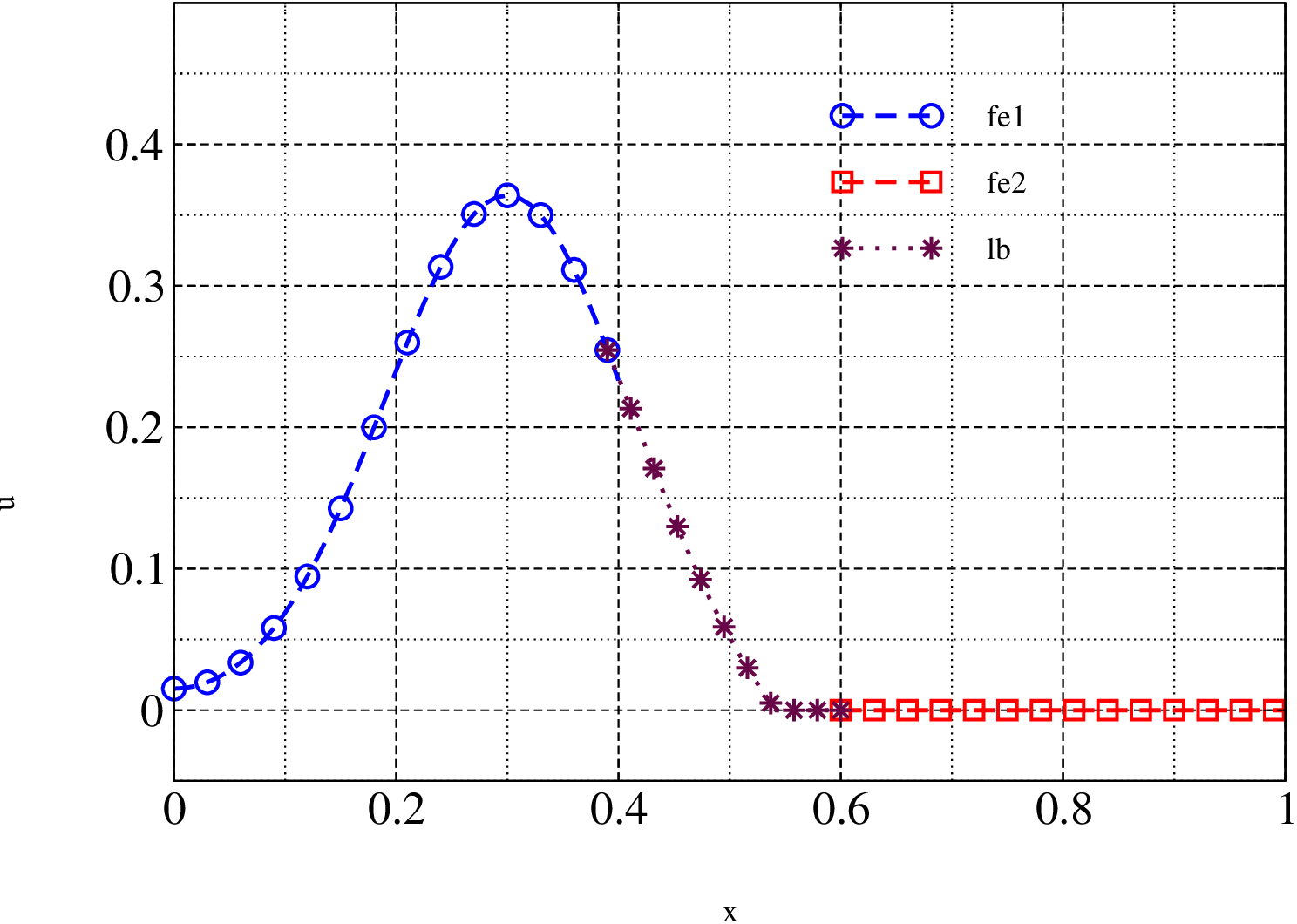}}
	\subfigure[$t = 0.25$]{\includegraphics[clip,scale=0.35]{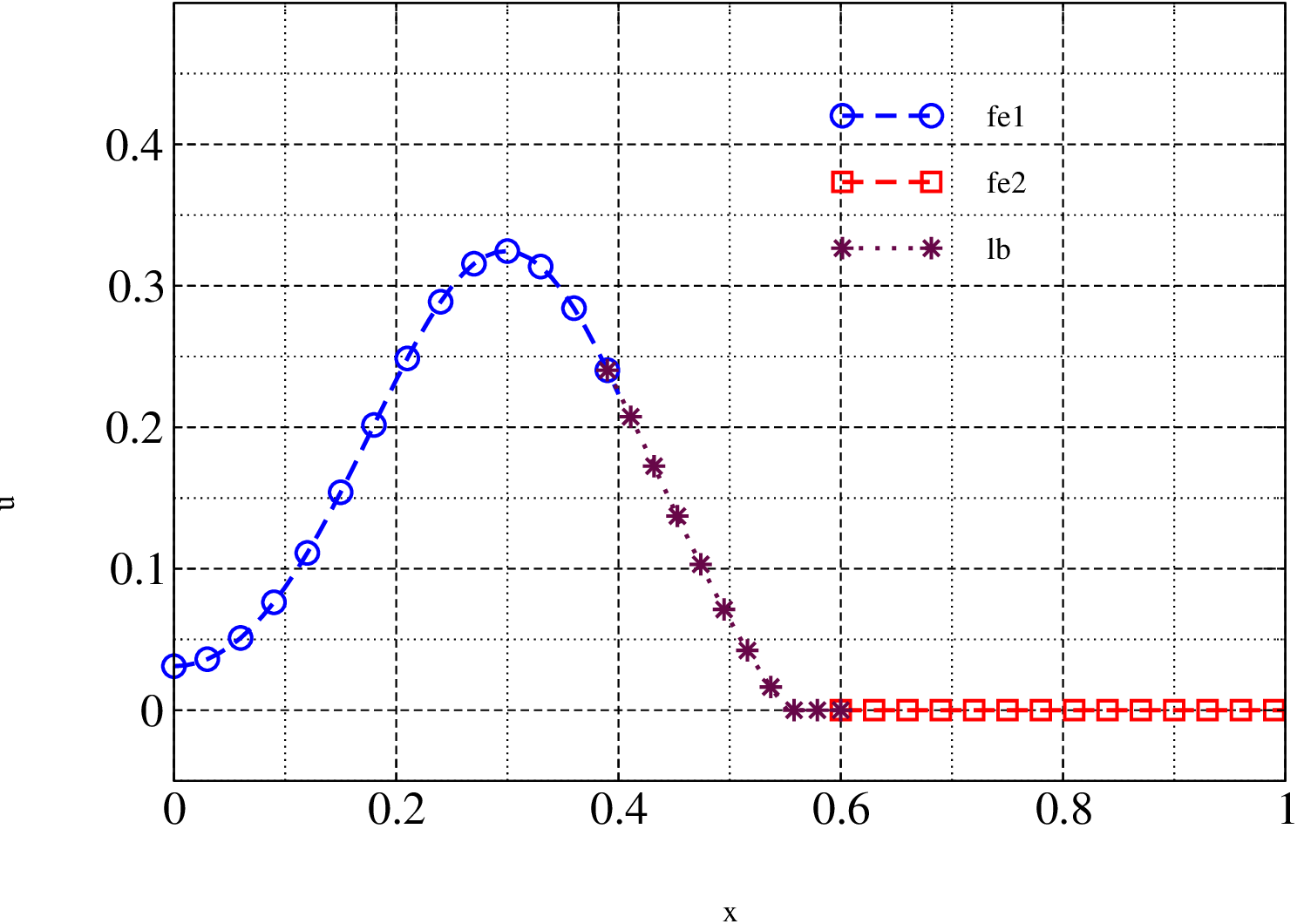}}
	\subfigure[$t = 0.5$]{\includegraphics[clip,scale=0.35]{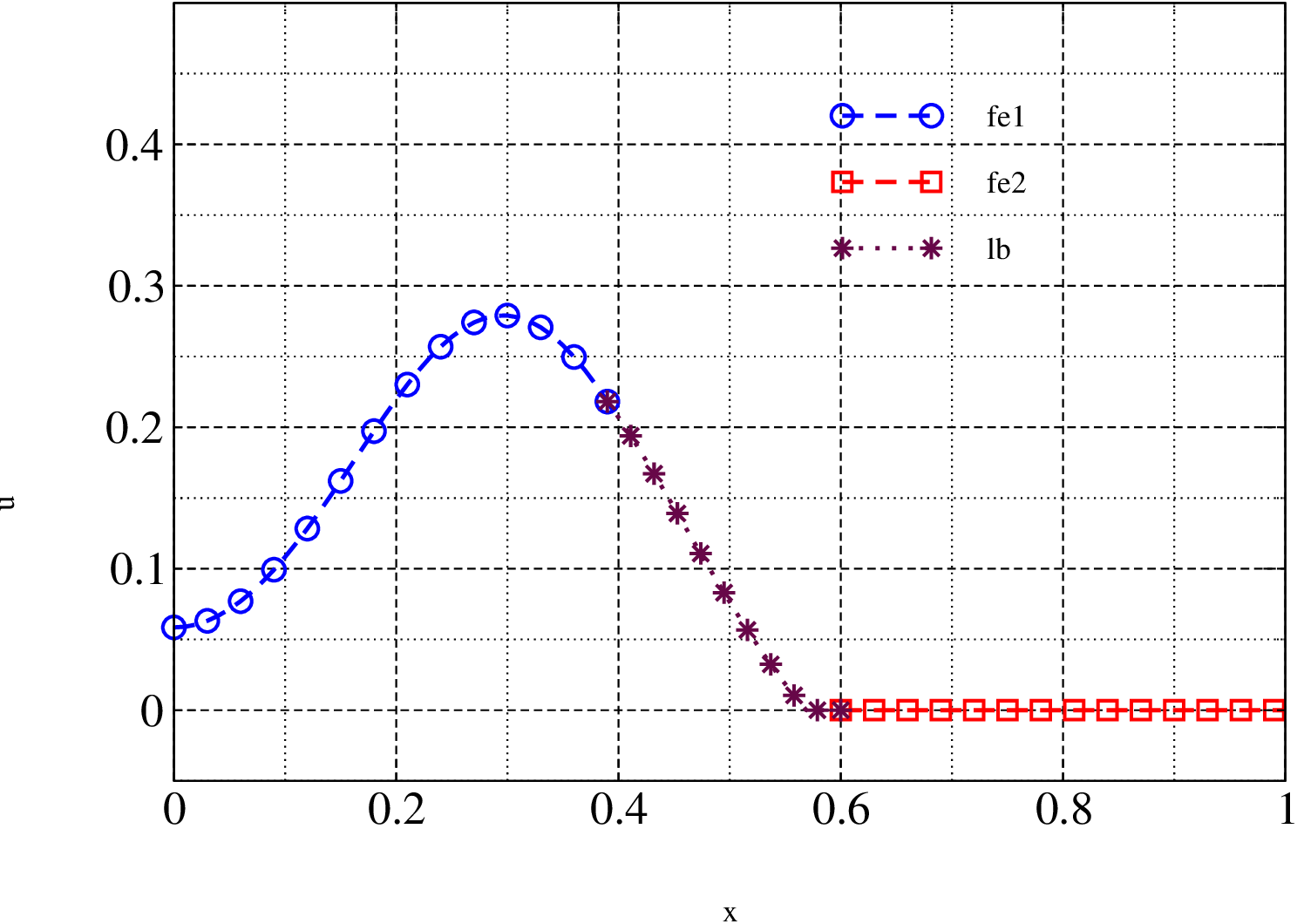}}
	\caption{\textsf{Fast bimolecular reaction in a one-dimensional domain}:~
		Concentration of chemical species $\mathrm{A}$ at different time-levels is shown.
	\label{Fig:1D_FR_cA}}
\end{figure}

\begin{figure}
	\centering
	\psfrag{x}{$x$-axis}
	\psfrag{u}{concentration}
	\psfrag{fe1}{$u_{\mathrm{c},1}$}
	\psfrag{fe2}{$u_{\mathrm{c},2}$}
	\psfrag{lb}{$u_{\mathrm{f}}$}
	\subfigure[$t = 0.1$]{\includegraphics[clip,scale=0.35]{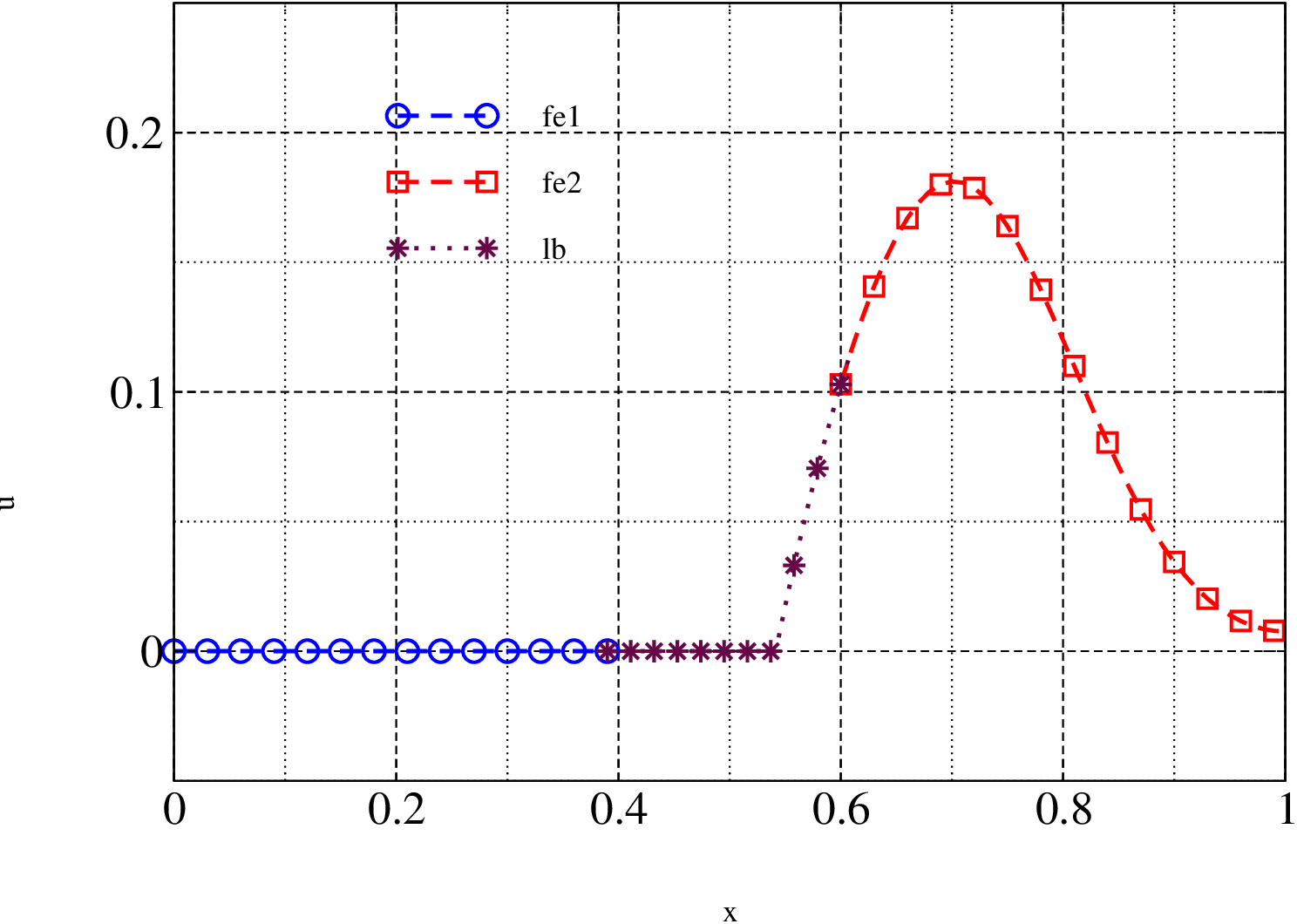}}
	\subfigure[$t = 0.25$]{\includegraphics[clip,scale=0.35]{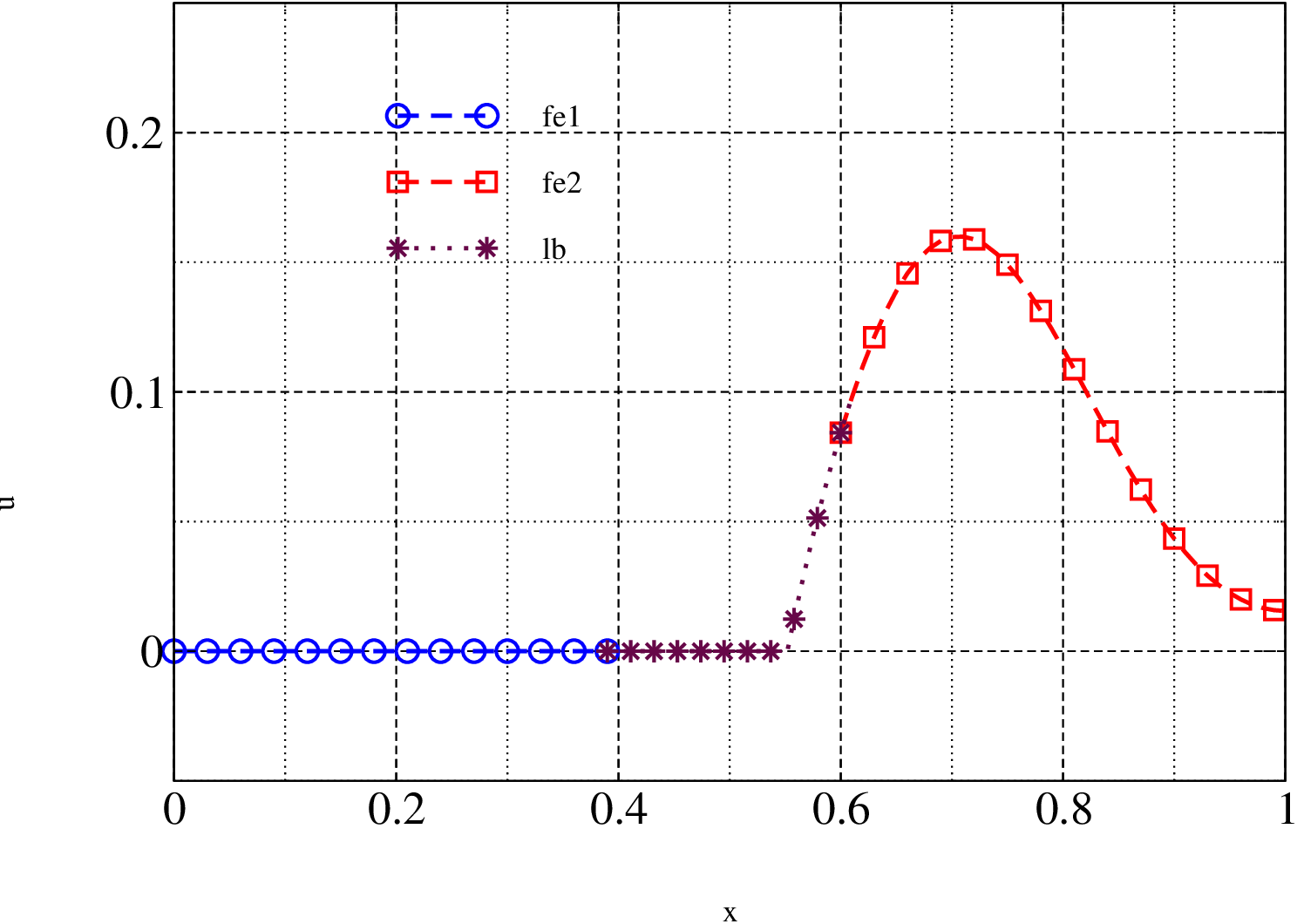}}
	\subfigure[$t = 0.5$]{\includegraphics[clip,scale=0.35]{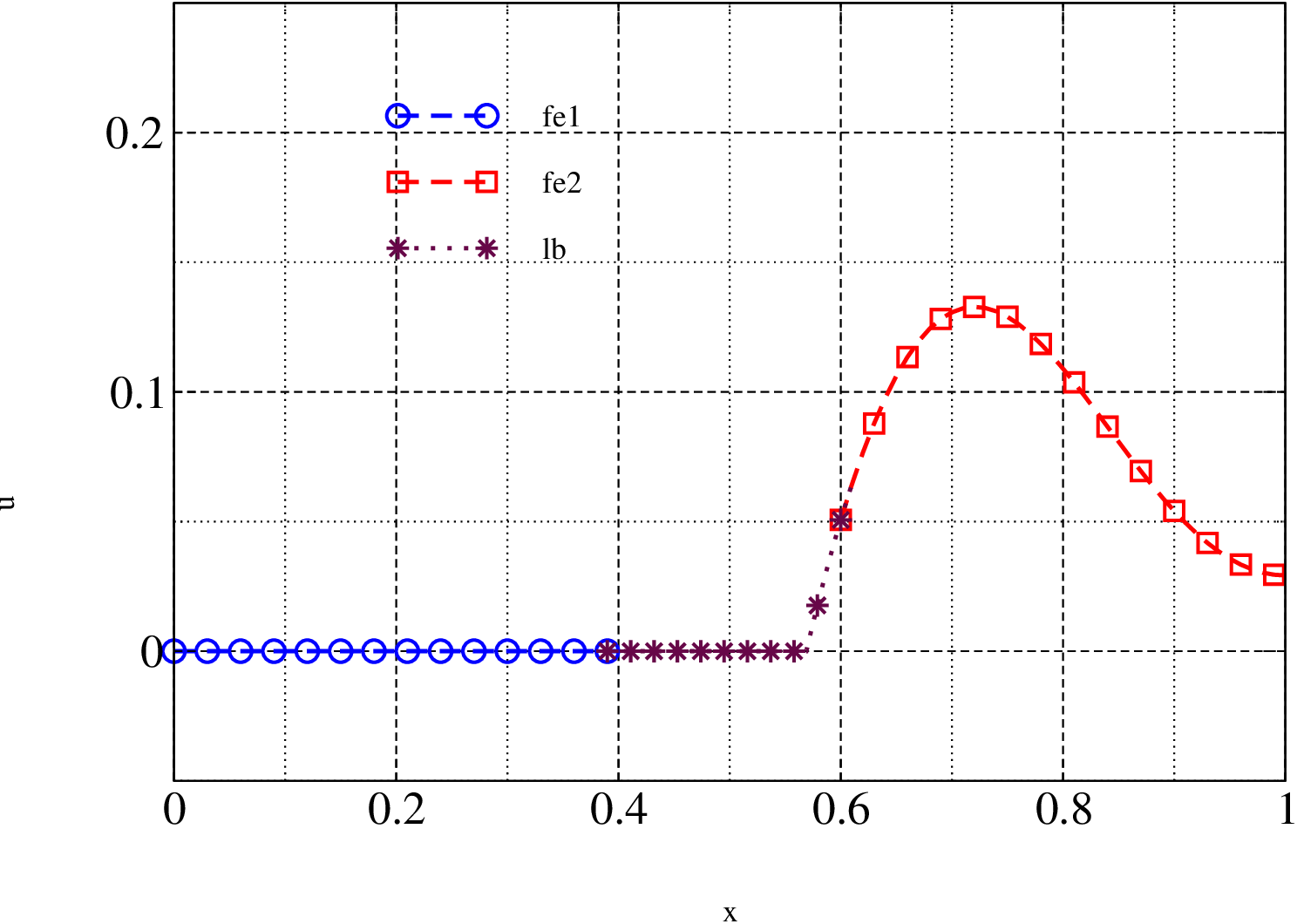}}
	\caption{\textsf{Fast bimolecular reaction in a one-dimensional domain}:~
		In this figure, concentration of species $\mathrm{B}$ is shown.
	\label{Fig:1D_FR_cB}}
\end{figure}

\begin{figure}
	\centering
	\psfrag{x}{$x$-axis}
	\psfrag{u}{concentration}
	\psfrag{fe1}{$u_{\mathrm{c},1}$}
	\psfrag{fe2}{$u_{\mathrm{c},2}$}
	\psfrag{lb}{$u_{\mathrm{f}}$}
	\subfigure[$t = 0.1$]{\includegraphics[clip,scale=0.35]{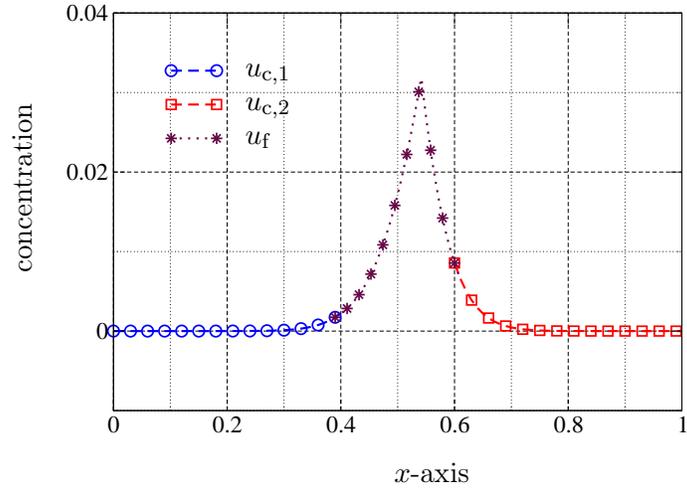}}
	\subfigure[$t = 0.25$]{\includegraphics[clip,scale=0.35]{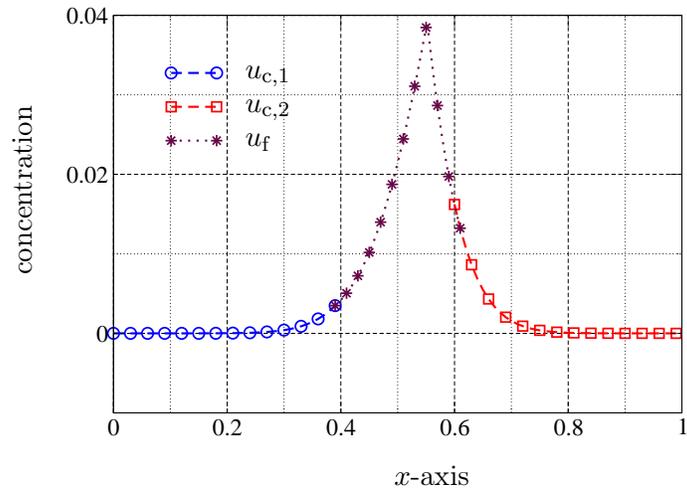}}
	\subfigure[$t = 0.5$]{\includegraphics[clip,scale=0.35]{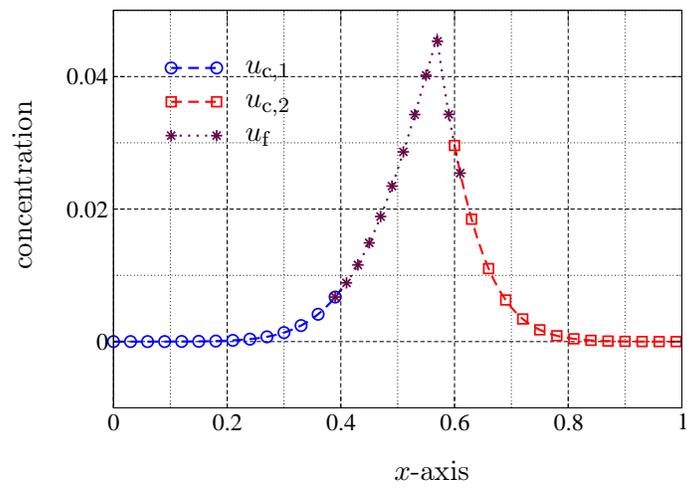}}
	\caption{\textsf{Fast bimolecular reaction in a one-dimensional domain}:~
		Concentration of species $\mathrm{C}$ is shown. The fine-scale subdomain is
		located near the region where majority of production occurs. 
	\label{Fig:1D_FR_cC}}
\end{figure}

Numerical results at various time-levels are presented
in figures \ref{Fig:1D_FR_cA}--\ref{Fig:1D_FR_cC}, which
show the concentrations of all the participating
chemical species from the coarse-scale subdomains
(which are denoted by $\mathrm{u}_{\mathrm{c},1}$ and
$\mathrm{u}_{\mathrm{c},2}$) and the fine-scale subdomain
(which is denoted by $\mathrm{u}_{\mathrm{f}}$). As evident
from these figures, the numerical solution is compatible
near and in the overlap region, and the proposed hybrid
framework has performed well. 

\subsection{Advection and diffusion in a homogeneous medium}
\begin{figure}
  \centering
  \includegraphics[scale=0.6,clip]{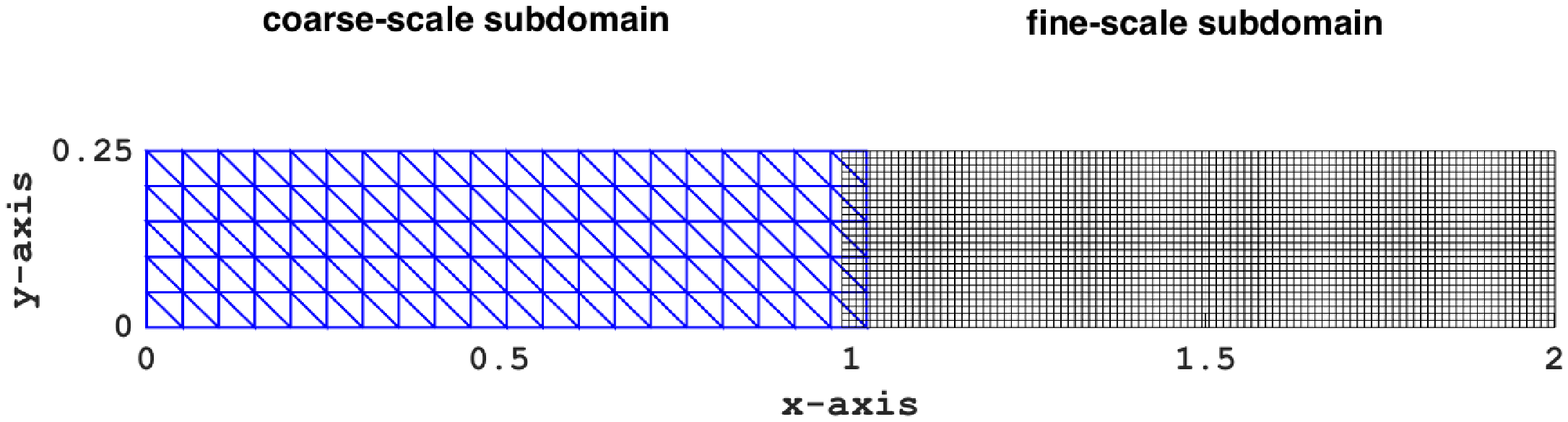}
  \caption{\textsf{Advection and diffusion in a homogeneous
      medium}:~This figure illustrates the overlapping domain
    decomposition as well as the non-matching grids for coarse-
    and fine-scale domains. The length of the overlap region is
    $L_{\mathrm{overlap}} = 4 \times 10^{-2}$.
    \label{Fig:2D_Homogeneous_demo}}
\end{figure}

\begin{figure}
  \centering
  \subfigure[$t = 20 \Delta t_{\mathrm{c}}$]{
    \includegraphics[scale=0.2,clip]{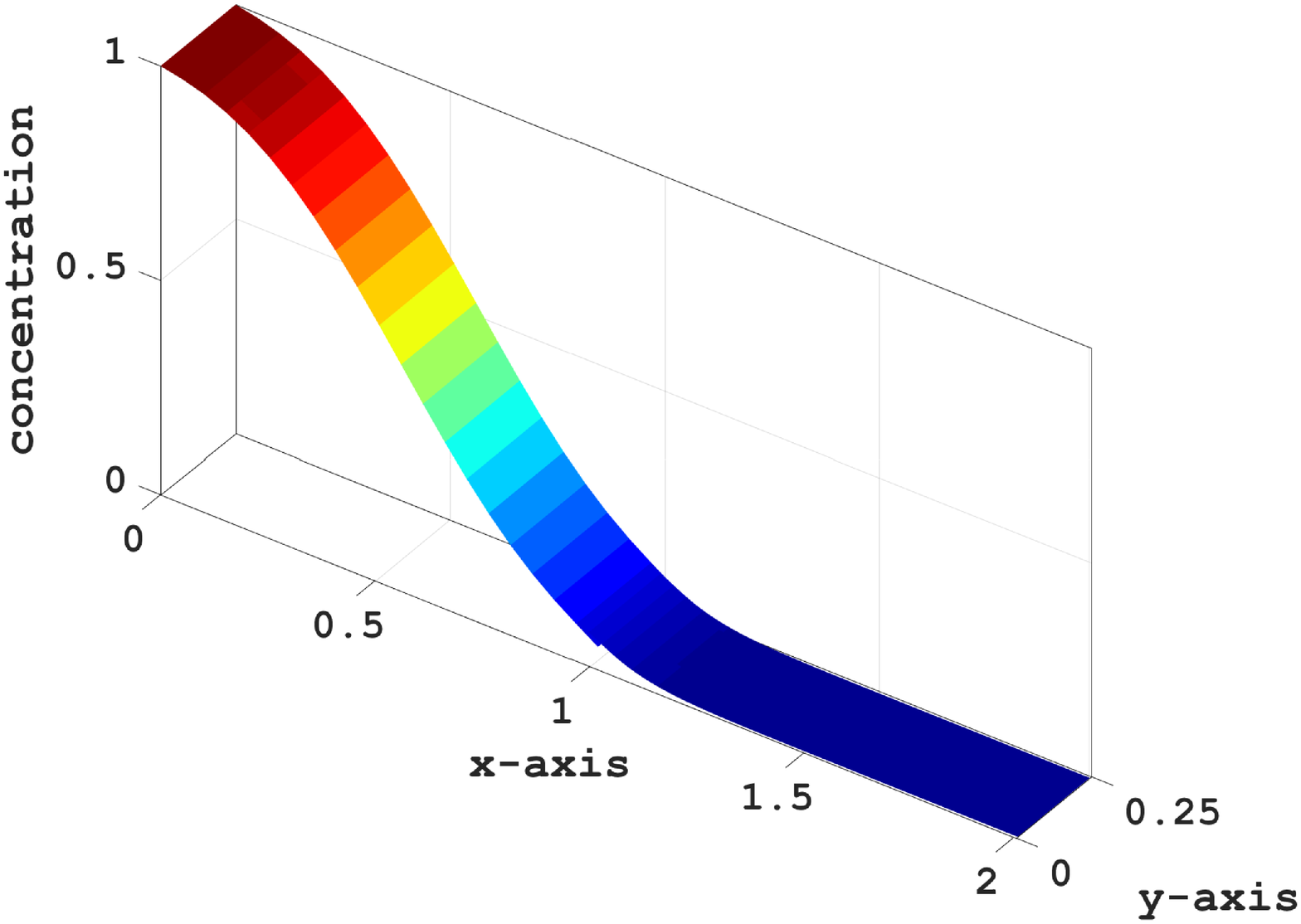}}
  \subfigure[$t = 40 \Delta t_{\mathrm{c}}$]{
    \includegraphics[scale=0.2,clip]{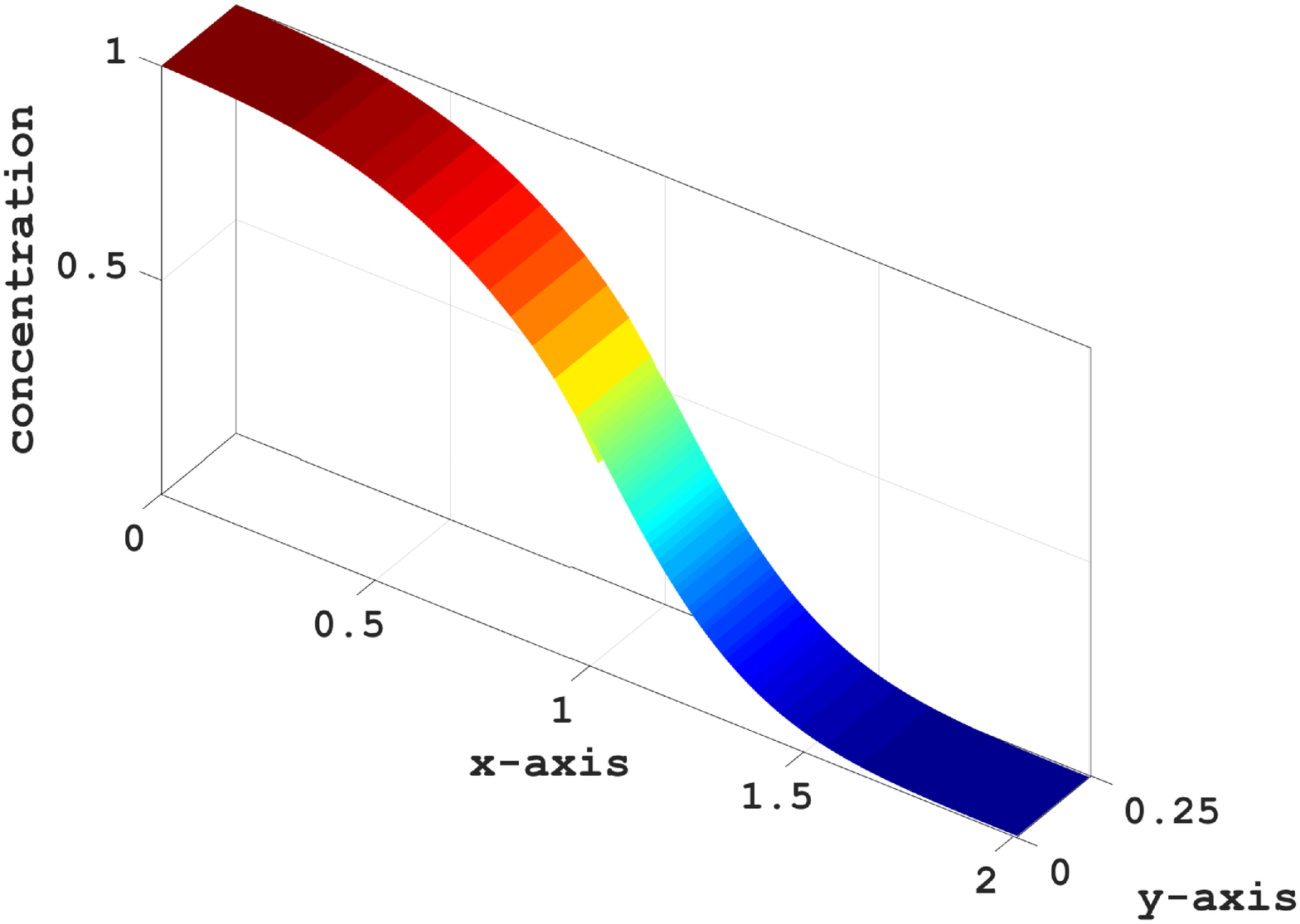}}
  \subfigure[$t = 60 \Delta t_{\mathrm{c}}$]{
    \includegraphics[scale=0.2,clip]{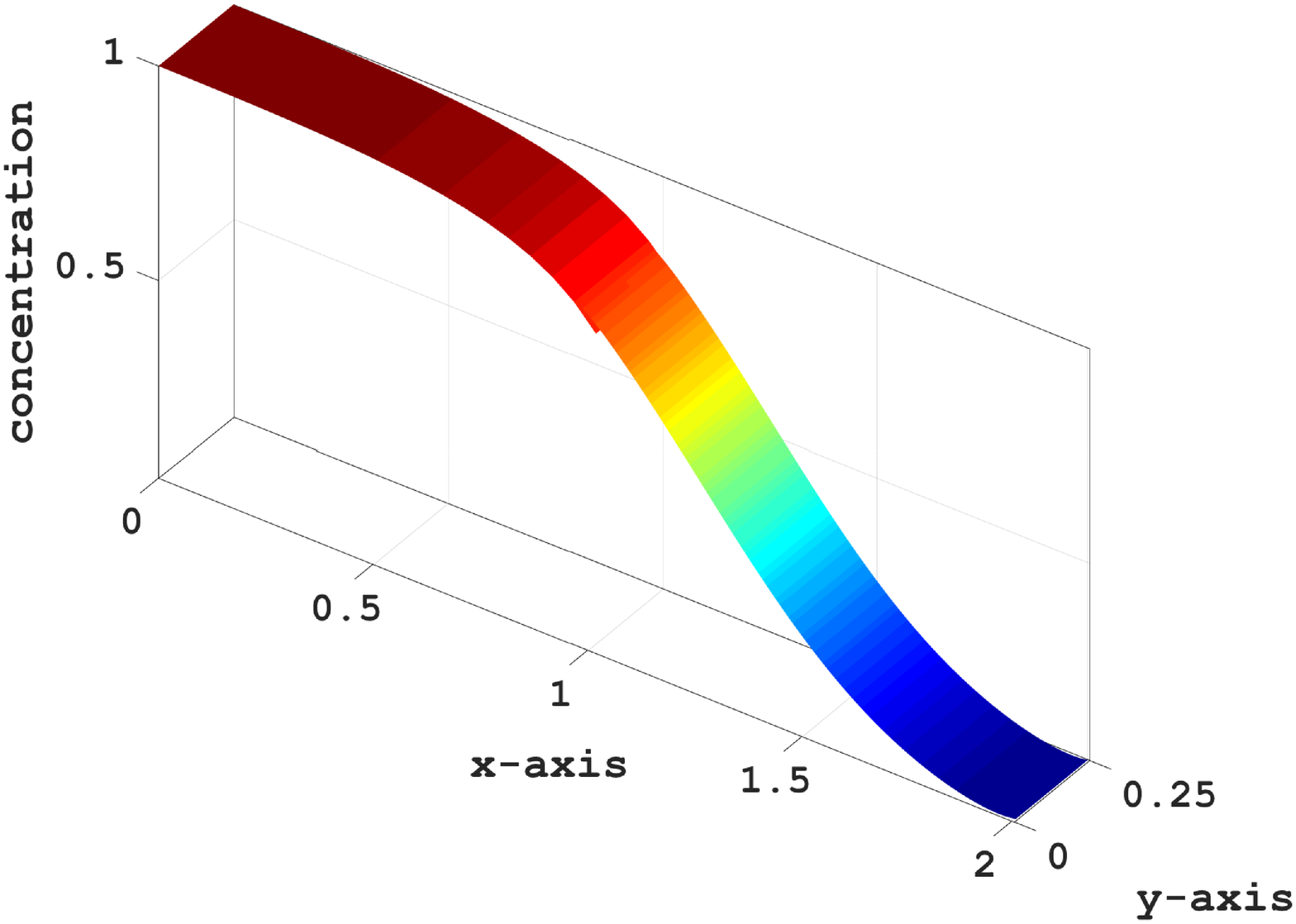}}
  \caption{\textsf{Advection and diffusion in a homogeneous medium}:~
    In this figure the concentration at different time-levels is shown.
    In this case P\'eclet number is $P = 20$. In each time-step, we 
    have employed 5 sub-iterations to ensure the compatibility of
    the solution in the overlap region. \label{Fig:2D_Homogeneous_s1}}
\end{figure}

\begin{figure}
	\centering
	\subfigure[$t = 10 \Delta t_{\mathrm{c}}$]{
		\includegraphics[scale=0.2,clip]{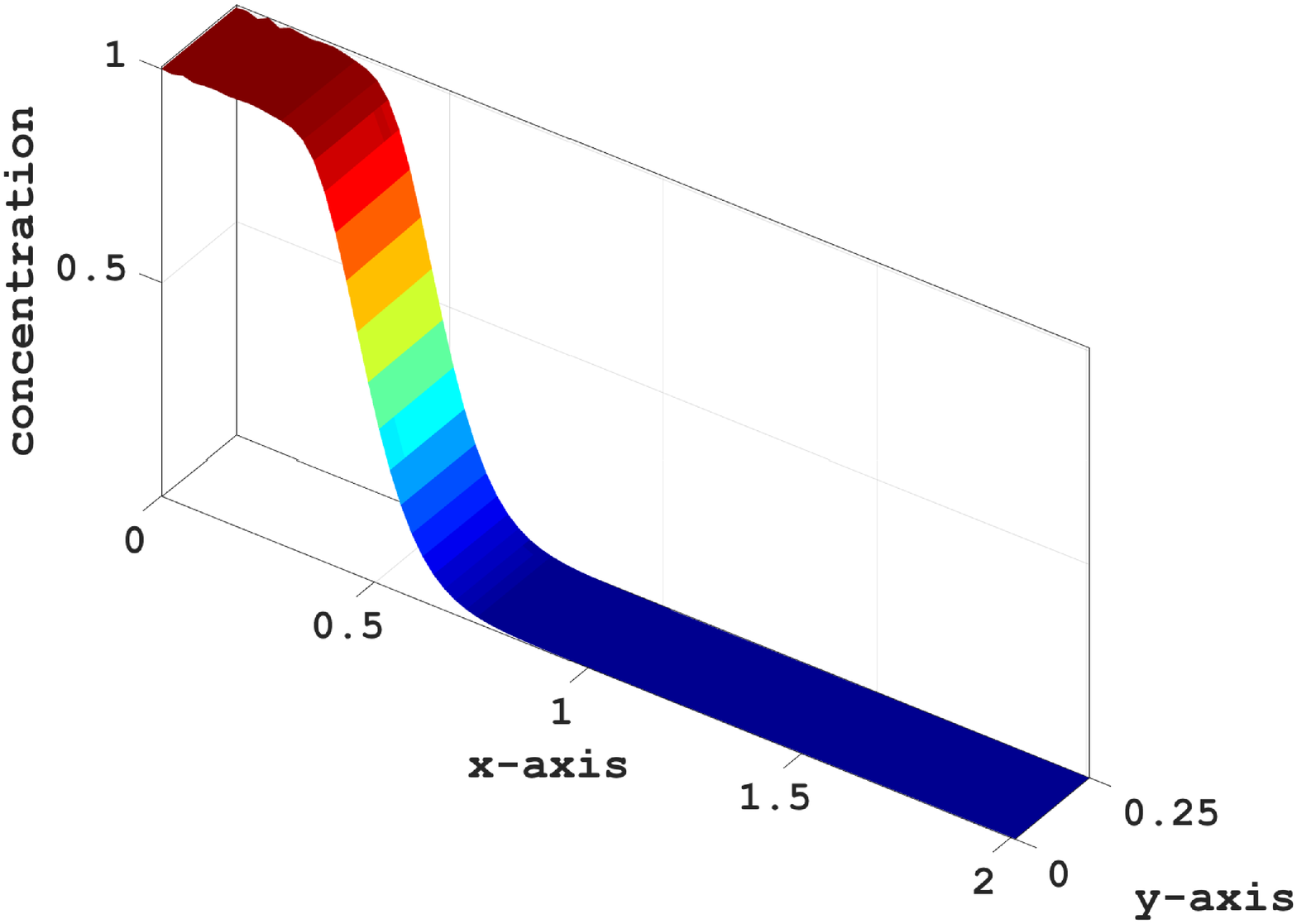}}
	\subfigure[$t = 20 \Delta t_{\mathrm{c}}$]{
		\includegraphics[scale=0.2,clip]{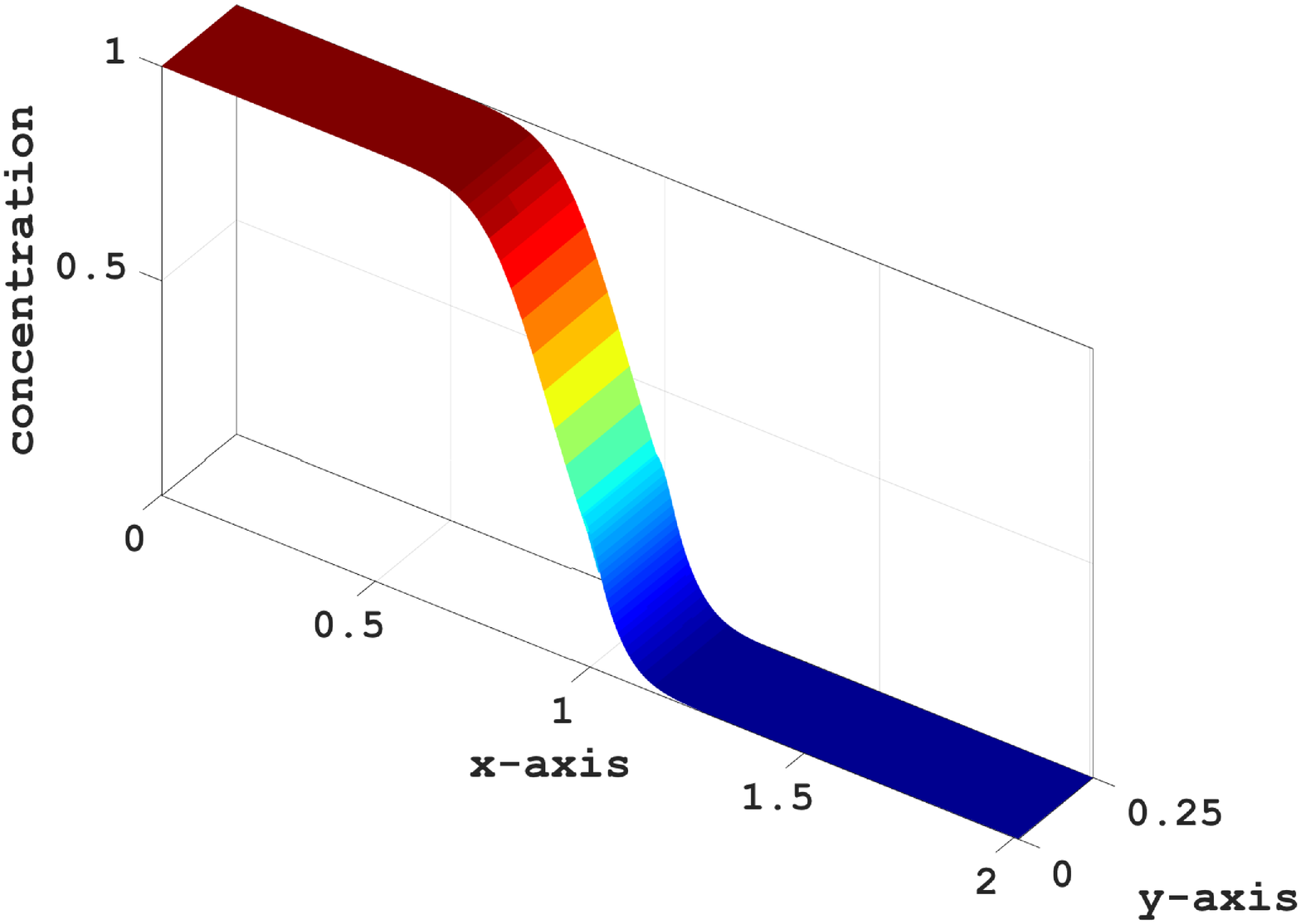}}
	\subfigure[$t = 30 \Delta t_{\mathrm{c}}$]{
		\includegraphics[scale=0.2,clip]{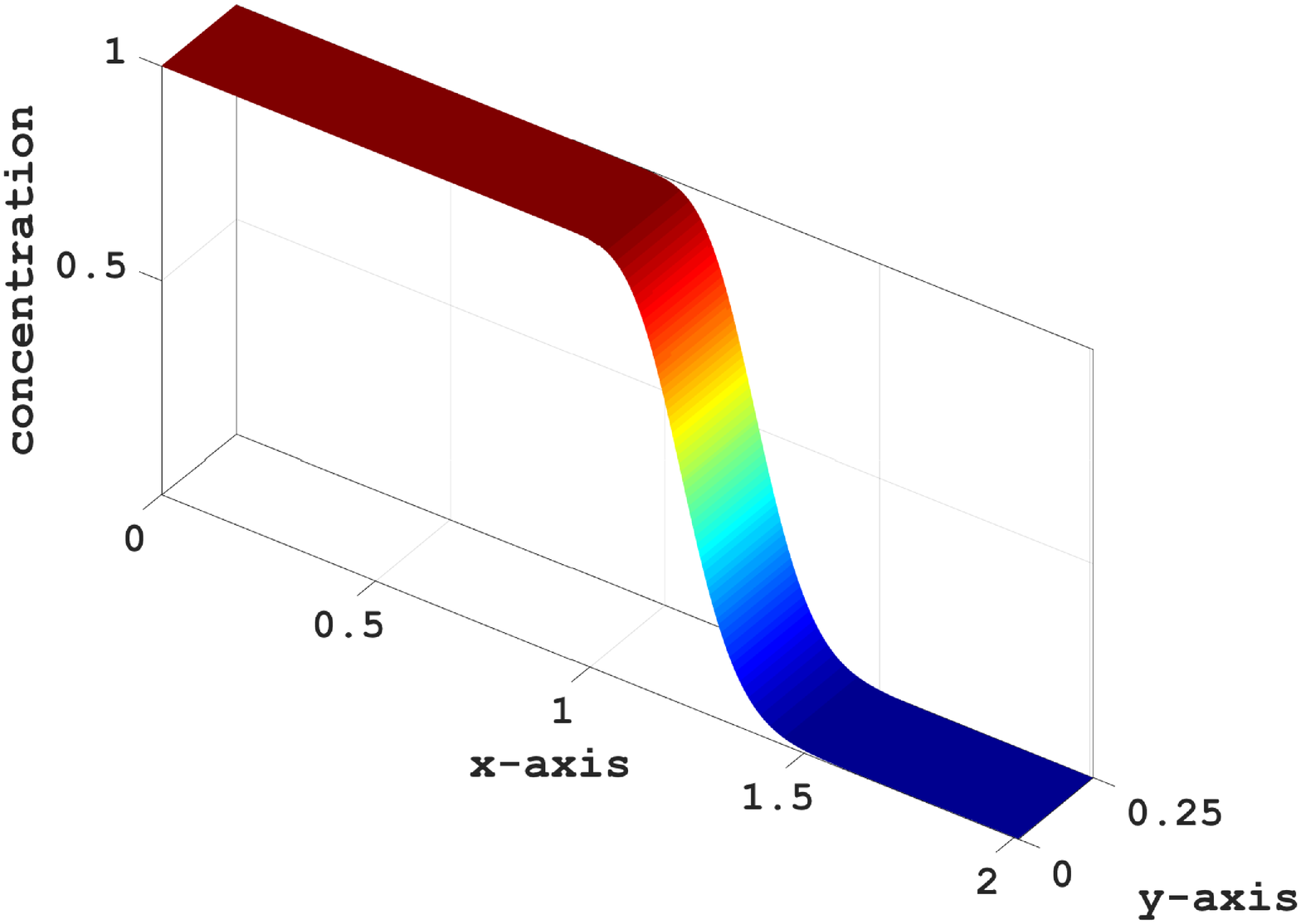}}
	\subfigure[$t = 40 \Delta t_{\mathrm{c}}$]{
		\includegraphics[scale=0.2,clip]{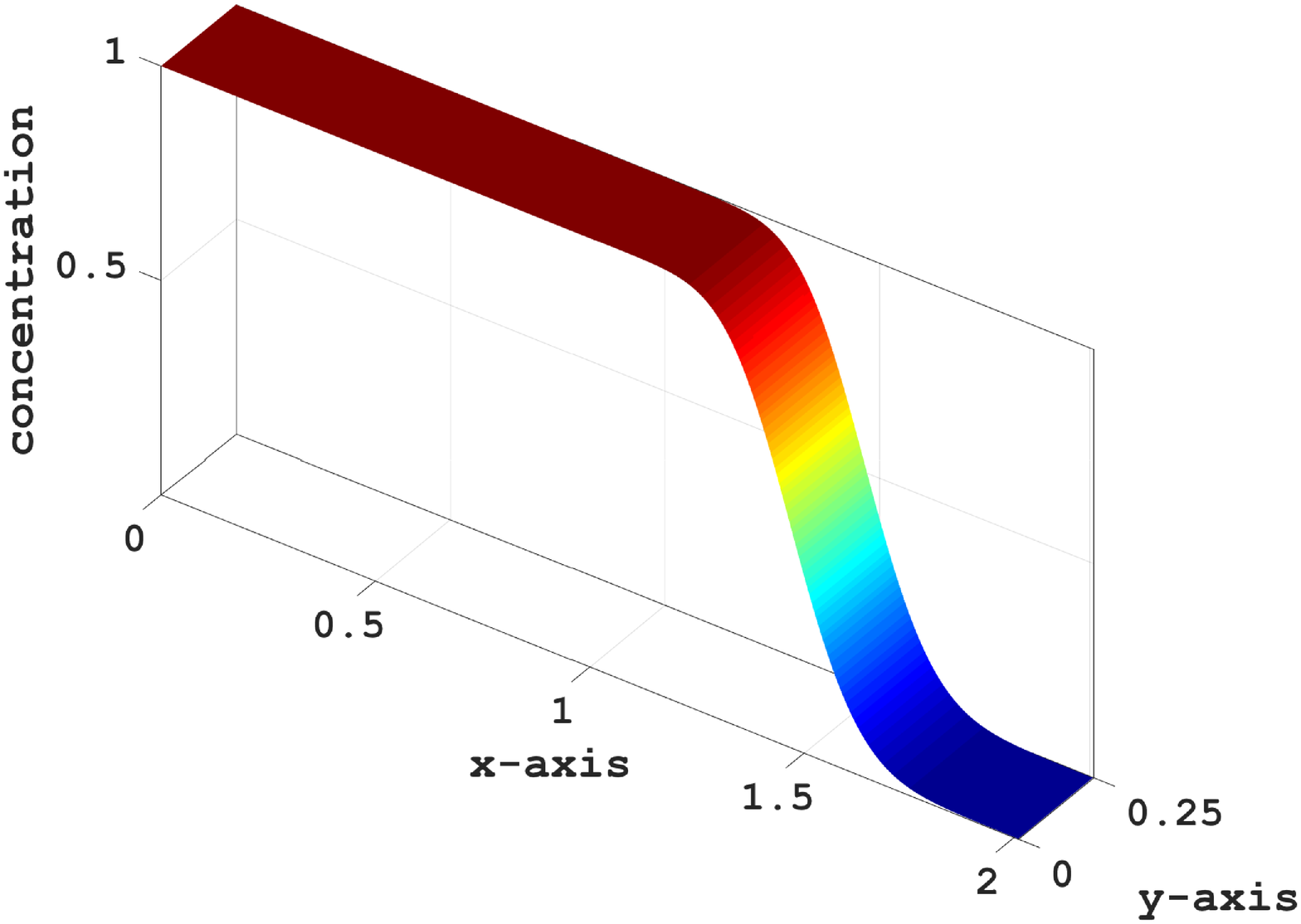}}
	\caption{\textsf{Advection and diffusion in a homogeneous medium}:~
	Concentration at different time-levels is shown. 
	In this case P\'eclet number is $P = 200$. 
	In each time-step we use 5 sub-iterations. In this 
	case, the coupling may cause minor disruptions in the 
	concentration profile when the front is passing through
	the interface. This can be improved by increasing the 
	number of sub iterations (MaxIter). The spurious 
	oscillations in the finite element method can be improved
	by mesh refinement. No spurious oscillations observed in the
	solution from LBM.
	\label{Fig:2D_Homogeneous_s2}}
\end{figure}

\begin{figure}
	\centering
	\subfigure[$t = 50 \Delta t_{\mathrm{c}}$]{
		\includegraphics[clip, scale=0.2]{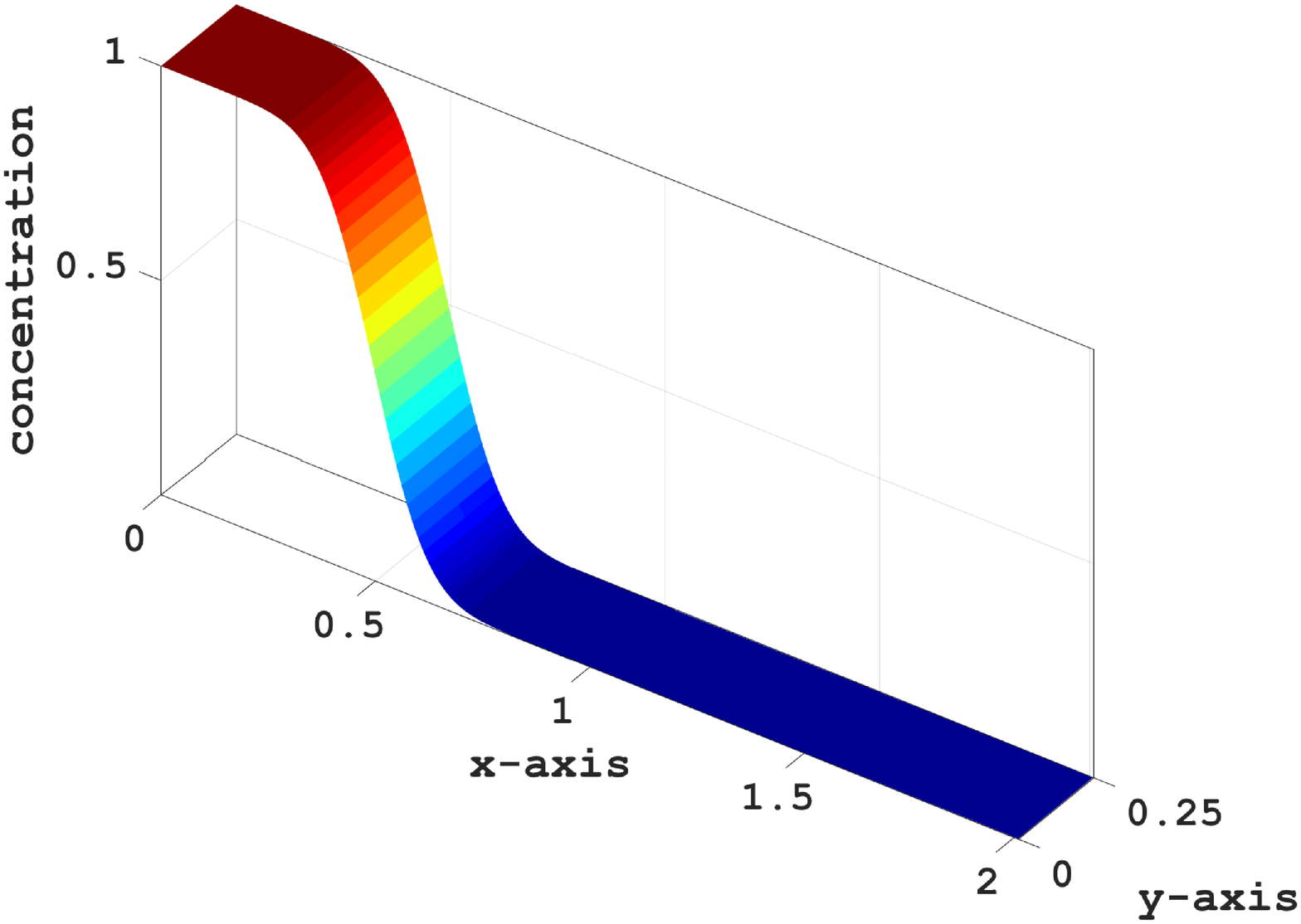}
	}
	\subfigure[$t = 100 \Delta t_{\mathrm{c}}$]{
		\includegraphics[clip, scale=0.2]{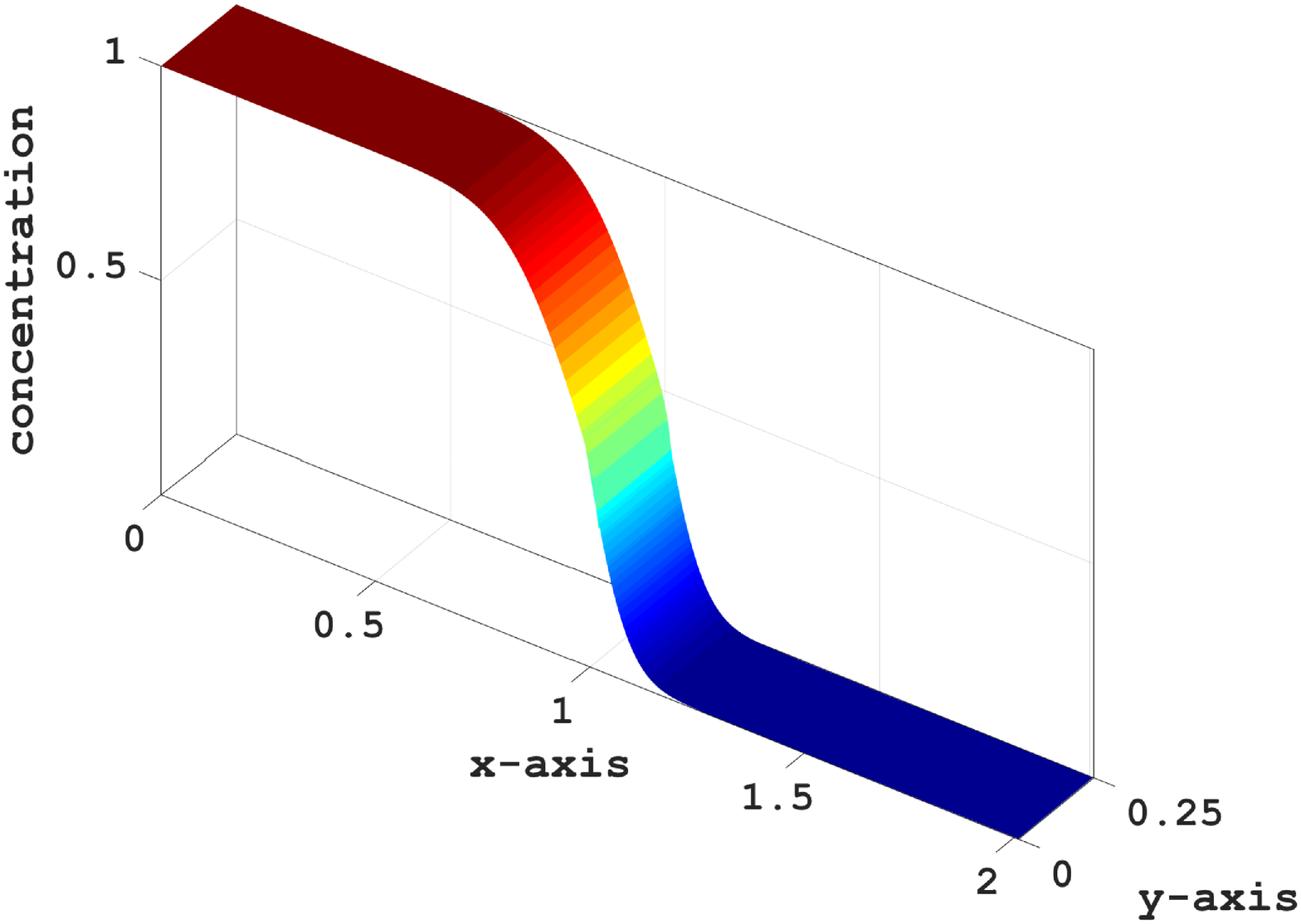}
	}
	\subfigure[$t = 150 \Delta t_{\mathrm{c}}$]{
		\includegraphics[clip, scale=0.2]{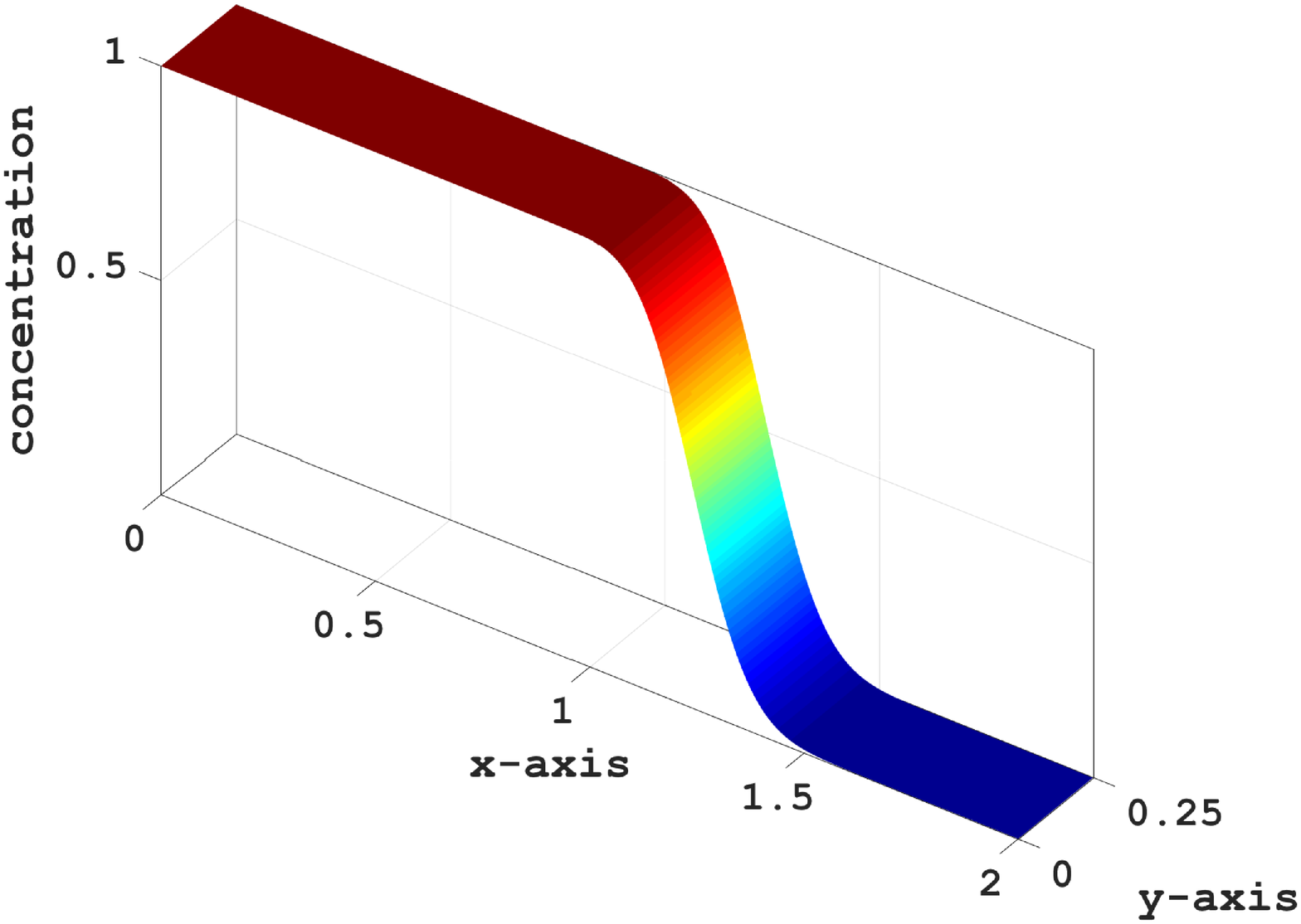}
	}
	\subfigure[$t = 200 \Delta t_{\mathrm{c}}$]{
		\includegraphics[clip, scale=0.2]{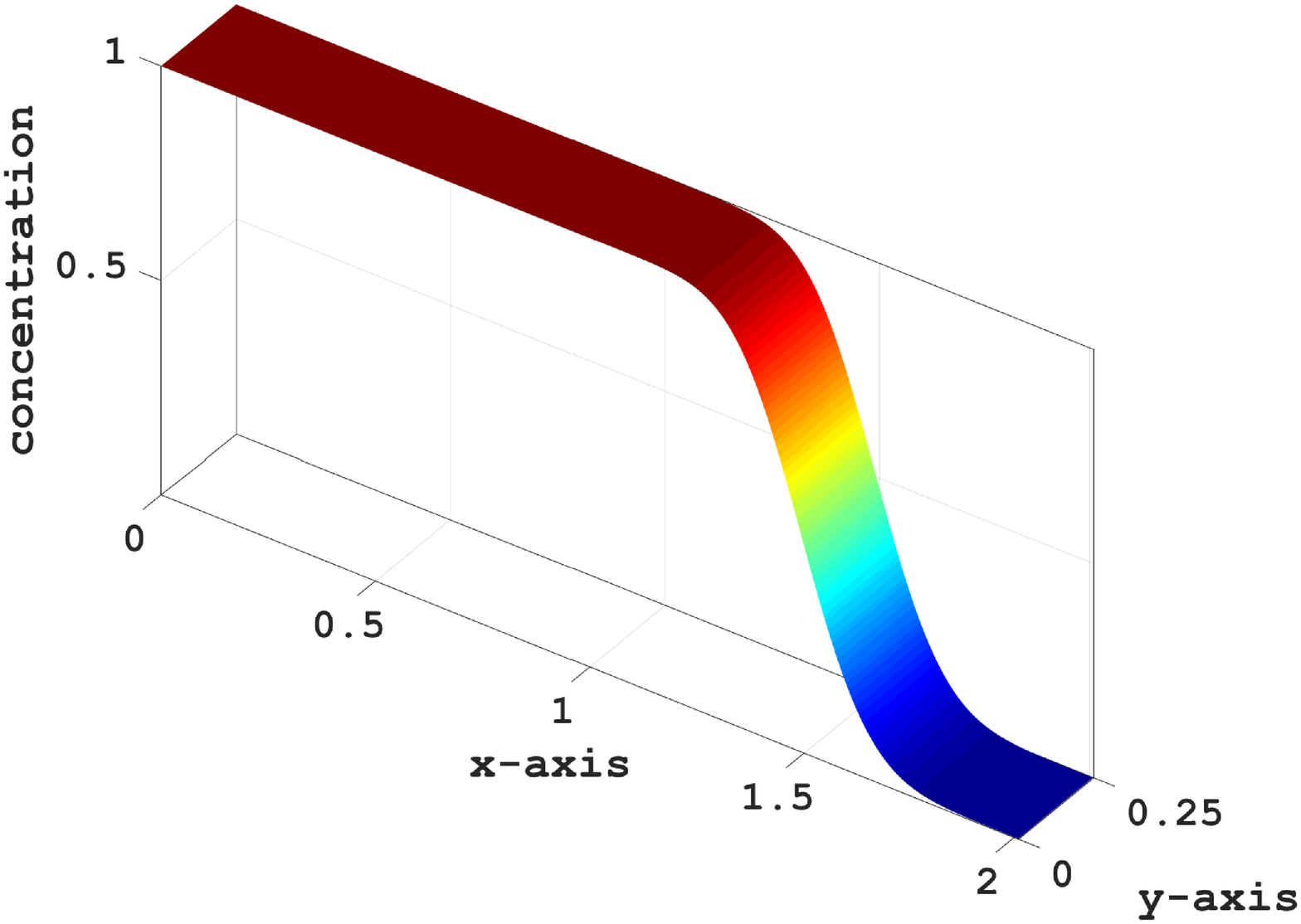}
	}
	\caption{\textsf{Advection and diffusion in a homogeneous medium}:~
		In this figure, concentration at different time-levels is 
		shown. Here, the space and time discretization is refined 
		in order to remove spurious oscillations. The number of 
		sub-iterations in each time-step is 10, which has helped 
		reduce the incompatibility in the overlap region.
	\label{Fig:2D_Homogeneous_s3}}
\end{figure}

Consider $\Omega = (0,2)\times(0,1/4)$ with 
$\Gamma^{\mathrm{D}} = \{0\}\times[0,1/4]$ and $\Gamma^{\mathrm{N}}
= \partial \Omega - \Gamma^{\mathrm{D}}$ corresponding to the
following boundary conditions
\begin{align}
	&\mathrm{u}^{\mathrm{p}}(\mathbf{x},\mathrm{t}) = 
	1 \quad \mathbf{x}\in\Gamma^{\mathrm{D}},\;\mathrm{t}\in\mathcal{I} \\
	&\mathrm{q}^{\mathrm{p}}(\mathbf{x},\mathrm{t}) = 0 \quad
	\mathbf{x}\in\Gamma^{\mathrm{N}},\;\mathrm{t}\in\mathcal{I}
\end{align}
where $\mathcal{I} = (0,T]$ is the time interval of interest.
The initial concentration in the entire domain is taken to be
$\mathrm{u}_{0}(\mathbf{x}) = 0$. The isotropic diffusion coefficient 
is $\mathrm{D} = 5 \times 10^{-3}$. Here, we shall use the proposed
framework to numerically solve this problem for different P\'eclet
numbers. We will define the coarse-scale domain $\Omega_{\mathrm{c}}$
and the fine-scale domain $\Omega_{\mathrm{f}}$ as follows:
\begin{align}
  \Omega_{\mathrm{c}} = \left(0,1 + \frac{L_{\mathrm{overlap}}}{2}\right)
  \times(0, 1/4),
  \quad \Omega_{\mathrm{f}} = \left(0,1 - \frac{L_{\mathrm{overlap}}}{2}\right)\times
	(0,1/4)
\end{align}
where we pick $L_{\mathrm{overlap}} = 4/100$. The SUPG
formulation \eqref{Eqn:SUPG} with linear three-node
triangular elements will be used in 
$\Omega_{\mathrm{c}}$. Numerical solution in $\Omega_{\mathrm{f}}$ 
will be sought for using lattice Boltzmann method with the 
$D2Q4$ lattice model. In figure \ref{Fig:2D_Homogeneous_demo} 
non-matching grid sized for finite element and lattice Boltzmann 
methods in the given domain is illustrated. We shall solve the 
problem for two different choices of advection velocity:
\begin{enumerate}[(i)]
\item \emph{Case 1}:~Considering the uniform advection velocity of 
	$\mathrm{v}_{\mathrm{x}} = 5 \times 10^{-2}$ and $\mathrm{v}_{\mathrm{y}} = 0$
	over domain $\Omega$, we find the P\'eclet number as $P = 20$.
	The element-size in the coarse-scale domain is $h_{\mathrm{f}} 
	\approx 7\times10^{-2}$, and the grid spacing for LBM is 
	$h_{\mathrm{f}} = 10^{-2}$. The time-steps in the coarse-scale 
	and fine-scale subdomains are $\Delta t_{\mathrm{c}} = 5.1
	\times10^{-1}$ and $\Delta t_{\mathrm{f}} = 10^{-2}$ respectively.
	Note that the ratio between the coarse and fine time-steps
	is $\eta = 51$. The number of iterations is MaxIter = 5. 
	The result is shown in figure 
	\ref{Fig:2D_Homogeneous_s1}. The numerical solution from FEM and 
	LBM retained good compatibility while the concentration front
	passed through the subdomain interfaces. The coupling of the 
	two methods did not result in any disruptions on the propagation
	of the chemical species in the domain.
\item \emph{Case 2}:~Here, we will take $\mathrm{v}_{\mathrm{x}} = 5\times10^{-1}$
	and $\mathrm{v}_{\mathrm{y}} = 0$. In this case the advection 
	velocity is much higher than the previous case, hence, the 
	P\'eclet number is $P = 200$. In this case, the gradient of 
	concentration near the front is steep. We take $h_{\mathrm{c}}
	\approx 2.5\times10^{-2}$, $h_{\mathrm{f}}=2\times10^{-3}$ 
	in coarse and fine-scale subdomains respectively. The time-steps
	are $\Delta t_{\mathrm{c}} = 10^{-1}$ and $\Delta t_{\mathrm{f}}
	= 4 \times 10^{-4}$. The ratio between the time-steps is 
	$\eta = 250$. Similar to the previous case, the number of 
	sub-iterations is MaxIter = 5. The numerical results are shown 
	in figure \ref{Fig:2D_Homogeneous_s2}. One of the numerical
	difficulties that can occur in this case is the spurious 
	oscillations in the concentration. It can be observed that 
	the numerical solution in the coarse-scale domain experiences
	some of this oscillations (see figure 
	\ref{Fig:2D_Homogeneous_s2}(a)), however, 
	it should be noted that this weak instability is not due to 
	the hybrid coupling and is an artifact of the finite element 
	formulation. With mesh refinement, these instabilities can be 
	removed. Note that when the front is reaching the interface
	of the subdomains, some minor incompatibility between the 
	numerical solution of different subdomains in the overlap
	region is seen (see figure \ref{Fig:2D_Homogeneous_s2}(b)). 
	This incompatibility can be alleviated by increasing the 
	number of sub-iterations in each time-step. As expected, 
	once the front leaves the 
	coarse-scale domain completely, no node-to-node 
	oscillations remain. In figure \ref{Fig:2D_Homogeneous_s3}, 
	the numerical solution using smaller time-steps and mesh 
	size is shown.  The time-step in the coarse-scale 
	domain is $\Delta t_{\mathrm{c}} = 2\times10^{-2}$
	and $\Delta t_{\mathrm{f}} = 4\times10^{-4}$ in the 
	fine-scale domain. The element-size in the coarse-scale
	subdomain is $h_{\mathrm{c}} \approx 1.8\times10^{-2}$
	and in the fine-scale subdomain is
        $h_{\mathrm{f}} = 2\times10^{-3}$. The number of 
	sub-iterations in each time-step is increased to 
	10. Hence, spurious oscillations and incompatibility 
	in the overlap region (while the front is passing 
	through the interface) are largely reduced.
\end{enumerate}
In this numerical experiment we conclude that in order to
capture interior/boundary layers more accurately, mere 
mesh or time-step refinement is not enough. One needs to 
increase the number of sub-iterations in each time-step.

\subsection{Hybrid simulation of dissolution of
  calcium carbonate in porous media} 
Calcium carbonate $\mathrm{CaCO}_{3}$ is a common
chemical compound found in the subsurface. The
dissolution of calcium carbonate is an important
geochemical equilibrium reaction, which arises
in a wide variety of subsurface applications
\citep{1988_Drever}. The chemical reaction
takes the following form:
\begin{align}
  \label{Eqn:Sol_CaCO3}
  \mathrm{CaCO}_{3} \rightleftharpoons
  \mathrm{Ca}^{2+} + \mathrm{CO}_{3}^{2-}
\end{align} 
For convenience, we shall use $\mathrm{u}_{1}$,
$\mathrm{u}_{2}$ and $\mathrm{u}_{3}$ to denote
the concentrations of $\mathrm{CaCO}_{3}$,
$\mathrm{Ca}^{2+}$ and $\mathrm{CO}_{3}^{2-}$,
respectively.
This chemical reaction is known to have a product
solubility constant $K_{\mathrm{sp}}$ of about $3.36
\times 10^{-9}$ at room temperature \citep{2002_WaterChem}. 
The product solubility for this chemical reaction can be
written as: 
\begin{align}
  \label{Eqn:Hybrid_Ksp}
  K_{\mathrm{sp}} = \frac{\mathrm{u}_{2} \mathrm{u}_{3}}{\mathrm{u}_{1}}
\end{align}
We introduce the following two reaction invariants:
\begin{subequations}
  \label{Eqn:Carbonate_uncoupled}
  \begin{align}
    \label{Eqn:Carbonate_uncoupled_psi1}
    &\psi_{1} = \mathrm{u}_{1} - \mathrm{u}_{2} \\
    \label{Eqn:Carbonate_uncoupled_psi2}
    &\psi_{2} = \mathrm{u}_{3} - \mathrm{u}_{2}
  \end{align}
\end{subequations}
It should be emphasized that $\psi_{1}$ and $\psi_{2}$
are not the concentrations of any real chemical species.
These invariants are introduced to simplify the problem,
as they decouple the governed equations and hence can be
solved for separately; for example, see
\citep{2013_Nakshatrala_JCP_v253_p278}. Once the values
of $\psi_{1}$ and $\psi_{2}$ are found, the concentration
of the species $\mathrm{Ca}^{2+}$ can be determined using
the following relation:
\begin{align}
  \mathrm{u}_{2} = \frac{1}{2}\left( -\left( \psi_2 + K_{\mathrm{sp}}\right)
  + \sqrt{\left( \psi_2 + K_{\mathrm{sp}}\right)^{2} + 4 K_{\mathrm{sp}} \psi_1}
  \right)
\end{align}
which is obtained by solving equations
\eqref{Eqn:Hybrid_Ksp}--\eqref{Eqn:Carbonate_uncoupled}
for $\mathrm{u}_{2}$. 
The values of $\mathrm{u}_{1}$ and $\mathrm{u}_{3}$
can then be determined using equations
\eqref{Eqn:Carbonate_uncoupled_psi1}--\eqref{Eqn:Carbonate_uncoupled_psi2}.
Here, we are interested in determining the fate of
the chemical species due to the chemical reaction
and transport. We employ the LBM to simulate the
transport problem at the pore-scale (fine-scale)
and the FEM at the continuum-scale.

\begin{figure}
  \centering
  \includegraphics[clip,scale=0.75]{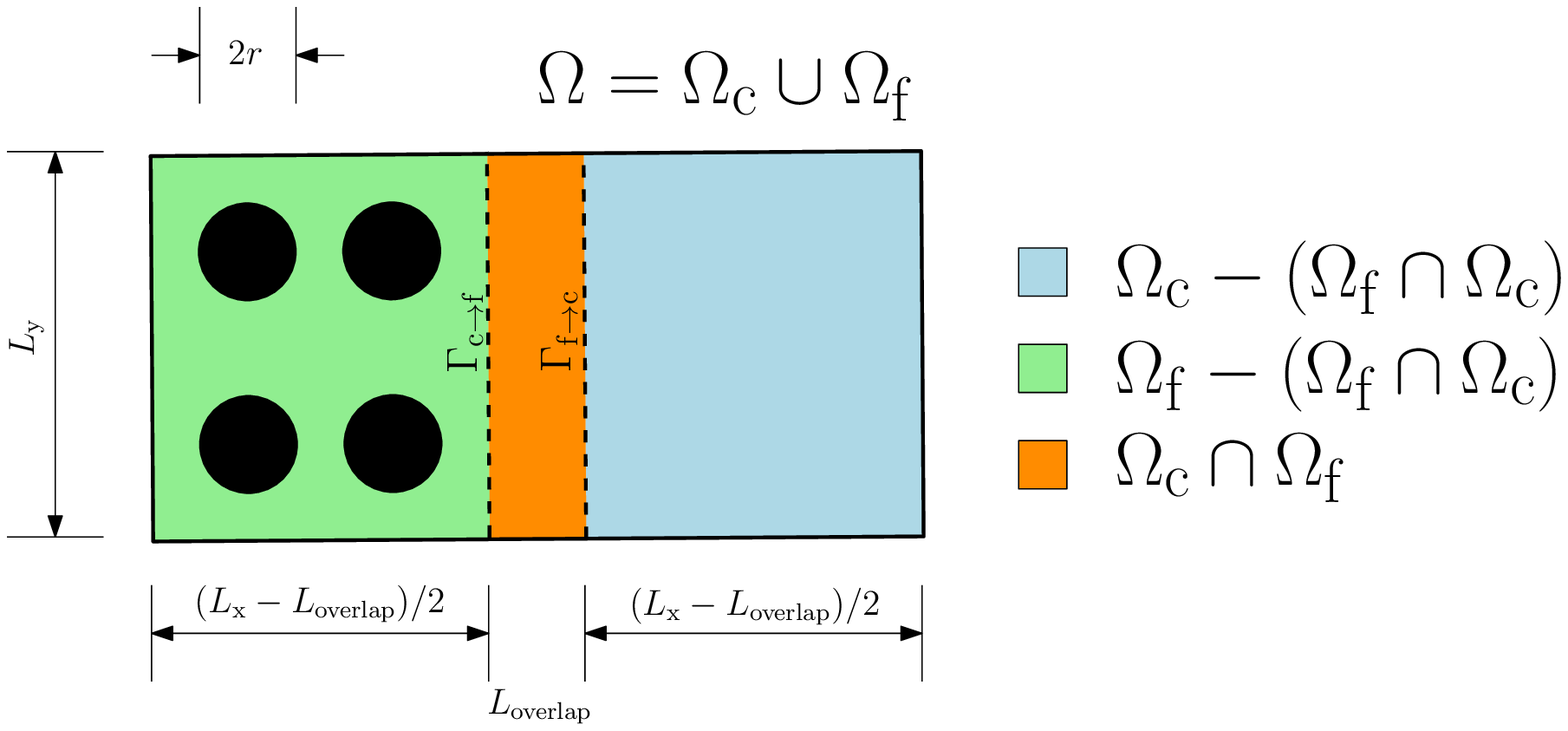}
  \caption{\textsf{Dissolution of calcite problem}:~Computational
    domain and its decomposition into fine and coarse-scale
    subdomains are shown. The black circles represent the
    solid phase in the porous medium. \label{Fig:2D_Carbon_domain}}
\end{figure}

\begin{figure}
	\centering
	\subfigure[velocity in the x-direction]{
		\includegraphics[clip,scale=0.3]{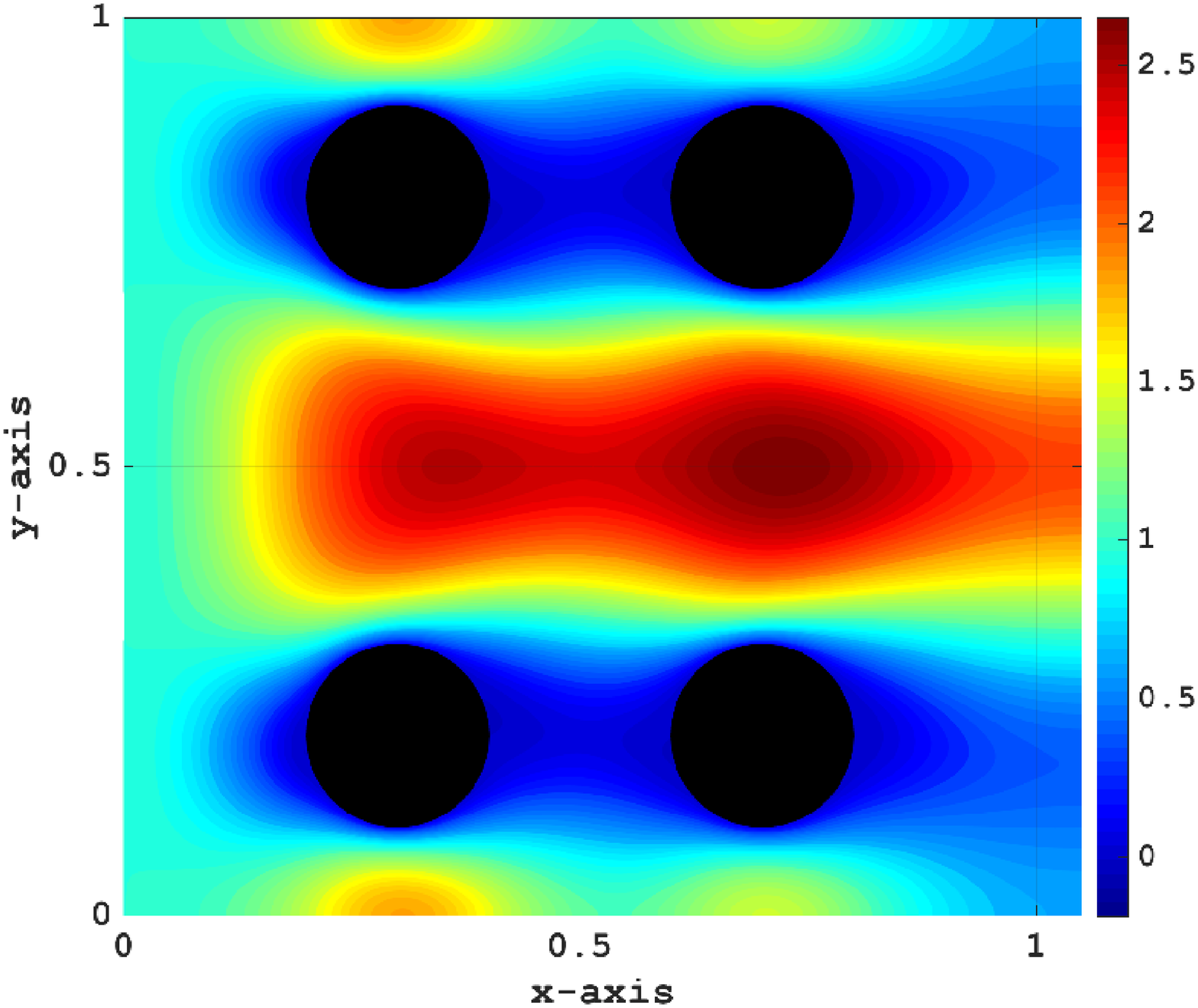}
	}
	\subfigure[velocity in the y-direction]{
		\includegraphics[clip,scale=0.3]{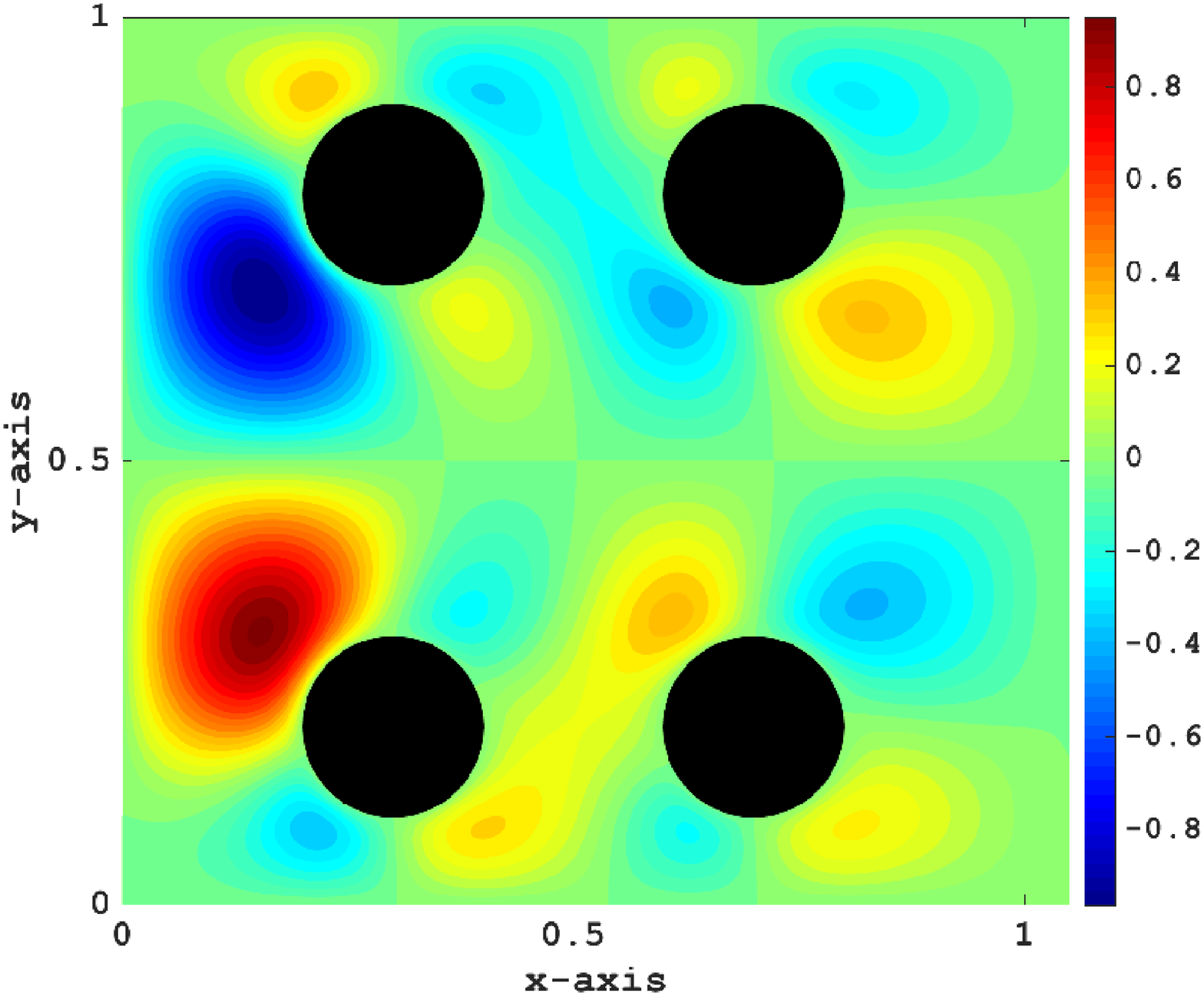}
	}
	\caption{\textsf{Two-dimensional problem}:~ The velocity field shown in this
		figure is obtained using a lattice Boltzmann simulation of incompressible
		Newtonian fluid. The black circles represent the solid obstacles in the 
		porous medium. 
	\label{Fig:2D_Carbon_velocity_LBM}}
\end{figure}

\begin{figure}
  \centering
  \includegraphics[clip, scale=1.0]{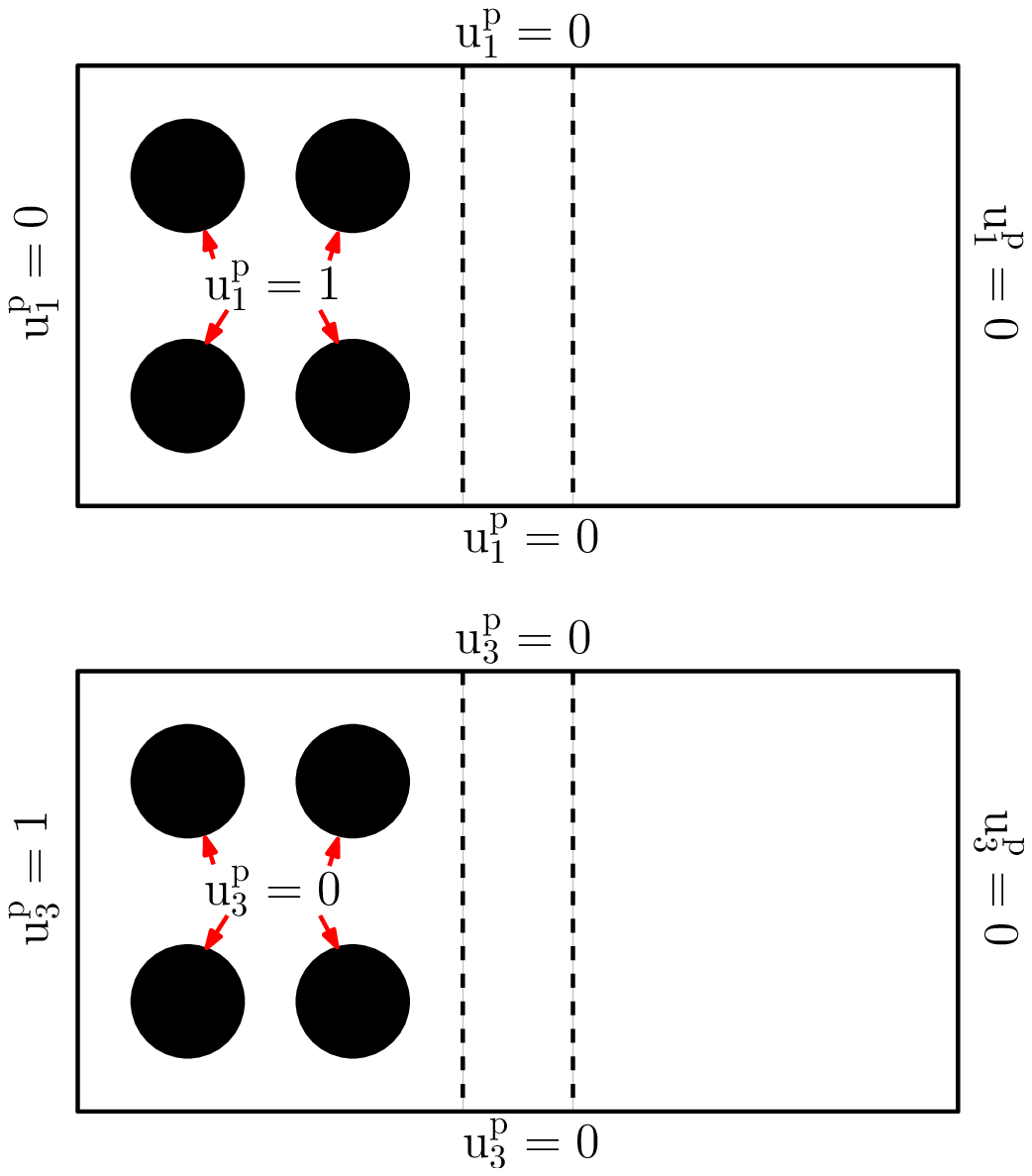}
  \caption{\textsf{Dissolution of calcite}:~The boundary
    conditions for the simulation of dissolution of
    calcite in the porous medium are shown.
    \label{Fig:2D_Carbon_BV}}
\end{figure}

\begin{figure}
  \centering
  \includegraphics[clip, scale=0.35]{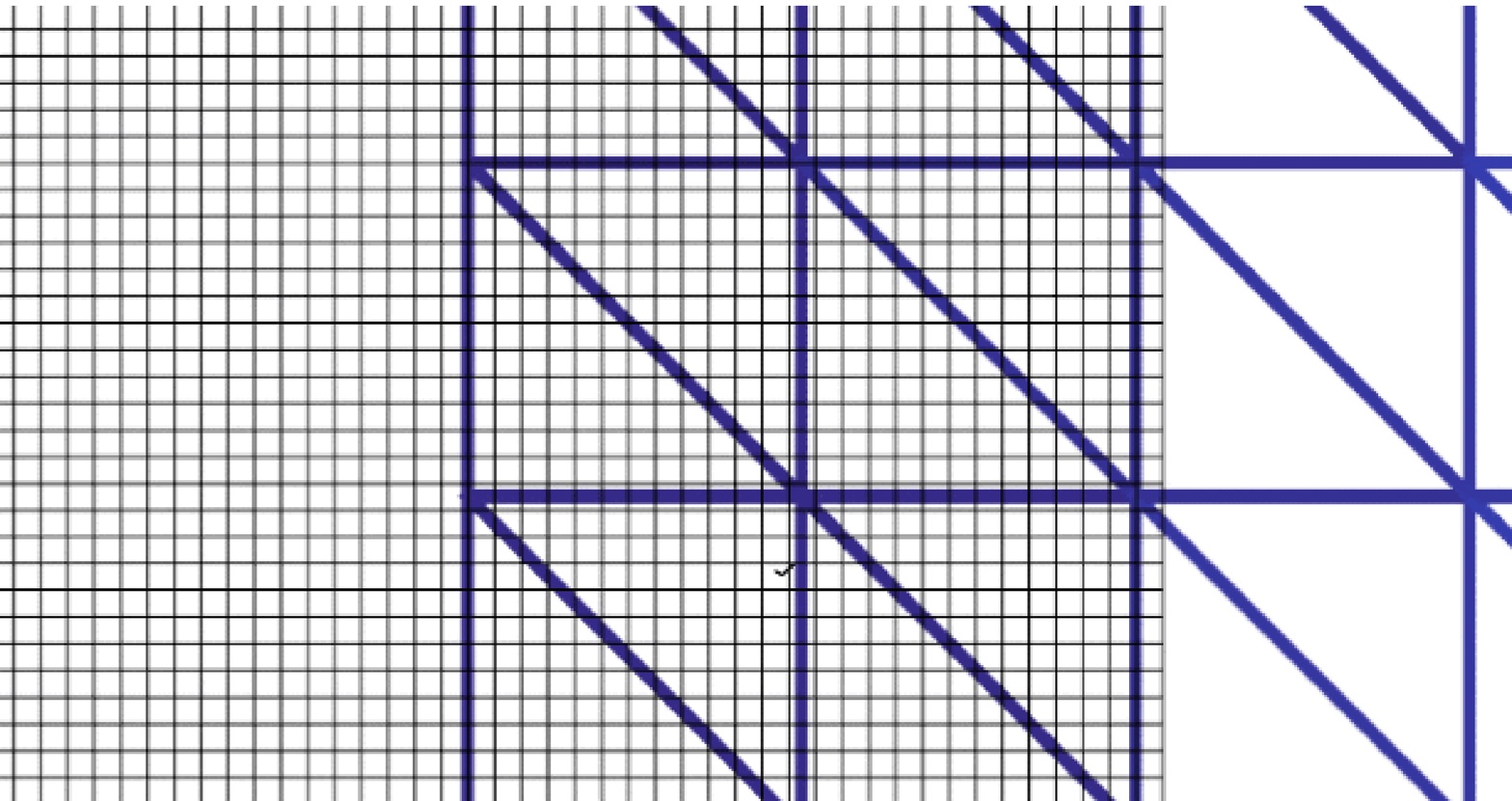}
  \caption{\textsf{Dissolution of calcite}:~This figure
    shows the finite element mesh (which is indicated
    using triangular elements) and the lattice for LBM
    analysis (which is indicated by square cells) near
    the overlapping interface. \label{Fig:2D_Carbon_mesh}}
\end{figure}

The computational domain is shown in figure
\ref{Fig:2D_Carbon_domain} where in $L_{\mathrm{x}}
= 2$ and $L_{\mathrm{y}} = 1$. The radius of the solid
obstacles in $\Omega_{\mathrm{f}}$ (the fine-scale problem)
is taken as $r = 10^{-1}$. The length of the overlap region is set to 
$L_{\mathrm{overlap}} = 10^{-1}$. Obviously, because of the geometry of 
$\Omega_{\mathrm{f}}$, a more detailed description of the flow is required.
We used LBM with a $D2Q9$ lattice model to solve the Navier-Stokes
equations in the fine-scale subdomain $\Omega_{\mathrm{f}}$
\citep{2001_Succi_LBM,2003_Yu_PAS}. 
The prescribed components on the inlet velocity on the boundary
$\mathrm{x} = 0$ are $\mathrm{v}_{\mathrm{x}} = 1$ and 
$\mathrm{v}_{\mathrm{y}} = 0$. 
The pressure on $\Gamma_{\mathrm{f}\rightarrow\mathrm{c}}$
is set to be zero and periodic boundary conditions are enforced on 
the boundaries located at $\mathrm{y} = 0$ and $\mathrm{y} = 1$ for
$0 < \mathrm{x} < (L_{\mathrm{x}} + L_{\mathrm{overlap}})/2$. The 
resulting velocity field is shown in figure \ref{Fig:2D_Carbon_velocity_LBM}
and will be used as the advection velocity for the fine-scale problem.
In the overlap region, the average velocity in the x-direction is 
close to 1 and the average velocity in the y-direction is close to 0.
Hence, the advection velocity in the coarse-scale domain is taken to
be $\mathrm{v}_{\mathrm{x}} = 1$ and $\mathrm{v}_{\mathrm{y}} = 0$.
The values of concentrations on the boundary of the domain are 
shown in figure \ref{Fig:2D_Carbon_BV} and the diffusion 
coefficient is taken to be $D = 10^{-1}$. For numerical 
simulation of the advection-diffusion problem, we will use 
$h_{\mathrm{c}} = 5.0\times10^{-2}$ and $h_{\mathrm{f}} = 4.0\times10^{-3}$. 
The time-steps are $\Delta t_{\mathrm{c}} = 10^{-1}$ and $\Delta t_{\mathrm{f}} 
= 4.0\times10^{-5}$ (the ratio between the time-steps is $\eta = 2500$). 
Furthermore, we will use the $D2Q9$ lattice model in the fine-scale 
domain (solved using LBM). The non-matching grid near the overlap 
region is shown in figure \ref{Fig:2D_Carbon_mesh}. 
Obviously, one of the advantages of the proposed coupling algorithm
is that fine-scale features (such as advection velocity within the
pores) can be accounted for without a noticeable overhead in the
computational cost. In this problem, fine-scale features are sought
after only in $\Omega_{\mathrm{f}}$, and a coarse estimate in 
$\Omega_{\mathrm{c}}$ is deemed enough.   

\begin{figure}
  \centering
  \subfigure[$t = \Delta t_{\mathrm{c}}$]{
    \includegraphics[clip, scale=0.55]{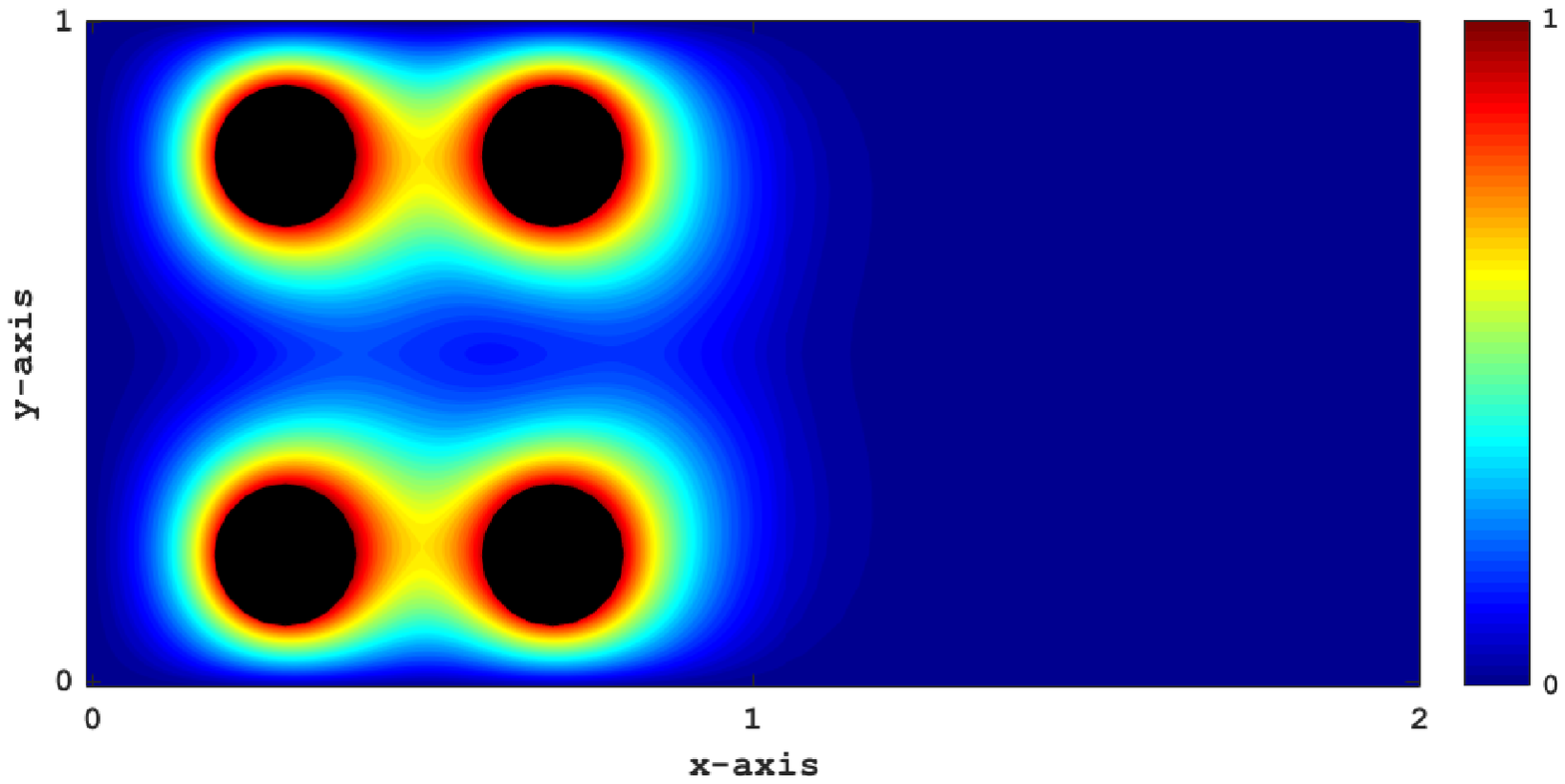}
  }
  \subfigure[$t = 5\Delta t_{\mathrm{c}}$]{
    \includegraphics[clip, scale=0.55]{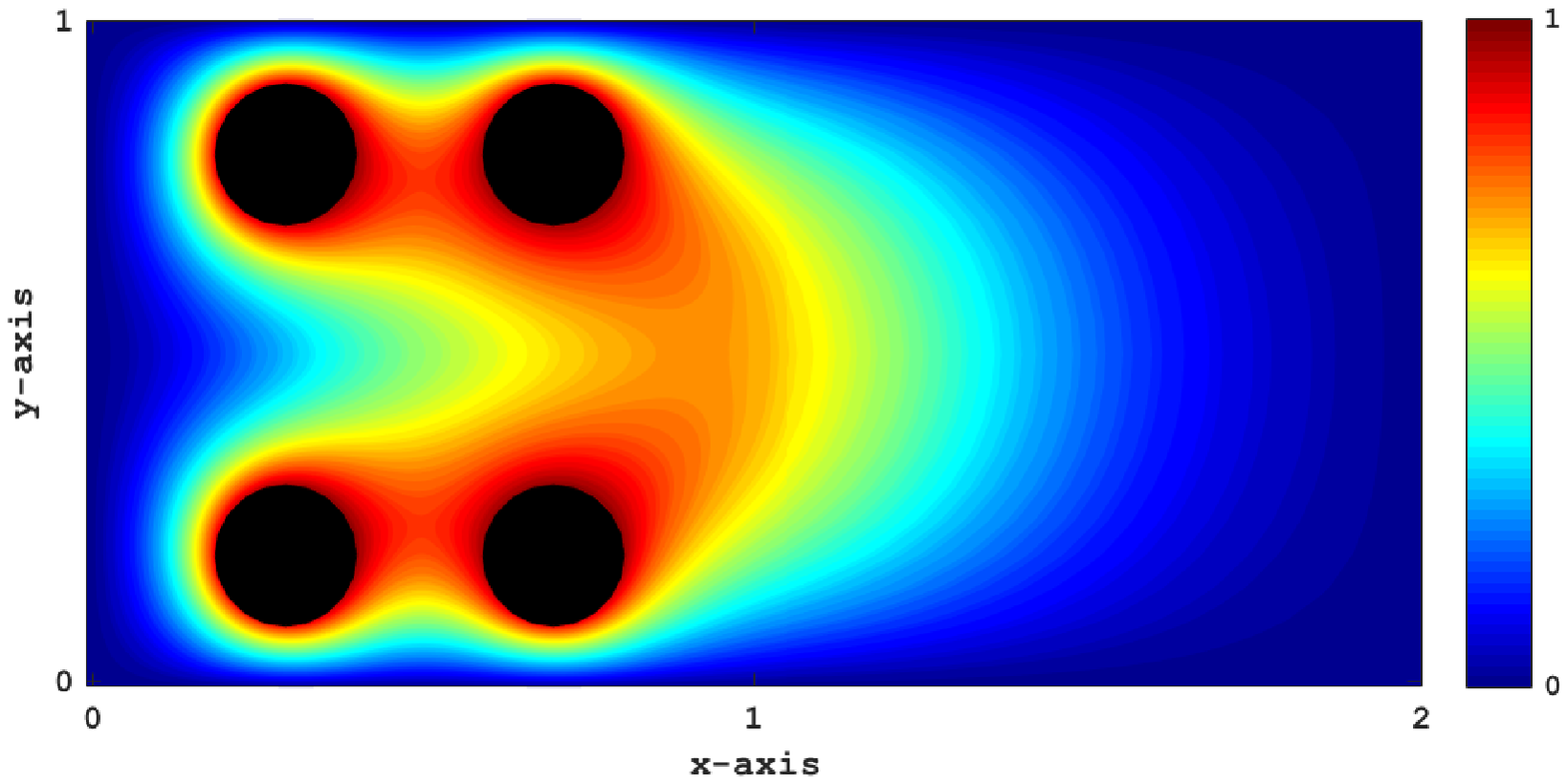}
  }
  \subfigure[$t = 10\Delta t_{\mathrm{c}}$]{
    \includegraphics[clip, scale=0.55]{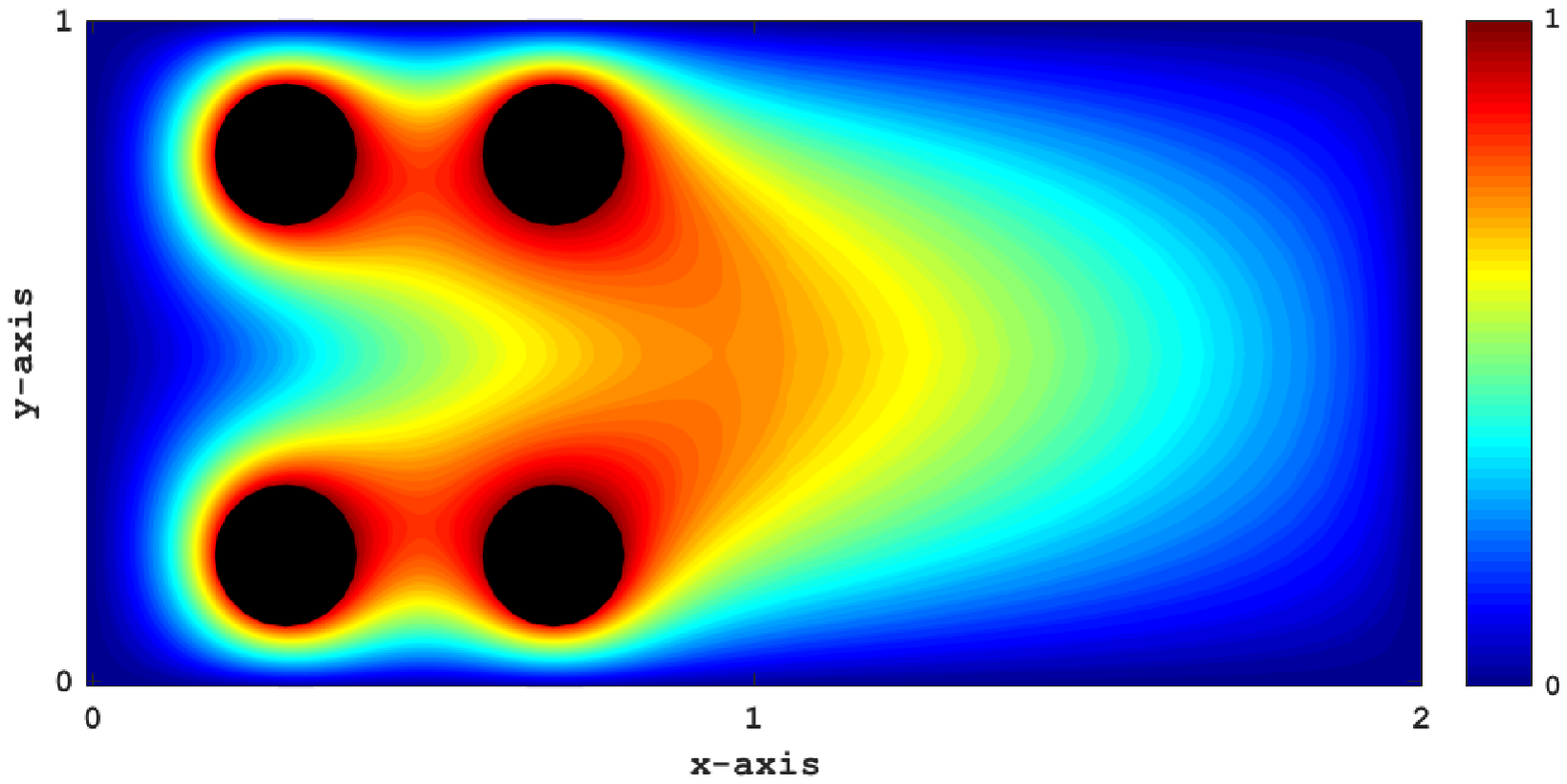}
  }
  \caption{\textsf{Dissolution of calcite}:~In this figure, concentration
  of calcite at different time-levels is shown. Initially, calcite is 
  concentrated near the solid obstacles and is transported throughout
  the domain at later times.
	\label{Fig:2D_Carbon_U_1}}
\end{figure}

\begin{figure}
	\centering
	\subfigure[$t = \Delta t_{\mathrm{c}}$]{
		\includegraphics[clip, scale=0.45]{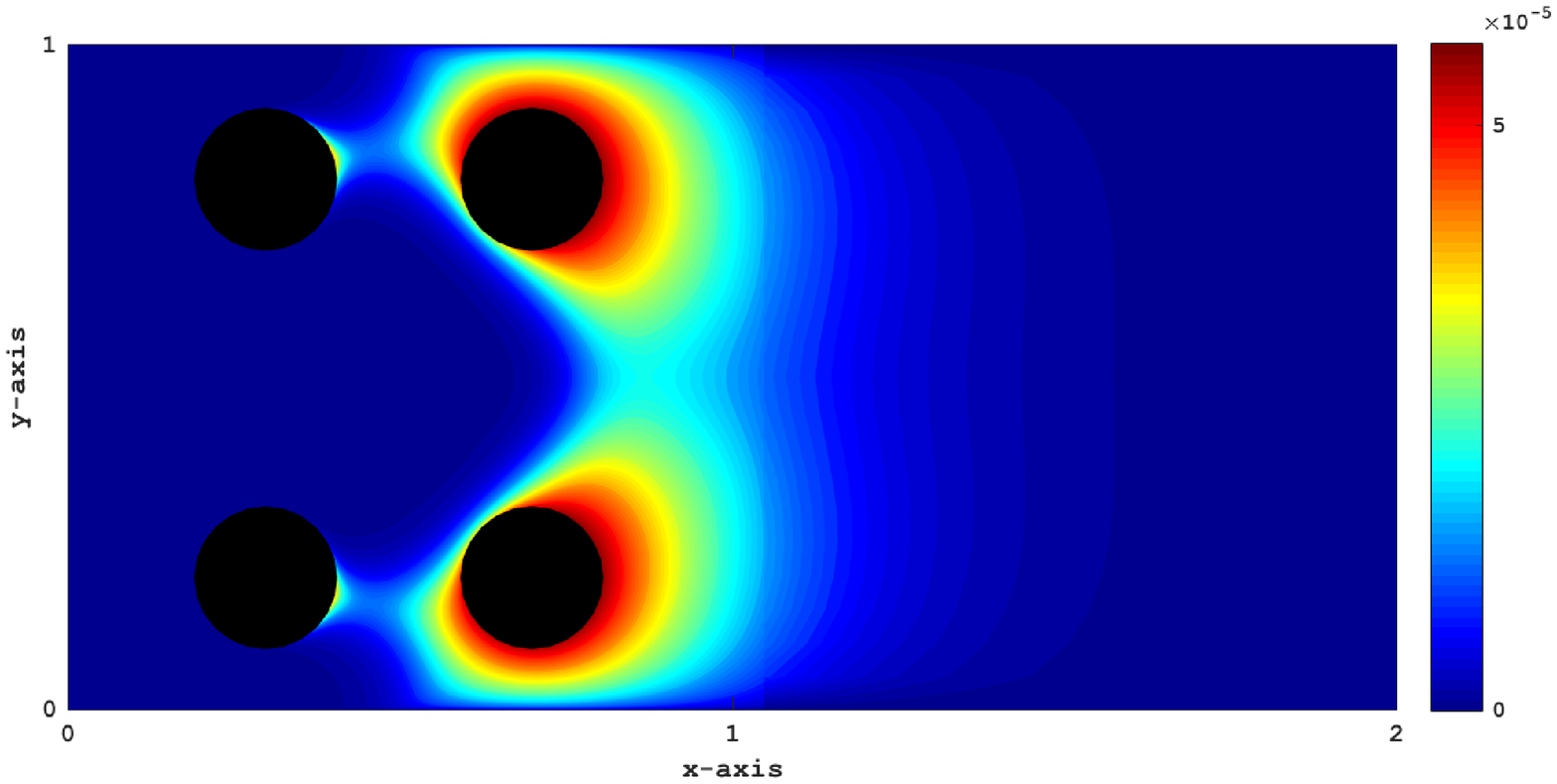}
	}
	\subfigure[$t = 2\Delta t_{\mathrm{c}}$]{
		\includegraphics[clip, scale=0.45]{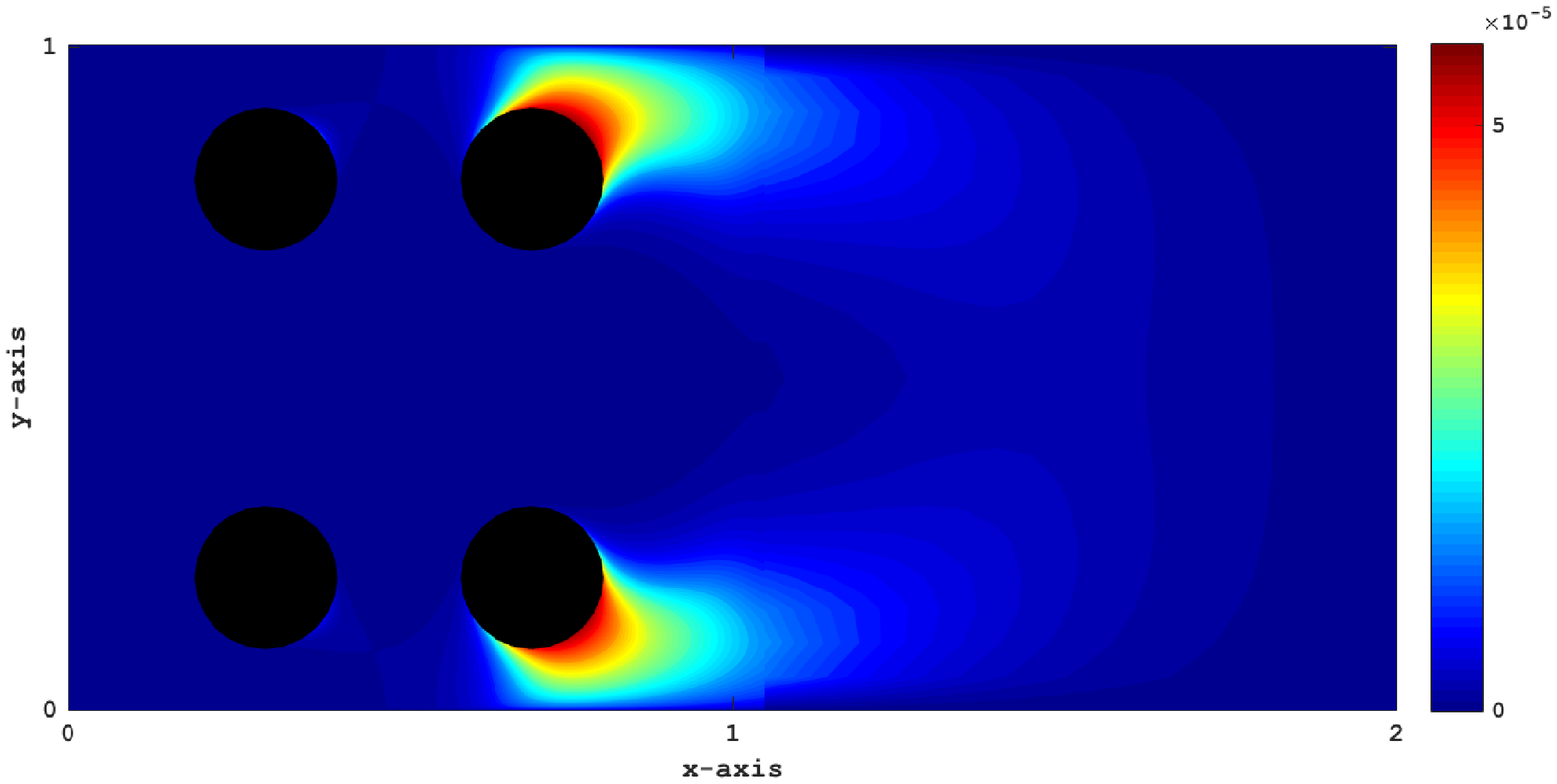}
	}
	\subfigure[$t = 3\Delta t_{\mathrm{c}}$]{
		\includegraphics[clip, scale=0.45]{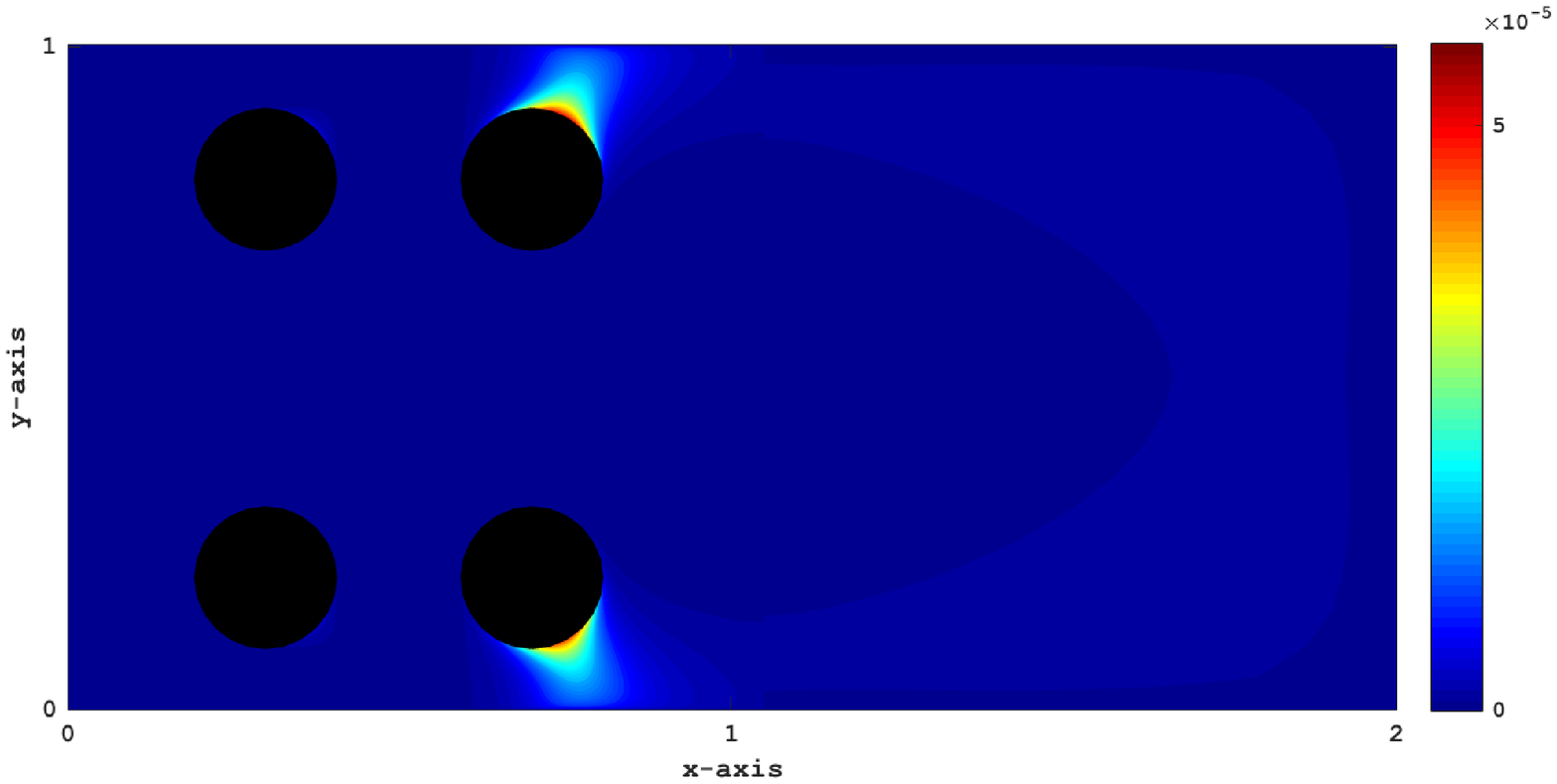}
	}
	\caption{\textsf{Dissolution of calcite}:~In this figure, concentration
	  of $\mathrm{Ca}^{2+}$ is shown. It can be observed that in the first time-steps,
	concentration of this chemical species is more spread out in the 
	spatial domain. At later time-levels, due to stronger presence of other
	participating chemical species, $\mathrm{Ca}^{2+}$ is largely dissolved and 
	precipitates near the solid obstacles (shown as black circles).
	\label{Fig:2D_Carbon_U_2}}
\end{figure}

\begin{figure}
  \centering
  \subfigure[$t = \Delta t_{\mathrm{c}}$]{
    \includegraphics[clip, scale=0.55]{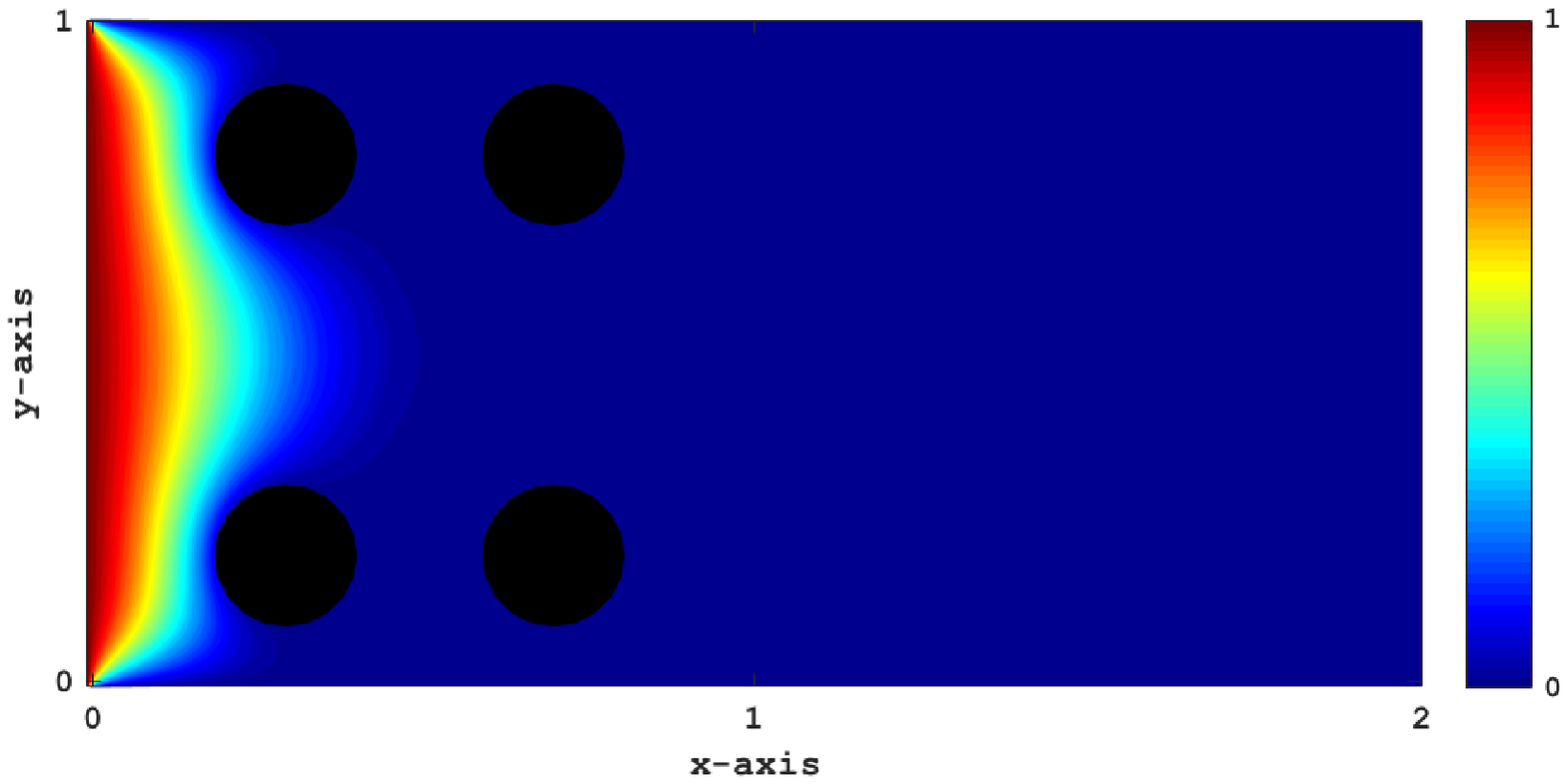}
  }
  \subfigure[$t = 5\Delta t_{\mathrm{c}}$]{
    \includegraphics[clip, scale=0.55]{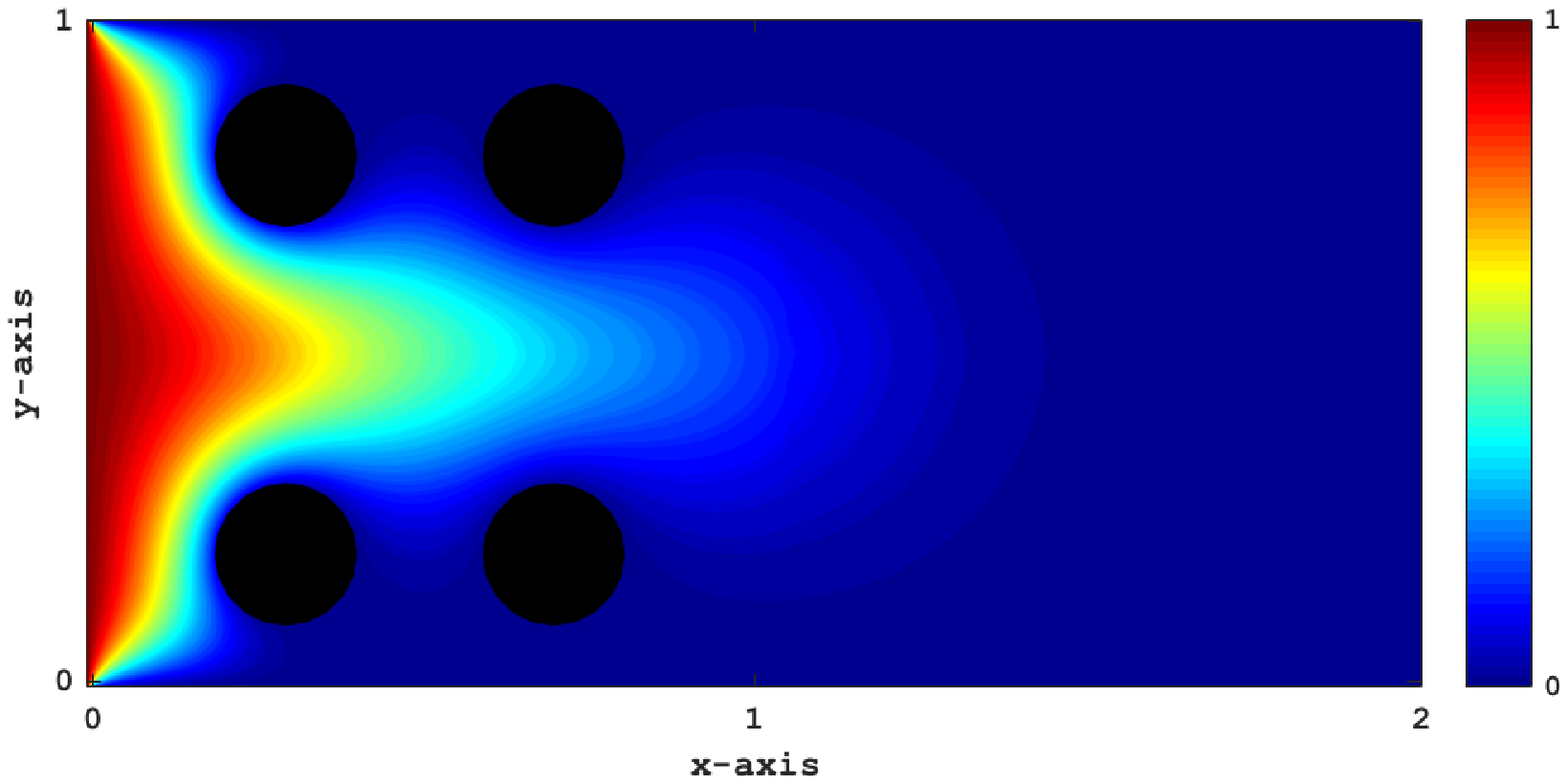}
  }
  \subfigure[$t = 10\Delta t_{\mathrm{c}}$]{
    \includegraphics[clip, scale=0.55]{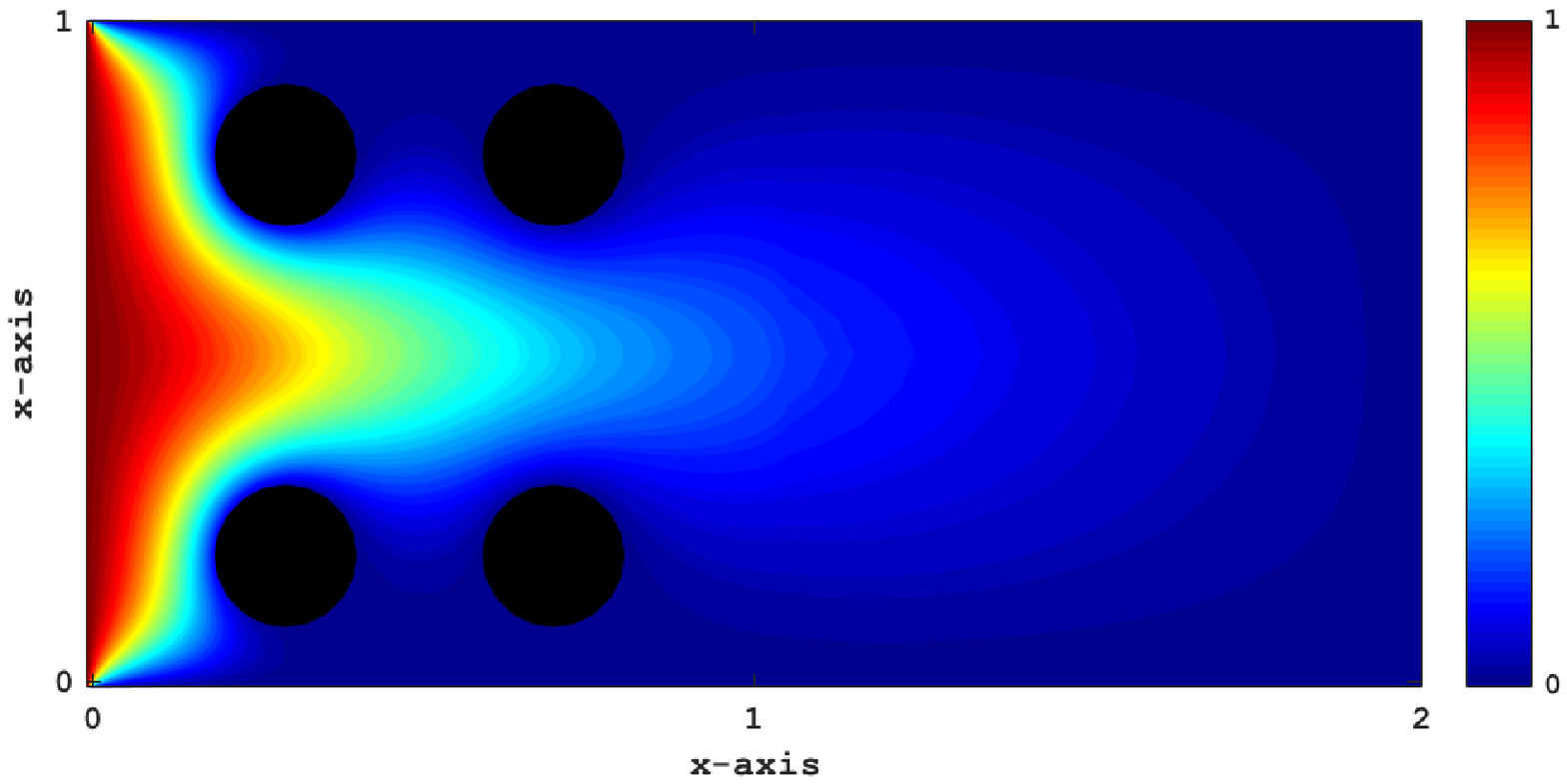}
  }
  \caption{\textsf{Dissolution of calcite}:~Concentration
    of the $\mathrm{CO}_{3}^{2-}$ is shown at different
    time-levels. This chemical species is often in solute
    form. As it can be seen in these figures, the transport
    within the pores is largely hindered due to presence of
    calcite, which is shown in figure
    \ref{Fig:2D_Carbon_U_1}. \label{Fig:2D_Carbon_U_3}}
\end{figure}

\begin{figure}
  \centering
  \psfrag{t}{$t/\Delta t_{\mathrm{c}}$}
  \psfrag{U}{$\mathcal{C}_{\mathrm{total}}(t)$}
  \psfrag{U1}{$\mathrm{CaCO}_{3}$}
  \psfrag{U2}{$\mathrm{Ca}^{2+}$}
  \psfrag{U3}{$\mathrm{CO}_{3}^{2-}$}
  \includegraphics[clip, scale=0.35]{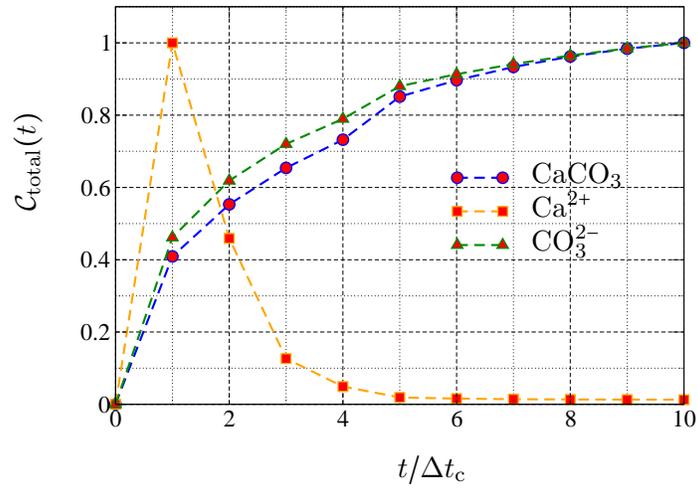}
  \caption{\textsf{Dissolution of calcite}:~This figure shows 
    the variation of the total concentration in the entire domain 
    of each participating chemical species with respect to time. 
    The concentration of $\mathrm{Ca}^{2+}$ decreases as the 
    concentrations of $\mathrm{CaCO}_{3}$ and $\mathrm{CO}_{3}^{2-}$ 
    increase in the domain. The values shown here are normalized 
    with respect to the maximum total concentration of respective 
    species. \label{Fig:IntU}}
\end{figure}

The concentrations of the participating chemical species are
shown in figures \ref{Fig:2D_Carbon_U_1}--\ref{Fig:2D_Carbon_U_3}.
The numerical simulation reveals that the concentrations of
$\mathrm{CaCO}_{3}$ and $\mathrm{CO}_{3}^{2-}$ inside the domain
increase with time. However, the evolution of $\mathrm{Ca}^{2+}$
cations is completely different from that of the other two
chemical species. At earlier time-levels, when the concentrations
of  $\mathrm{CaCO}_{3}$ and $\mathrm{CO}_{3}^{2-}$ are low within
the domain, $\mathrm{Ca}^{2+}$ has a more noticeable presence
throughout the domain. At later time-levels, as a consequence
of increasing concentration of $\mathrm{CO}_{3}^{2-}$ anions,
$\mathrm{Ca}^{2+}$ disappears from much of the domain and
gathers in the regions where the concentration of
$\mathrm{CO}_{3}^{2-}$ is low.
Figure \ref{Fig:IntU} further corroborates this finding,
in which the normalized total concentrations of chemical
species are plotted against time.
The total concentration in the entire domain,
$\mathcal{C}_{\mathrm{total}}$, is defined as 
\begin{align}
  \mathcal{C}_{\mathrm{total}}(t) = \int_{\Omega} u_{i}(\boldsymbol{x},t) \; \mathrm{d}\Omega ,
  \quad i=1,2,3
\end{align}
The normalization for each chemical species is
done with respect to the corresponding maximum
in the time interval of interest. That is,  
\begin{align}
  \max_{t} \; \mathcal{C}_{\mathrm{total}}(t)
\end{align}

In this example, we have demonstrated how to use the
proposed multi-time-step hybrid coupling framework for
the analysis of geochemical processes by simultaneously
incorporating both pore and continuum models. A detailed
pore geometry and complex transport processes can be
accounted for in the fine-scale domain, whereas a rough
approximation can be sought in the coarse-scale domain.

\section{CONCLUDING REMARKS}
\label{Sec:Hybrid_CR}
Simulation of transport of chemical species in porous media 
poses several challenges. These include disparate mathematical 
scales in space and time, not all the essential physical and
chemical processes can be upscaled from the pore-scale to the
meso-scale, high computational cost to solve realistic problem; 
just to name a few. In this paper, we have presented a 
computational framework that can make multi-scale simulation 
of transport in porous media feasible even for realistic problems. 
The framework allows to take into account the features and processes 
at the pore-scale and still be able to solve problems at the field-scale 
with manageable computational cost. The findings and advances 
made in this paper can be listed as follows:
\begin{enumerate}[(i)]
\item \textsf{Simulation of advection and diffusion using LBM}:~
	The lattice Boltzmann method for simulation of transport 
	is outlined. A drawback of LBM in such simulations can be 
	the possibility of discrete distributions attaining unphysical
	(negative) values. To rectify this issue, we presented a 
	bound on discretization parameters under LBM that guarantees 
	non-negativity of discrete populations. Furthermore, new 
	methods for enforcing macroscopic boundary values, in the 
	form of Neumann or Dirichlet conditions, on the numerical
	solution from the LBM are proposed. These methods are based 
	on entropy principles and warrant non-negative values for 
	discrete populations.
\item \textsf{Information transfer across non-matching grids}:~
	Methods for transferring information from one computational
	grid to another non-matching grid were documented. Accuracy
	of these methods with respect to grid size in different 
	domains is also explored.
      \item \textsf{Governing equations for hybrid simulation}:~Time
        and space continuous partial differential equations for
        coupled analysis are presented. These equations provide
        a precise mathematical framework for further developments
        in this area of research.
      \item \textsf{Hybrid coupling computational framework}:~
        A numerical framework, based on domain decomposition,
        was presented that can employ different numerical
        methods (e.g., finite element method and lattice
        Boltzmann method) in different subdomains. This
        framework can account for pore-scale processes
	as well as continuum scale models. Also, disparate spatial
	and temporal discretization can be incorporated. Hence, the 
	primary factor in choosing grid size and time-steps in each 
	subdomain is the accuracy in that subdomain. The hybrid coupling
	framework poses no restriction on the discretization parameters in
	different subdomains. Furthermore, various chemical reaction 
	dynamics among the present chemical species can be included
	using LBM and other approximations of the same phenomena in 
	the finite element solver. In all of the numerical experiments,
	this framework was numerically stable and accurate. Interior 
	layers can be captured accurately and typical weak instabilities
	in the solution can be suppressed using appropriate numerical
	techniques (such as stabilized finite element formulations)
	in those subdomains. We also demonstrated application of this
	framework in assessing the fate of chemical species in a 
	sample geochemical reaction problem.
	As a courtesy of its domain decomposition 
	basis, this framework provides the user with great flexibility in 
	distributing the computational workload onto different 
	processors and possibly in a heterogeneous GPU-CPU 
	computing setup. For instance, the subdomains solved using
	the lattice Boltzmann method can be transferred to a 
	GPU, while other subdomains where the finite element method
	is used can be solved for using a different processing 
	environment. This computational framework can handle 
        multiple subdomains using the multiplicative Schwartz 
        methods.
\end{enumerate}

We shall conclude the paper by outlining some possible
future research directions. 
\begin{enumerate}[(R1)]
\item A good research endeavor can be towards a comprehensive 
mathematical analysis (i.e., stability, accuracy and convergence 
properties) of the proposed computational framework.
\item One can implement the proposed computational framework
  in a combined GPU-CPU computing environment, and study
  the numerical performance of such an implementation.
  %
  %
\item Substantial progress in development of hybrid methods
  can result from extension of the proposed computational
  framework to fully coupled thermal-flow-transport processes,
  including precipitation at the solid-fluid interface (in
  pore-scale) and application of such methods to simulation
  of viscous fingering and other physical instabilities.
\end{enumerate}

\bibliographystyle{plainnat}
\bibliography{References}

\begin{thebibliography}{76}
\providecommand{\natexlab}[1]{#1}
\providecommand{\url}[1]{\texttt{#1}}
\expandafter\ifx\csname urlstyle\endcsname\relax
  \providecommand{\doi}[1]{doi: #1}\else
  \providecommand{\doi}{doi: \begingroup \urlstyle{rm}\Url}\fi

\bibitem[Albuquerque et~al.(2004)Albuquerque, Alemani, Chopard, and
  Leone]{2004_Albuquerque_ICCS}
P.~Albuquerque, D.~Alemani, B.~Chopard, and P.~Leone.
\newblock Coupling a lattice {B}oltzmann and a finite difference scheme.
\newblock In \emph{Computational Science-ICCS 2004}, pages 540--547. Springer,
  2004.

\bibitem[Arbogast et~al.(2007)Arbogast, Pencheva, Wheeler, and
  Yotov]{2007_Arbogast_SIAM}
T.~Arbogast, G.~Pencheva, M.~F. Wheeler, and I.~Yotov.
\newblock A multiscale mortar mixed finite element method.
\newblock \emph{Multiscale Modeling \& Simulation}, 6\penalty0 (1):\penalty0
  319--346, 2007.

\bibitem[Astorino et~al.(2014)Astorino, Chouly, and
  Quarteroni]{2014_Astorino_PhDThesis}
M.~Astorino, F.~Chouly, and A.~Quarteroni.
\newblock {A time-parallel framework for coupling finite element and lattice
  {B}oltzmann methods}.
\newblock Research report, {LMB - CMCS - MOX}, 2014.
\newblock URL \url{https://hal.archives-ouvertes.fr/hal-00746942}.

\bibitem[Augustin et~al.(2011)Augustin, Caiazzo, Fiebach, Fuhrmann, John,
  Linke, and Umla]{2011_Augustin_CMAME}
M.~Augustin, A.~Caiazzo, A.~Fiebach, J.~Fuhrmann, V.~John, A.~Linke, and
  R.~Umla.
\newblock An assessment of discretizations for convection-dominated
  convection--diffusion equations.
\newblock \emph{Computer Methods in Applied Mechanics and Engineering},
  200\penalty0 (47):\penalty0 3395--3409, 2011.

\bibitem[Babu{\v{s}}ka and Suri(1987)]{1987_Babuska_SIAM}
I.~Babu{\v{s}}ka and M.~Suri.
\newblock The optimal convergence rate of the p-version of the finite element
  method.
\newblock \emph{SIAM {J}ournal on {N}umerical {A}nalysis}, 24\penalty0
  (4):\penalty0 750--776, 1987.

\bibitem[Balhoff et~al.(2008)Balhoff, Thomas, and Wheeler]{2008_Balhoff_CG}
M.~T. Balhoff, S.~G. Thomas, and M.~F. Wheeler.
\newblock Mortar coupling and upscaling of pore-scale models.
\newblock \emph{Computational {G}eosciences}, 12\penalty0 (1):\penalty0 15--27,
  2008.

\bibitem[Battiato et~al.(2009)Battiato, Tartakovsky, Tartakovsky, and
  Scheibe]{2009_Battiato_AWR}
I.~Battiato, D.~M. Tartakovsky, A.~M. Tartakovsky, and T.~Scheibe.
\newblock On breakdown of macroscopic models of mixing-controlled heterogeneous
  reactions in porous media.
\newblock \emph{Advances in {W}ater {R}esources}, 32\penalty0 (11):\penalty0
  1664--1673, 2009.

\bibitem[Battiato et~al.(2011)Battiato, Tartakovsky, Tartakovsky, and
  Scheibe]{2011_Battiato_AWR}
I.~Battiato, D.~M. Tartakovsky, A.~M. Tartakovsky, and T.~D. Scheibe.
\newblock Hybrid models of reactive transport in porous and fractured media.
\newblock \emph{Advances in Water Resources}, 34\penalty0 (9):\penalty0
  1140--1150, 2011.

\bibitem[Bekri et~al.(1995)Bekri, Thovert, and Adler]{1995_Adler_CES}
S.~Bekri, J.~F. Thovert, and P.~M. Adler.
\newblock Dissolution of porous media.
\newblock \emph{Chemical Engineering Science}, 50\penalty0 (17):\penalty0
  2765--2791, 1995.

\bibitem[Benjamin(2002)]{2002_WaterChem}
M.~M. Benjamin.
\newblock \emph{Water {C}hemistry}.
\newblock Waveland {P}ress {I}nc., {L}ong {G}rove, 2002.

\bibitem[Bhatnagar et~al.(1954)Bhatnagar, Gross, and Krook]{1954_BGK}
P.~L. Bhatnagar, E.~P. Gross, and M.~Krook.
\newblock A model for collision processes in gases. {I}. small amplitude
  processes in charged and neutral one-component systems.
\newblock \emph{Physical Review}, 94\penalty0 (3):\penalty0 511, 1954.

\bibitem[Bridges and Reich(2001)]{2001_Reich_PLA}
T.~Bridges and S.~Reich.
\newblock Multi-symplectic integrators: {N}umerical schemes for {H}amiltonian
  {P}{D}{E}s that conserve symplecticity.
\newblock \emph{Physics {L}etters {A}}, 284\penalty0 (4):\penalty0 184--193,
  2001.

\bibitem[Cai and Saad(1996)]{1996_Cai_NLAA}
X.~C. Cai and Y.~Saad.
\newblock Overlapping domain decomposition algorithms for general sparse
  matrices.
\newblock \emph{Numerical {L}inear {A}lgebra with {A}pplications}, 3\penalty0
  (3):\penalty0 221--237, 1996.

\bibitem[Cercignani(1988)]{1988_Cercignani_book}
C.~Cercignani.
\newblock \emph{The {B}oltzmann {E}quation and its {A}pplications}.
\newblock Springer, {N}ew {Y}ork, 1988.

\bibitem[Cercignani and Lampis(1971)]{1971_Cercignani_Lampis}
C.~Cercignani and M.~Lampis.
\newblock Kinetic models for gas-surface interactions.
\newblock \emph{Transport Theory and Statistical Physics}, 1\penalty0
  (2):\penalty0 101--114, 1971.

\bibitem[Cercignani et~al.(2013)Cercignani, Illner, and
  Pulvirenti]{2013_Cercignani_book}
C.~Cercignani, R.~Illner, and M.~Pulvirenti.
\newblock \emph{The {M}athematical {T}heory of {D}ilute {G}ases}, volume 106.
\newblock Springer {S}cience \& {B}usiness {M}edia, {N}ew {Y}ork, 2013.

\bibitem[Chang et~al.(2015)Chang, Tai, Wang, and Yang]{2015_Chang_SIAM}
H.~Chang, X.~Tai, L.~Wang, and D.~Yang.
\newblock Convergence rate of overlapping domain decomposition methods for the
  {R}udin--{O}sher--{F}atemi model based on a dual formulation.
\newblock \emph{SIAM {J}ournal on {I}maging {S}ciences}, 8\penalty0
  (1):\penalty0 564--591, 2015.

\bibitem[Chen et~al.(2013)Chen, Zhang, and Zhang]{2013_Chen_PRE}
Q.~Chen, X.~Zhang, and J.~Zhang.
\newblock Improved treatments for general boundary conditions in the lattice
  {B}oltzmann method for convection-diffusion and heat transfer processes.
\newblock \emph{Physical {R}eview {E}}, 88\penalty0 (3):\penalty0 033304, 2013.

\bibitem[Chen and Doolen(1998)]{1998_Chen_ARFM}
S.~Chen and G.~D. Doolen.
\newblock Lattice {B}oltzmann method for fluid flows.
\newblock \emph{Annual Review of Fluid Mechanics}, 30\penalty0 (1):\penalty0
  329--364, 1998.

\bibitem[Chesshire and Henshaw(1990)]{1990_Henshaw_JCP}
G.~Chesshire and W.~D. Henshaw.
\newblock Composite overlapping meshes for the solution of partial differential
  equations.
\newblock \emph{Journal of {C}omputational {P}hysics}, 90\penalty0
  (1):\penalty0 1--64, 1990.

\bibitem[Constantinescu and Sandu(2013)]{2013_Sandu_JSciComp}
E.~M. Constantinescu and A.~Sandu.
\newblock Extrapolated multirate methods for differential equations with
  multiple time scales.
\newblock \emph{Journal of {S}cientific {C}omputing}, 56\penalty0 (1):\penalty0
  28--44, 2013.

\bibitem[de~Boer et~al.(2007)de~Boer, Zuijlen, and Bijl]{2007_deBoer_CMAME}
A.~de~Boer, A.~H.~Van Zuijlen, and H.~Bijl.
\newblock Review of coupling methods for non-matching meshes.
\newblock \emph{Computer {M}ethods in {A}pplied {M}echanics and {E}ngineering},
  196\penalty0 (8):\penalty0 1515--1525, 2007.

\bibitem[Dellacherie(2014)]{2014_Dellacherie_AAM}
S.~Dellacherie.
\newblock Construction and analysis of lattice {B}oltzmann methods applied to a
  1{D} convection-diffusion equation.
\newblock \emph{Acta {A}pplicandae {M}athematicae}, 131\penalty0 (1):\penalty0
  69--140, 2014.

\bibitem[Drever(1988)]{1988_Drever}
J.~I. Drever.
\newblock \emph{The {G}eochemistry of {N}atural {W}aters}.
\newblock Prentice {H}all, {N}ew {J}ersey, 1988.

\bibitem[Evans(1998)]{Evans_PDE}
L.~C. Evans.
\newblock \emph{{P}artial {D}ifferential {E}quations}.
\newblock American {M}athematical {S}ociety, Providence, {R}hode {I}sland,
  1998.

\bibitem[Fatt(1956)]{1956_Fatt}
I.~Fatt.
\newblock The network model of porous media.
\newblock \emph{Petroleum {T}ransactions}, 207:\penalty0 144--181, 1956.

\bibitem[Franca et~al.(1992)Franca, Frey, and Hughes]{1992_Franca_CMAME}
L.~P. Franca, S.~L. Frey, and T.~J.~R. Hughes.
\newblock Stabilized finite element methods: I. application to the
  advective-diffusive model.
\newblock \emph{Computer Methods in Applied Mechanics and Engineering},
  95\penalty0 (2):\penalty0 253--276, 1992.

\bibitem[Gramling et~al.(2002)Gramling, Harvey, and Meigs]{2002_Gramling_EST}
C.~M. Gramling, C.~F. Harvey, and L.~C. Meigs.
\newblock Reactive transport in porous media: A comparison of model prediction
  with laboratory visualization.
\newblock \emph{Environmental Science \& Technology}, 36\penalty0
  (11):\penalty0 2508--2514, 2002.

\bibitem[Gresho and Sani(2000)]{2000_Gresho_V1}
P.~M. Gresho and R.~L. Sani.
\newblock \emph{{Incompressible Flow and the Finite Element Method:
  Advection-Diffusion}}, volume~1.
\newblock John Wiley \& Sons, Inc., Chichester, UK, 2000.

\bibitem[G{\"u}nther et~al.(2001)G{\"u}nther, Kvaern{\o}, and
  Rentrop]{2001_Gunther}
M.~G{\"u}nther, A.~Kvaern{\o}, and P.~Rentrop.
\newblock Multirate partitioned {R}unge-{K}utta methods.
\newblock \emph{BIT {N}umerical {M}athematics}, 41\penalty0 (3):\penalty0
  504--514, 2001.

\bibitem[Haslam et~al.(2008)Haslam, Crouch, and Sea{\"\i}d]{2008_Haslam_CMAME}
I.~W. Haslam, R.~S. Crouch, and M.~Sea{\"\i}d.
\newblock Coupled finite element--lattice {B}oltzmann analysis.
\newblock \emph{Computer Methods in Applied Mechanics and Engineering},
  197\penalty0 (51):\penalty0 4505--4511, 2008.

\bibitem[He and Luo(1997)]{1997_He_Luo_PRE}
X.~He and L.~Luo.
\newblock Theory of the lattice {B}oltzmann method: From the {B}oltzmann
  equation to the lattice {B}oltzmann equation.
\newblock \emph{Physical {R}eview {E}}, 56\penalty0 (6):\penalty0 6811, 1997.

\bibitem[Henshaw and Schwendeman(2003)]{2003_Henshaw_JCP}
W.~D. Henshaw and D.~W. Schwendeman.
\newblock An adaptive numerical scheme for high-speed reactive flow on
  overlapping grids.
\newblock \emph{Journal of {C}omputational {P}hysics}, 191\penalty0
  (2):\penalty0 420--447, 2003.

\bibitem[Hou and Wu(1997)]{1997_Hou_JCP}
T.~Y. Hou and X.~H. Wu.
\newblock A multiscale finite element method for elliptic problems in composite
  materials and porous media.
\newblock \emph{Journal of computational physics}, 134\penalty0 (1):\penalty0
  169--189, 1997.

\bibitem[Hughes et~al.(1998)Hughes, Feij{\'o}o, Mazzei, and
  Quincy]{1998_Hughes_CMAME}
T.~J.~R. Hughes, G.~R. Feij{\'o}o, L.~Mazzei, and J.~Quincy.
\newblock The variational multiscale method?a paradigm for computational
  mechanics.
\newblock \emph{Computer {M}ethods in {A}pplied {M}echanics and {E}ngineering},
  166\penalty0 (1):\penalty0 3--24, 1998.

\bibitem[Hundsdrofer and Verwer(2007)]{Hundsdorfer_Verwer}
W.~H. Hundsdrofer and J.~G. Verwer.
\newblock \emph{{Numerical {S}olution of {T}ime-{D}ependent
  {A}dvection-{D}iffusion-{R}eaction {E}quations}}.
\newblock Springer-Verlag, New York, USA, 2007.

\bibitem[Jaiman et~al.(2006)Jaiman, Jiao, Geubelle, and
  Loth]{2006_Geubelle_JCP}
R.~K. Jaiman, X.~Jiao, P.~H. Geubelle, and E.~Loth.
\newblock Conservative load transfer along curved fluid--solid interface with
  non-matching meshes.
\newblock \emph{Journal of {C}omputational {P}hysics}, 218\penalty0
  (1):\penalty0 372--397, 2006.

\bibitem[Jiao and Heath(2004)]{2004_Heath_IJNME}
X.~Jiao and M.~T. Heath.
\newblock Common-refinement-based data transfer between non-matching meshes in
  multiphysics simulations.
\newblock \emph{International {J}ournal for {N}umerical {M}ethods in
  {E}ngineering}, 61\penalty0 (14):\penalty0 2402--2427, 2004.

\bibitem[Karimi and Nakshatrala(2014)]{2014_Karimi_JCP}
S.~Karimi and K.~B. Nakshatrala.
\newblock On multi-time-step monolithic coupling algorithms for elastodynamics.
\newblock \emph{Journal of Computational Physics}, 273:\penalty0 671--705,
  2014.

\bibitem[Karimi and Nakshatrala(2015{\natexlab{a}})]{2015_Karimi_CMAME}
S.~Karimi and K.~B. Nakshatrala.
\newblock A monolithic multi-time-step computational framework for first-order
  transient systems with disparate scales.
\newblock \emph{Computer Methods in Applied Mechanics and Engineering},
  283:\penalty0 419--453, 2015{\natexlab{a}}.

\bibitem[Karimi and Nakshatrala(2015{\natexlab{b}})]{2015_Karimi_LBM}
S.~Karimi and K.~B. Nakshatrala.
\newblock Do current lattice {B}oltzmann methods for diffusion and
  diffusion-type equations respect maximum principles and the non-negative
  constraint?
\newblock \emph{arXiv:1503.08360}, 2015{\natexlab{b}}.

\bibitem[Karlin et~al.(2006)Karlin, Ansumali, Frouzakis, and
  Chikatamarla]{2006_Karlin}
I.~V. Karlin, S.~Ansumali, C.~E. Frouzakis, and S.~S. Chikatamarla.
\newblock Elements of the lattice {B}oltzmann method i: {L}inear advection
  equation.
\newblock \emph{{C}ommunications in {C}omputational {P}hysics}, 1\penalty0
  (4):\penalty0 616--655, 2006.

\bibitem[Lallemand and Luo(2000)]{2000_Lallemand_PRE}
P.~Lallemand and L.~S. Luo.
\newblock Theory of the lattice {B}oltzmann method: {D}ispersion, dissipation,
  isotropy, {G}alilean invariance, and stability.
\newblock \emph{Physical {R}eview {E}}, 61\penalty0 (6):\penalty0 6546, 2000.

\bibitem[Leimkuhler and Reich(2004)]{2004_Reich_book}
B.~Leimkuhler and S.~Reich.
\newblock \emph{Simulating {H}amiltonian {D}ynamics}.
\newblock Cambridge {U}niversity {P}ress, Cambridge, U.K., 2004.

\bibitem[Lew et~al.(2004)Lew, Marsden, Ortiz, and West]{2004_Marsden_IJNME}
A.~Lew, J.~E. Marsden, M.~Ortiz, and M.~West.
\newblock Variational time integrators.
\newblock \emph{International Journal for Numerical Methods in Engineering},
  60\penalty0 (1):\penalty0 153--212, 2004.

\bibitem[Li et~al.(2004)Li, Shock, Zhang, and Chen]{2004_Li_JFM}
Y.~Li, R.~Shock, R.~Zhang, and H.~Chen.
\newblock Numerical study of flow past an impulsively started cylinder by the
  lattice-{B}oltzmann method.
\newblock \emph{Journal of {F}luid {M}echanics}, 519:\penalty0 273--300, 2004.

\bibitem[Lions(1988)]{1988_Lions}
P.~L. Lions.
\newblock On the {S}chwarz alternating method. i.
\newblock In \emph{First {I}nternational {S}ymposium on {D}omain
  {D}ecomposition {M}ethods for {P}artial {D}ifferential {E}quations}, pages
  1--42. Paris, {F}rance, 1988.

\bibitem[Logg et~al.(2012)Logg, Mardal, and Wells]{FENICS}
A.~Logg, K.~Mardal, and G.~Wells.
\newblock \emph{Automated {S}olution of {D}ifferential {E}quations by the
  {F}inite {E}lement {M}ethod:~{T}he {F}{E}ni{C}{S} {B}ook}.
\newblock Springer, {B}erlin {H}eidelberg, 2012.

\bibitem[Mathew(2008)]{Mathew_book}
T.~Mathew.
\newblock \emph{Domain {D}ecomposition {M}ethods for the {N}umerical {S}olution
  of {P}artial {D}ifferential {E}quations}.
\newblock Springer, {B}erlin {H}eidelberg, 2008.

\bibitem[Mehmani and Balhoff(2014)]{2014_Balhoff_MSModSim}
Y.~Mehmani and M.~T. Balhoff.
\newblock Bridging from pore to continuum: A hybrid mortar domain decomposition
  framework for subsurface flow and transport.
\newblock \emph{Multiscale Modeling \& Simulation}, 12\penalty0 (2):\penalty0
  667--693, 2014.

\bibitem[Melenk and Babu{\v{s}}ka(1996)]{1996_Babuska_CMAME}
J.~M. Melenk and I.~Babu{\v{s}}ka.
\newblock The partition of unity finite element method: {B}asic theory and
  applications.
\newblock \emph{Computer {M}ethods in {A}pplied {M}echanics and {E}ngineering},
  139\penalty0 (1):\penalty0 289--314, 1996.

\bibitem[Monaghan(2005)]{2005_Monaghan_RPP}
J.~J. Monaghan.
\newblock Smoothed particle hydrodynamics.
\newblock \emph{Reports on Progress in Physics}, 68\penalty0 (8):\penalty0
  1703, 2005.

\bibitem[Murphy and Ginn(2000)]{2000_Ginn_HJ}
E.~M. Murphy and T.~R. Ginn.
\newblock Modeling microbial processes in porous media.
\newblock \emph{Hydrogeology Journal}, 8\penalty0 (1):\penalty0 142--158, 2000.

\bibitem[Nakshatrala et~al.(2008)Nakshatrala, Hjelmstad, and
  Tortorelli]{2008_Nakshatrala_IJNME}
K.~B. Nakshatrala, K.~D. Hjelmstad, and D.~A. Tortorelli.
\newblock A {F}{E}{T}{I}-based domain decomposition technique for
  time-dependent first-order systems based on a {D}{A}{E} approach.
\newblock \emph{International Journal for Numerical Methods in Engineering},
  75\penalty0 (12):\penalty0 1385--1415, 2008.

\bibitem[Nakshatrala et~al.(2013)Nakshatrala, Mudunuru, and
  Valocchi]{2013_Nakshatrala_JCP_v253_p278}
K.~B. Nakshatrala, M.~K. Mudunuru, and A.~J. Valocchi.
\newblock A numerical framework for diffusion-controlled bimolecular-reactive
  systems to enforce maximum principles and the non-negative constraint.
\newblock \emph{Journal of Computational Physics}, 253:\penalty0 278--307,
  2013.

\bibitem[Nataf and Nier(1997)]{1997_Nataf_NM}
F.~Nataf and F.~Nier.
\newblock Convergence rate of some domain decomposition methods for overlapping
  and nonoverlapping subdomains.
\newblock \emph{Numerische {M}athematik}, 75\penalty0 (3):\penalty0 357--377,
  1997.

\bibitem[Noorden and Pop(2008)]{2008VanNoorden_IMA}
T.~L.~Van Noorden and I.~S. Pop.
\newblock A {S}tefan problem modeling crystal dissolution and precipitation.
\newblock \emph{IMA Journal of Applied Mathematics}, 73\penalty0 (2):\penalty0
  393--411, 2008.

\bibitem[Olsson and Petersson(1996)]{1996_Olsson_CF}
F.~Olsson and N.~A. Petersson.
\newblock Stability of interpolation on overlapping grids.
\newblock \emph{Computers and {F}luids}, 25\penalty0 (6):\penalty0 583--605,
  1996.

\bibitem[Quarteroni and Valli(1999)]{Quarteroni_book}
A.~Quarteroni and A.~Valli.
\newblock \emph{Domain {D}ecomposition {M}ethods for {P}artial {D}ifferential
  {E}quations}.
\newblock Oxford {U}niversity {P}ress, {O}xford, 1999.

\bibitem[Scheibe et~al.(2015)Scheibe, Murphy, Chen, Rice, Carroll, Palmer,
  Tartakovsky, Battiato, and Wood]{2015_Scheib_Groundwater}
T.~D. Scheibe, E.~M. Murphy, X.~Chen, A.~K. Rice, K.~C. Carroll, B.~J. Palmer,
  A.~M. Tartakovsky, I.~Battiato, and B.~D. Wood.
\newblock An analysis platform for multiscale hydrogeologic modeling with
  emphasis on hybrid multiscale methods.
\newblock \emph{Groundwater}, 53\penalty0 (1):\penalty0 38--56, 2015.

\bibitem[Shapiro and Brenner(1986)]{1986_Shapiro_CES}
M.~Shapiro and H.~Brenner.
\newblock Taylor dispersion of chemically reactive species: irreversible
  first-order reactions in bulk and on boundaries.
\newblock \emph{Chemical Engineering Science}, 41\penalty0 (6):\penalty0
  1417--1433, 1986.

\bibitem[Skordos(1993)]{1993_Skordos_PRE}
P.~A. Skordos.
\newblock Initial and boundary conditions for the lattice {B}oltzmann method.
\newblock \emph{Physical {R}eview {E}}, 48\penalty0 (6):\penalty0 4823, 1993.

\bibitem[Succi(2001)]{2001_Succi_LBM}
S.~Succi.
\newblock \emph{The Lattice {B}oltzmann {E}quation}.
\newblock Oxford {U}niversity {P}ress, {O}xford, 2001.

\bibitem[Sun et~al.(2012)Sun, Mehmani, and Balhoff]{2012_Sun_EaF}
T.~Sun, Y.~Mehmani, and M.~T. Balhoff.
\newblock Hybrid multiscale modeling through direct substitution of pore-scale
  models into near-well reservoir simulators.
\newblock \emph{Energy \& Fuels}, 26\penalty0 (9):\penalty0 5828--5836, 2012.

\bibitem[Tang et~al.(2015)Tang, Valocchi, and Werth]{2015_Tang_WRR}
Y.~Tang, A.~J. Valocchi, and C.~J. Werth.
\newblock A hybrid pore-scale and continuum-scale model for solute diffusion,
  reaction, and biofilm development in porous media.
\newblock \emph{Water {R}esources {R}esearch}, 51\penalty0 (3):\penalty0
  1846--1859, 2015.

\bibitem[Tartakovsky et~al.(2008)Tartakovsky, Tartakovsky, Scheibe, and
  Meakin]{2008_Tartakovsky_SIAM}
A.~M. Tartakovsky, D.~M. Tartakovsky, T.~D. Scheibe, and P.~Meakin.
\newblock Hybrid simulations of reaction-diffusion systems in porous media.
\newblock \emph{SIAM Journal on Scientific Computing}, 30\penalty0
  (6):\penalty0 2799--2816, 2008.

\bibitem[Toselli and Widlund(2005)]{Toselli_book}
A.~Toselli and O.~Widlund.
\newblock \emph{Domain {D}ecomposition {M}ethods: {A}lgorithms and {T}heory},
  volume~3.
\newblock Springer, {H}eidelberg, 2005.

\bibitem[van~der Walt et~al.(2011)van~der Walt, Colbert, and Varoquaux]{Numpy}
S.~van~der Walt, S.~C. Colbert, and G.~Varoquaux.
\newblock The {N}um{P}y array: A structure for efficient numerical computation.
\newblock \emph{Computing in {S}cience \& {E}ngineering}, 13:\penalty0 22--30,
  2011.

\bibitem[Wohlmuth(2000)]{2000_Wohlmuth_SIAM}
B.~I. Wohlmuth.
\newblock A mortar finite element method using dual spaces for the {L}agrange
  multiplier.
\newblock \emph{SIAM Journal on Numerical Analysis}, 38\penalty0 (3):\penalty0
  989--1012, 2000.

\bibitem[Wood et~al.(2007)Wood, Radakovich, and Golfier]{2007_Wood_AWR}
B.~D. Wood, K.~Radakovich, and F.~Golfier.
\newblock Effective reaction at a fluid--solid interface: applications to
  biotransformation in porous media.
\newblock \emph{Advances in Water Resources}, 30\penalty0 (6):\penalty0
  1630--1647, 2007.

\bibitem[Wood(1990)]{1990_Wood}
W.~L. Wood.
\newblock \emph{Practical {T}ime-stepping {S}chemes}.
\newblock Clarendon {P}ress, {O}xford, 1990.

\bibitem[Xian and Takayuki(2011)]{2011_Wnag_PC}
W.~Xian and A.~Takayuki.
\newblock Multi-{G}{P}{U} performance of incompressible flow computation by
  lattice {B}oltzmann method on {G}{P}{U} cluster.
\newblock \emph{Parallel {C}omputing}, 37\penalty0 (9):\penalty0 521--535,
  2011.

\bibitem[Yu et~al.(2003)Yu, Mei, Luo, and Shyy]{2003_Yu_PAS}
D.~Yu, R.~Mei, L.~Luo, and W.~Shyy.
\newblock Viscous flow computations with the method of lattice {B}oltzmann
  equation.
\newblock \emph{Progress in {A}erospace {S}ciences}, 39\penalty0 (5):\penalty0
  329--367, 2003.

\bibitem[Ziegler(1993)]{1993_Ziegler_JSP}
D.~P. Ziegler.
\newblock Boundary conditions for lattice {B}oltzmann simulations.
\newblock \emph{Journal of {S}tatistical {P}hysics}, 71:\penalty0 1171--1177,
  1993.

\bibitem[Zienkiewicz and Taylor(1977)]{1977_Zienkiewicz_Book}
O.~C. Zienkiewicz and R.~L. Taylor.
\newblock \emph{The {F}inite {E}lement {M}ethod}.
\newblock McGraw-{H}ill, {L}ondon, 1977.

\bibitem[Zou and He(1997)]{1997_Zou_PF}
Q.~Zou and X.~He.
\newblock On pressure and velocity boundary conditions for the lattice
  {B}oltzmann {B}{G}{K} model.
\newblock \emph{Physics of {F}luids}, 9\penalty0 (6):\penalty0 1591--1598,
  1997.

\end{thebibliography}

\end{document}